\documentclass[11pt]{elsarticle}
\usepackage{epsfig,graphics}
\usepackage{amssymb,amsmath,amsthm,bm, amscd, amsfonts}
\usepackage{booktabs}  
\usepackage{threeparttable}  
\usepackage{multirow}
\usepackage{mathabx}

\usepackage{float}
\biboptions{numbers,sort&compress}

\newtheorem{pro}{Proposition}

\newtheorem{mdef}{Definition}

\newtheorem{alg}{Algorithm}

\def\ZZ{{\mathbb Z}}
\def\RR{{\mathbb R}}

\def\CC{{\mathbb C}}

\def\R{{\mathbb R}}

\def\CC{{\mathbb C}}

\def\gga{\gamma}
\def\gth{\theta}

\def\gO{\Omega}
\def\gT{\Theta}

\def\gd{\delta}

\def\gs{\sigma}
\def\gl{\lambda}
\def\gL{\Lambda}
\def\wt{\widetilde}

\def\ep{\epsilon}
\def\vep{\varepsilon}

\def\gt{\triangle}
\def\gD{\Delta}
\def\gp{{\prime}}
\def\b0{{\bf 0}}
\def\1{{\bf 1}}

\def\cH{\mathcal H}

\def\cT{\mathcal T}
\def\cU{\mathcal U}

\def\wh{\widehat}
\def\wt{\widetilde}

\makeatletter
\def\widebreve{\mathpalette\wide@breve}
\def\wide@breve#1#2{\sbox\z@{$#1#2$}%
     \mathop{\vbox{\m@th\ialign{##\crcr
\kern0.08em\brevefill#1{0.99\wd\z@}\crcr\noalign{\nointerlineskip}%
                    $\hss#1#2\hss$\crcr}}}\limits}
\def\brevefill#1#2{$\m@th\sbox\tw@{$#1($}%
  \hss\resizebox{#2}{\wd\tw@}{\rotatebox[origin=c]{90}{\upshape(}}\hss$}
\makeatletter
\def\wb{\widebreve}

\begin{document}

\begin{frontmatter}

\title{Synchrosqueezed  X-Ray Wavelet-Chirplet Transform\\
for Accurate Chirp Rate Estimation and Retrieval of Modes from Multicomponent Signals with Crossover  Instantaneous Frequencies
}
%\thanks{This work was partially supported by the National Key Research and Development Program of China (No. 2022YFA1005703) and the National Natural Science Foundation of China (No. U21A20455)} }

\author[university1]{Qingtang Jiang}
\ead{jiangq@zjnu.edu.cn}
%\author[university1]{Shuixin Li\corref{corresponding}}
\author[university1]{Shuixin Li}
\ead{lishuixin@zjnu.edu.cn}
\author[university1]{Jiecheng Chen}
\ead{jcchen@zjnu.cn}
\author[university2]{Lin Li}
\ead{lilin@xidian.edu.cn}

%\cortext[corresponding]{Corresponding author.}

\address[university1]{School of Mathematical Sciences, Zhejiang Normal University, Jinhua 321004, China}
\address[university2]{School of Electronic Engineering, Xidian University, Xi'an 710071, China}

%\vskip -0.5cm

%% Abstract
\begin{abstract}
Recent advances in the chirplet transform and wavelet-chirplet transform (WCT) have enabled the estimation of instantaneous frequencies (IFs) and chirprates, as well as mode retrieval from multicomponent signals with crossover IF curves. However, chirprate estimation via these approaches remains less accurate than IF estimation, primarily due to the slow decay of the chirplet transform or WCT along the chirprate direction. To address this, the synchrosqueezed chirplet transform (SCT) and multiple SCT methods were proposed, achieving moderate improvements in IF and chirprate estimation accuracy. Nevertheless, a novel approach is still needed to enhance the transform’s decay along the chirprate direction.

This paper introduces an X-ray transform–based wavelet-chirprate transform, termed the X-ray wavelet-chirplet transform (XWCT), which exhibits superior decay along the chirprate direction compared to the WCT. Furthermore, third-order synchrosqueezed variants of the WCT and XWCT are developed to yield sharp time–frequency–chirprate representations of signals. Experimental results demonstrate that the XWCT achieves significantly faster decay along the chirprate axis, while the third-order synchrosqueezed XWCT enables accurate IF and chirprate estimation, as well as mode retrieval, without requiring multiple synchrosqueezing operations.   

\end{abstract}

%% Keywords
\end{frontmatter}

\section{Introduction}

Many real-world signals are represented as a superposition of locally band-limited, amplitude- and frequency-modulated oscillatory modes:
\begin{equation}
\label{AHM0}
x(t) = A_0(t) + \sum_{k=1}^K x_k(t), \quad x_k(t) = A_k(t)\cos\big(2\pi \phi_k(t)\big),
\end{equation}
where \(A_k(t) > 0\) and \(\phi_k'(t) > 0\) are functions with \(A_k(t)\) evolving slowly. To analyze phenomena governed by \eqref{AHM0}, recovering the component modes \(x_k(t)\) is essential.

The empirical mode decomposition (EMD) \cite{huang1998empirical} is a widely used method for decomposing non-stationary signals in the form of \eqref{AHM0}. Time-frequency analysis offers an alternative approach: the synchrosqueezing transform (SST) \cite{daubechies2011synchrosqueezed} rigorously recovers components of multicomponent signals, with extensions to short-time Fourier transform (STFT)-based variants \cite{thakur2011synchrosqueezing,oberlin2014fourier}. Since \cite{daubechies2011synchrosqueezed}, numerous SST variants have been developed, including second-order SST \cite{oberlin2015second,oberlin2017second,behera2018theoretical,lu2021second}, higher-order SST \cite{pham2017high}, synchroextracting transform (SET) \cite{yu2017synchroextracting}, multiple-SST (MSST) \cite{yu2018multisynchrosqueezing}, demodulated SST \cite{li2012generalized,wang2013matching,meignen2017demodulation,jiang2017instantaneous,wang2018matching,jiang2022instantaneous}, and adaptive SST with parameter-varying window functions \cite{sheu2017entropy,berrian2017adaptive,li2020adaptive,li2020adaptivestft,li2018time,lu2020analysis,cai2021analysis}. Fractional lower-order SST for instantaneous frequency (IF) estimation has also been studied \cite{long2017applications,li2022synchrosqueezing}. SST mitigates limitations of EMD \cite{meignen2012new,auger2013time} and finds many applications including machine fault diagnosis (refer to the references in \cite{snyder2025integrating})
and medical data analysis (see \cite{wu2020current} and references therein). Additionally, the signal separation operator (SSO) scheme \cite{chui2016signal} and linear chirp-based SSO models \cite{li2022direct,chui2021analysis,chui2021signal} offer direct time-frequency separation approaches.

While SST and its variants provide sharp time-frequency representations and capture IF dynamics, they struggle to resolve transient features. To address this, time-reassigned SST (TSST) was proposed \cite{he2019time,fourer2019second,he2020gaussian}. For complex signals containing both harmonic and impulsive components, recent techniques have emerged, including MSST combined with time-assignment MSST \cite{dong2023time}, bidirectional SST \cite{ma2024high}, synchroextracting with transient-extracting transforms \cite{ma2024synchro}, and IF equation-based methods \cite{zhu2025if}.

Accurate component recovery using methods such as EMD, SST, SET, MSST, SSO, TSST, and the recent techniques described in \cite{dong2023time, ma2024high, ma2024synchro, zhu2025if} typically requires well-separated instantaneous frequencies (IFs) in the time-frequency plane. However, certain real-world signals exhibit overlapping time-frequency representations, where the IF curves of components cross one another. For instance, in radar signal processing, micro-Doppler effects—generated by target micro-motions (e.g., mechanical vibrations, rotations, tumbling, or coning motions)—produce highly non-stationary signals with intersecting frequency curves \cite{chen2006micro, stankovic2013compressive}.

To recover components with crossover instantaneous frequencies (IFs), the SSO scheme has been extended from the 2D time-frequency (or time-scale respectively) plane to 3D spaces of time–frequency–
chirprate (or time–scale–chirprate respectively) using the chirplet transform (CT) \cite{li2022chirplet, chui2023analysis} 
(time-scale-chirprate transform \cite{chui2021time} respectively). The time–scale–chirprate transform is herein referred to as the wavelet-chirplet transform (WCT) for simplicity, and ``chirprate" is used interchangeably with ``chirp rate" throughout this paper.

As far as the CT for time-frequency analysis is concerned, it has been used in \cite{yu2016general, abratkiewicz2020double} for sharp time-frequency representations and \cite{zhu2019multiple} proposed a (one-direction) multiple squeezed CT with the squeeze acting along the frequency direction only and treating the chirprate as a parameter to be selected so that the resulting CT has the largest amplitude. 
The three-dimensional extracting transform, introduced in \cite{zhu2021three} generates a sharp 3D representation of a signal using the CT, facilitating accurate IF and chirprate estimation. In addition, several other approaches have been put forward recently for estimating IF and/or retrieving the modes of multicomponent signals with overlapping IFs. For instance, in \cite{bruni2020radon}, time-frequency analysis and the Radon transform were employed, while \cite{bruni2021pde} introduced a spectrogram time-frequency evolution law for estimating both IF and the chirprate. Additionally, \cite{stankovic2013compressive} developed a compressive sensing approach, and \cite{chen2017nonlinear} proposed a variational nonlinear chirp mode decomposition method for mode retrieval. Readers interested in other relevant methods are directed to refer to \cite{li2022chirplet}.

The CT-based and WCT-based SSO methods proposed in \cite{li2022chirplet, chui2023analysis, chui2021time} offer a significantly more effective approach for recovering components with crossover instantaneous frequencies (IFs) compared to EMD, SST, and conventional SSO. However, further improvements are needed: component recovery accuracy hinges on the precision of IF and chirprate estimation, yet chirprate estimates lag behind IF estimates due to the slow decay of the CT along the chirprate direction. To address this, synchrosqueezed CT (SCT) \cite{zhu2020frequency, chen2023disentangling}—which squeezes along both frequency and chirprate axes—was developed to enhance IF and chirprate estimation.More recently, \cite{chen2024multiple} proposed multiple synchrosqueezed CT (MSCT) to improve chirprate estimation. Motivated by decay analysis in \cite{chen2023disentangling}, this approach employs two distinct window functions: a Gaussian window \(g_\sigma(t)\) for IF reference functions and \(t^2 g_\sigma(t)\) for chirprate reference functions.   Furthermore,   authors in \cite{chen2024composite} proposed to use the 3rd-order IF reference function, but somehow they did  not use the 3rd-order chirprate reference function though they provide a formula for it there. This is probably due to the complicated expression of  the 3rd-order chirprate reference function obtained there. However, for some cases  MSCT or the multiple synchrosqueezed WCT does not provide desirable IFs and chirprate estimates. 

This paper proposes an X-ray transform–based wavelet-chirplet transform, termed the X-ray wavelet-chirplet transform (XWCT), which exhibits faster decay along the chirprate direction than WCT. Additionally, third-order synchrosqueezed variants of WCT and XWCT are developed, featuring a third-order chirprate reference function simplified compared to that in \cite{chen2024composite}. Experimental results demonstrate that XWCT achieves significant decay along the chirprate axis, while the third-order synchrosqueezed XWCT enables accurate IF and chirprate estimation, as well as mode retrieval, without requiring multiple synchrosqueezing operations.

The remainder of this paper is structured as follows. Section 2 briefly reviews the WCT-based scheme for recovering signal components with crossover IFs. Section 3 proposes the X-ray wavelet-chirplet transform (XWCT). Section 4 formulates the third-order synchrosqueezed WCT and XWCT. Section 5 presents algorithm implementations and numerical experimental results. Finally, Section 6 summarizes the conclusions and outlines potential directions for future work.

For simplicity and without loss of generality, the following complex-form version of \eqref{AHM0}—omitting the trend term \(A_0(t)\)—is adopted throughout the remainder of this paper:
\begin{equation}
\label{AHM}
x(t) = \sum_{k=1}^K x_k(t) = \sum_{k=1}^K A_k(t) e^{i2\pi\phi_k(t)},
\end{equation}
where \(A_k(t) > 0\) and \(\phi_k'(t) > 0\).

\section {Wavelet-chirplet transform for signal mode retrieval}
EMD, SST, SET, MSST, and SSO 
require components of $x(t)$ in \eqref{AHM} 
be well-separated in the time-frequency plane:   
 
\begin{equation}
\label{freq_resolution}
\frac{\phi_k'(t)-\phi'_{k-1}(t)}{\phi_k'(t)+\phi'_{k-1}(t)}\ge \gt, 
\; \hbox{or} \;  \phi^\gp_k(t)-\phi^\gp_{k-1}(t)\ge 2\gt,
 t\in \R, \; 2\le k\le K. 
\end{equation}
\eqref{freq_resolution} is  called the well-separated condition with resolution $\gt$. 

As mentioned above, some real-world signals are overlapping in the time-frequency plane.
Two signal components $x_k(t)$ and $x_\ell(t)$ of a multicomponent signal $x(t)$ governed by \eqref{AHM0} or \eqref{AHM} are said to overlap in the time-frequency plane at $t=t_0$ or their IFs are crossover at $t=t_0$, provided that 
$$
\phi_k^\gp(t_0) = \phi_\ell^\gp(t_0),  \quad \phi_k^\gp(t)\not = \phi_\ell^\gp(t), t\in (t_0-c_1, t_0)\cup (t_0, t_0+c_2)
$$
for some  positive numbers $c_1, c_2$, which may depend on $t_0$. 

To extract and separate the (unknown) signal components with crossover IFs from the multicomponent signal governed by \eqref{AHM0} or  \eqref{AHM}, \cite{chui2021time} proposed 
the following transform:
	\begin{eqnarray}
\label{def_TSC0}
	U^g_x(a, b, \gl) \hskip -0.6cm  &&:= \int_{-\infty}^\infty x(t) \frac 1{a}g\big(\frac{t-b}{a}\big)
	e^{-i2\pi \mu  \frac{t-b}{a}}
	  e^{ -i\pi \lambda (t-b)^2} dt\\
\label{def_TSC}	&&= \int_{\RR} x(b+at)  g(t) e^{-i2\pi \mu  t -i\pi \gl a^2 t^2}dt,
	\end{eqnarray}
where $g(t)$ is a  window function, $\mu$ is a positive constant.   
When $\lambda=0$,  $U^g_x(a, b, \gl)$ is reduced to the continuous wavelet(-like) transform of $x(t)$ with the Morlet(-like) wavelet $\psi(t)=\overline {g(t)} e^{i2\pi \mu  t}$ given by 
$$
W_x(a, b)= \langle x, \psi_{a, b}\rangle =
\int_{-\infty}^\infty x(t) \overline{\psi_{a, b}(t)}dt, 
$$
where $\psi_{a,b}(t)=\frac 1a \psi\big(\frac {t-b}a\big)$.  Denote
\begin{equation}
\label{def_chirplet}
\psi^\gl(t):=\psi(t) e^{i\pi \gl t^2}=\overline{g(t)} e^{i2\pi \mu  t} e^{i\pi \gl t^2}. 
\end{equation}
Then $U^g_x(a, b, \gl)$ is essentially the CWT of  $x(t)$ with 
$\psi^\gl(t)$:
\begin{eqnarray}
 \nonumber 
	\wt U^g_x(a, b, \gl)\hskip -0.6cm  &&:= \langle x, \psi^\gl_{a, b}\rangle =
\int_{-\infty}^\infty x(t) \overline{\psi^{\gl}_{a, b}(t)}dt\\
&&\label{def_TSC_orig} =\int_{-\infty}^\infty x(t) \frac 1{a}g \big(\frac{t-b}{a}\big)
	e^{-i2\pi \mu  \frac{t-b}{a}}
	  e^{ -i\pi \lambda (\frac {t-b}a)^2} dt. 
\end{eqnarray}
%Compared with $\wt U^g_x(a, b, \gl)$, 
Note that $e^{ -i\pi \lambda ({t-b})^2}$ is used in \eqref{def_TSC0} for $U^g_x(a, b, \gl)$ instead of  $e^{ -i\pi \lambda (\frac {t-b}a)^2}$ as in \eqref{def_TSC_orig} for $\wt U^g_x(a, b, \gl)$. Overall there is no significant differenece between  $U^g_x(a, b, \gl)$ and  $\wt U^g_x(a, b, \gl)$ either in the theoretical derivation or applications. The variable $\gl$ in $U^g_x(a, b, \gl)$ will provide a more direct relation with the chirprate of $x(t)$ than $\wt U^g_x(a, b, \gl)$, see \cite{chui2021time} or \eqref{approx00} below. 

A chirp ``wavelet'' $\psi^\gl(t)$ in \eqref{def_chirplet}  was coined by Mann and Haykin the term ``chirplet'' in \cite{mann1991chirplet} and the transform $\wt U^g_x(a, b, \gl)$ was called the chirplet transform, which a special case of the chirplet transform with more parameters further introduced by them in \cite{mann1992adaptive, mann1995chirplet}. Another special type of chirplet transform which is associated with the STFT is called the Guassian chirplet transform in \cite{mann1995chirplet}:
\begin{eqnarray}
 \label{def_CT}
% \begin{array}{ll}
 Q_x(t,\eta,\lambda)\hskip -0.6cm && := \int_{\RR} x(\tau) g\big(\tau-t\big) e^{-i2\pi\eta(\tau-t) -i\pi \lambda (\tau-t)^2} d\tau, 
 \end{eqnarray}
where $g(t)$ is a window function. Note that when$\gl=0$, $Q_x(t, \eta, \gl)$  is reduced 
to the (modified) STFT. 

Nowadays, compared with  $\wt U^g_x(a, b, \gl)$ or $U^g_x(a, b, \gl)$, $Q_x(t, \eta, \lambda)$ is 
more commonly used and it is often called the chireplet transform (CT) in the literature. 
$U^g_x(a, b, \gl)$ is a function defined on a 3D space of time $b$, scale $a$, and chirprate $\gl$, and hence it is called the time-scale-chirprate transform of $x(t)$ in \cite{chui2021time}. In this paper $U^g_x(a, b, \gl)$ is called  the ``{wavelet-based chirplet transform}" or the ``{wavelet-chirplet transform}" ({WCT}) to disguish it from the STFT-based CT $Q_x(t, \eta, \lambda)$ defined by \eqref{def_CT}. 

The importance of WCT and CT is that when the IF curves of two components $x_k(t)$ and $x_\ell(t)$ cross each other, they may be well-separated in the 3D space by WCT or CT transform, provided that  $\phi_k''(t)\not = \phi_\ell''(t)$ for $t$ in some neighborhood of the crossover time instant $t_0$.  Thus, a multicomponent signal $x(t)$ with certain signal components that have the same IF values at certain time instances can be extracted and well-separated in the 3D time-scale-chirprate or time-frequency-chirprate space. Hence, it is feasible to reconstruct signal components by WCT or CT as shown in \cite{li2022chirplet,chui2023analysis,chui2021time}. This paper focuses on WCT. 

As in \cite{chui2021time}, it is assumed in  this paper any two IF curves of the signal components $x_k$ and $x_{\ell}$ satisfy 
\begin{equation*}
%\label{def_sep_cond_cros}
 \hbox{either} ~~	\frac {|\phi'_{k}(t)-\phi'_{\ell}(t)|}{\phi'_{k}(t)+\phi'_{\ell}(t)}\ge \frac \gt \mu,  \; t\in \RR,  ~~  \hbox{or} ~~ |\phi''_{k}(t)-\phi''_{\ell}(t)| \ge 2 \gt_1, \; t\in \RR,  
	\end{equation*}
where $\gt, \gt_1$ are positive constants with $\gt<\mu$.  

For a fixed $b$, and a positive number  $\wt \ep_1$, let $\cH_b$ and $\cH_{b, k}$ denote the sets defined by 
\begin{equation*}
%\label{def_cGk}
\cH_b:=\big\{(a, \gl):  |U^g_x(a, b, \gl)|>\wt \ep_1\big\},  
\cH_{b, k}:=\Big\{(a, \gl) \in \cH_b:  
|\mu-a \phi'_k(b)|<\gt,  |\gl-\phi''_k(b)| <  \gt_1\Big\}.  
\end{equation*}
Then under certain conditions on the $A_k(t)$ and $\phi_k(t)$ (see \cite{chui2021time}),  $\cH_{b, k}, 1\le k\le K$ form a disjoint partition of $\cH_b$ and  $\big(\frac{\mu}{\phi'_k(b)}, \phi''_k(b)\big) \in \cH_{b, k}, k=1, \cdots, K$.

\begin{alg}\label{alg1} {\bf (Wavelet-chirplet transform-based SSO scheme \cite{chui2021time})} \; Let $x(t)$ be a multicomponent signal give by \eqref{AHM}, $U^g_x(a, b, \gl)$ be the WCT of $x(t)$ with a window function $g$ satisfying $\int_\R g(t) dt=1$.

{\bf Step 1.} Calculate  $\check a_\ell(b)$ and $\check \gl_\ell(b)$ by 
\begin{equation}
\label{def_max_eta}
 (\check a_\ell, \check \gl_\ell) =(\check a_\ell(b), \check \gl_\ell(b)):={\rm argmax}_{(a, \gl) \in\cH_{b, \ell}  }|U^g_x(a, b,  \gl)|, ~~ \ell=1, \cdots,  K, 
\end{equation}

{\bf Step 2.} Obtain IF and chirprate estimates by 
\begin{equation}
\label{IF_estimate}
\phi'_\ell(b) \approx \frac \mu{\check a_\ell(b)}, \quad 
\phi^{\gp\gp}_\ell(b) \approx \check \gl_\ell(b), 
\end{equation} 

%\item[] 
{\bf Step 3.} Obtain the recovered $\ell$-th component by $x_\ell(b)\approx \wt x_\ell(b)$, where  
\begin{equation}
\label{comp_recover}
\wt x_\ell(b):= U^g_x(\check a_\ell(b), b,  \check \gl_\ell(b)).  
\end{equation}
\end{alg}

Below the authors present the idea behind the WCT scheme in Algorithm 1. Suppose 
$A_k(t)$ and $\phi_k(t)$ satisfy
\begin{eqnarray*}
\label{cond_A}
&&	|A_k(t+\tau)-A_k(t)|\leq \vep_1 |\tau|A_{k}(t),~~	|\phi'''_{k}(t)| \leq \vep_2, ~~t\in \RR, \; k=1,\cdots, K,
	\end{eqnarray*}
for some small $\vep_1, \vep_2>0$. Then for $x_k(t)=A_k(t)e^{i2\pi \phi_k(t)}$, 
$$
x_k(b+a t)\approx A_k(b)e^{i2\pi\big(\phi_k(b)+\phi'_k(b)at +\frac 12 \phi''_k(b) (at)^2\big)}=x_k(b)e^{i2\pi \big(\phi'_k(b) a t +\frac 12\phi''_k(b) (at)^2\big) }.
$$
Thus 
\begin{eqnarray}
\nonumber U^g_{x_k}(a, b, \gl)&&\hskip -0.6cm \approx 
 \int_{\RR} x_k(b)e^{i2\pi (\phi'_k(b) a t +\frac 12\phi''_k(b) (at)^2) } g(t)  e^{-i2\pi \mu t-i\pi \lambda a^2 t^2} dt
\\
\label{Ux_approx} && = x_k(b) \wb g\big(\mu -a \phi'_k(b), a^2(\gl - \phi''_k(b))\big), 
\end{eqnarray}
where $\wb g (\eta, \gl)$ is the 2nd-order  polynomial Fourier transform of $g(t)$ defined by (see %\cite{katkovnik1995new, li2011local,stankovic2013time})  
\cite{stankovic2013time})  
\begin{equation}
\label{def_PFT}
\wb g (\eta, \gl):=\int_{\RR} g(t) e^{-i2\pi\eta t-i\pi \gl t^2}dt. 
\end{equation}
Therefore, one has 
\begin{equation}
\label{approx00}
 U^g_x(a, b, \gl) \approx \sum_{k=1}^K x_k(b) \wb g\big(\mu -a \phi'_k(b), a^2(\gl - \phi''_k(b)\big),  
\end{equation}
which leads to 
\begin{equation}
\label{approx0}
 U^g_x\big(\check a_\ell(b), b, \check \gl_\ell(b)\big) \approx \sum_{k=1}^K x_k(b) \wb g\big(\mu -\check a_\ell(b) \phi'_k(b), 
\check a_\ell(b)^2( \check \gl_\ell(b) - \phi''_k(b))\big), \; \ell=1, 2, \cdots, K.   
\end{equation}
Hence, when $\mu/\check a_\ell(b)$ and $\check \gl_\ell(b)$ are accurate approxiamtions to  $\phi'_k(b)$ and $ \phi''_k(b)$ resp., then 
\begin{equation}
\label{approx1}
\begin{array}{l}
\wb g\big(\mu -\check a_\ell(b) \phi'_k(b), \check a_\ell(b)^2( \check \gl_\ell(b) - \phi''_k(b))\big) \approx 0, \hbox{for $\ell\not=k$}; \\
 \wb g\big(\mu -\check a_\ell(b) \phi'_\ell(b), 
\check a_\ell(b)^2( \check \gl_\ell(b) - \phi''_\ell(b))\big)\approx \wb g(0, 0)=1. 
\end{array}
\end{equation}
Finally \eqref{approx0} and \eqref{approx1} lead to \eqref{comp_recover}. 

In addition, replacing  $\phi'_k(b)$ and $ \phi''_k(b)$  in \eqref{approx0} by 
$\mu/\check a_\ell(b)$ and $\check \gl_\ell(b)$ reps., one has
\begin{equation*}
\label{approx2}
 U^g_x\big(\check a_\ell(b), b, \check \gl_\ell(b)\big) \approx \sum_{k=1}^K c_{\ell, k} \; x_k(b), \; 1\le \ell \le K; 
\end{equation*}
or 
\begin{equation*}
 \left [
\begin{array}{c}
U^g_x\big(\check a_1(b), b, \check \gl_1(b)\big)\\
U^g_x\big(\check a_2(b), b, \check \gl_2(b)\big)\\
\vdots\\
U^g_x\big(\check a_K(b), b, \check \gl_K(b)\big)
\end{array}
\right]
 \approx  C
 \left [
\begin{array}{c}
x_1(b)\\
x_2(b)\\
\vdots\\
x_K(b)
\end{array}
\right],  \; C:=\big[c_{\ell, k}\big]_{1\le \ell, k\le K}= \left[ 
\begin{array}{cccc}
1 &c_{1, 2} &\cdots &c_{1, K}\\
c_{2, 1}&1 &\cdots &c_{2, K}\\
\vdots &\vdots &\ddots&\vdots\\
c_{K, 1}&c_{K, 2} &\cdots &1\\
\end{array}
\right], 
\end{equation*}
where 
\begin{equation}
\label{def_clk}
c_{\ell,k}:=\wb g\Big(\mu\big(1 -\frac {\check a_\ell(b)}{\check a_k(b)}\big), 
\check a_\ell(b)^2\big( \check \gl_\ell(b) - \check \gl_k(b)\big)\Big), \; 1 \le \ell, k \le K.   
\end{equation}
Therefore, one reaches the following mode retrieval formula. 

\begin{alg}\label{alg2} {\bf (Group wavelet-chirplet transform-based SSO scheme)} \; Obtain $\check a_k(b), \check \gl_k(b)$ as in Algorithm \ref{alg1} and define $c_{\ell, k}$ by \eqref{def_clk}. Then   
 $\wt {\wt x}_\ell(b)$ obtained in \eqref{approx_group} provides an estimate for the  $\ell$-th component 
$x_\ell(b)$:  
\begin{equation}
\label{approx_group}
 \left [
\begin{array}{c}
\wt {\wt x}_1(b)\\
\vdots\\
\wt {\wt x}_K(b)
\end{array}
\right]:= 
 C^{-1}
 \left [
\begin{array}{c}
U^g_x\big(\check a_1(b), b, \check \gl_1(b)\big)\\
\vdots\\
U^g_x\big(\check a_K(b), b, \check \gl_K(b)\big)
\end{array}
\right], 
\end{equation}
if $C:=\big[c_{\ell, k}\big]_{1\le \ell, k\le K}$ is nonsingular; otherwise $C^{-1}$ in \eqref{approx_group} is replaced by the pseudo-inverse of $C$. 
\end{alg}

In general, the recovery formula \eqref{approx_group} provides a more accurate approximation to $x_k(t)$ than \eqref{comp_recover}. 
The authors observe that the accuracy of mode retrieval depends on the degree of accuracy in the estimation of the IFs and chirprates.
In addtion, condition in \eqref{approx1} implies it is desirable that   $\wb g(\eta, \gl)$ decays fast or at least decays fairly along both the frequency $\eta$ and chirprate $\gl$ directions. 

\bigskip

\begin{figure}[th]
	\centering
	\begin{tabular}{ccc}
		\resizebox {1.5in}{1.1in} {\includegraphics{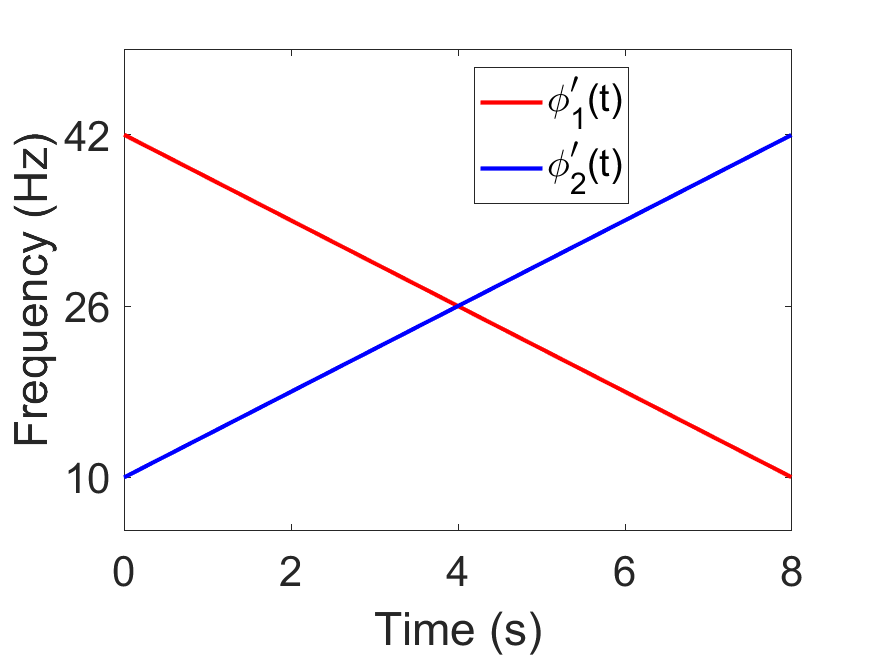}}
\quad &\quad \resizebox {1.8in}{1.2in} {\includegraphics{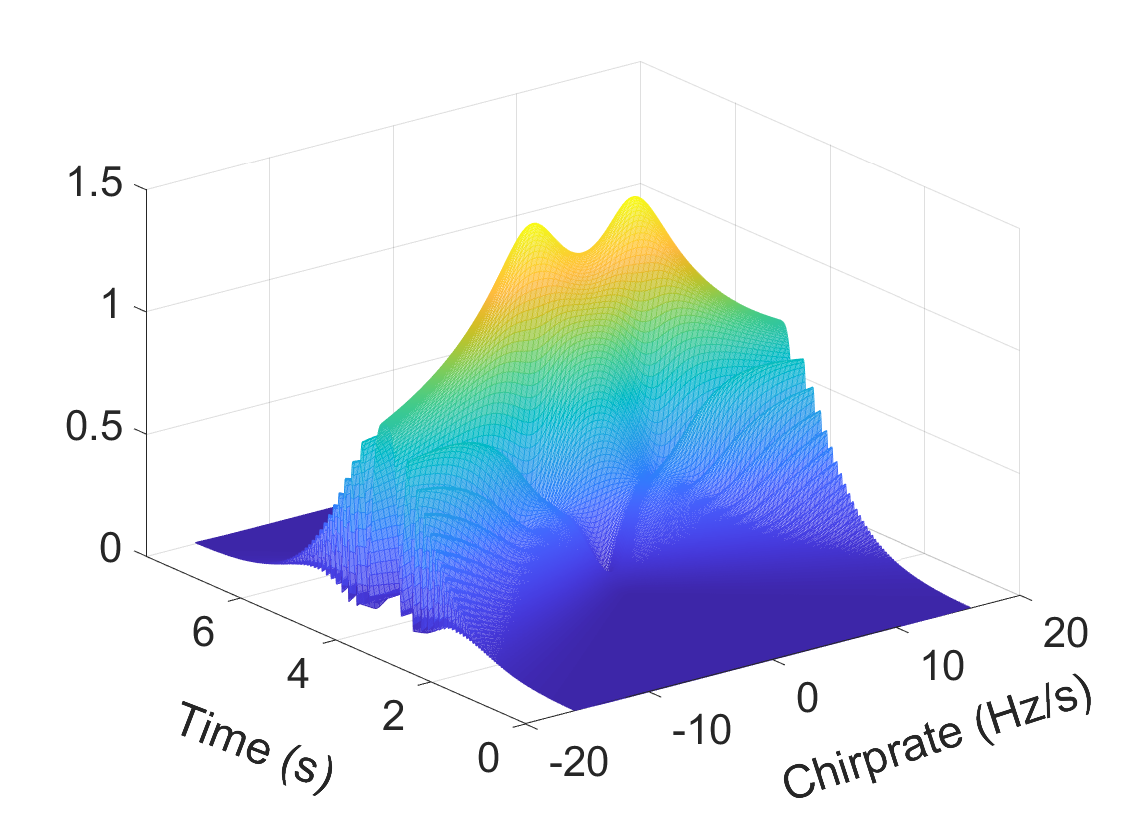}}
\quad &\quad
\resizebox {1.6in}{1.2in} {\includegraphics{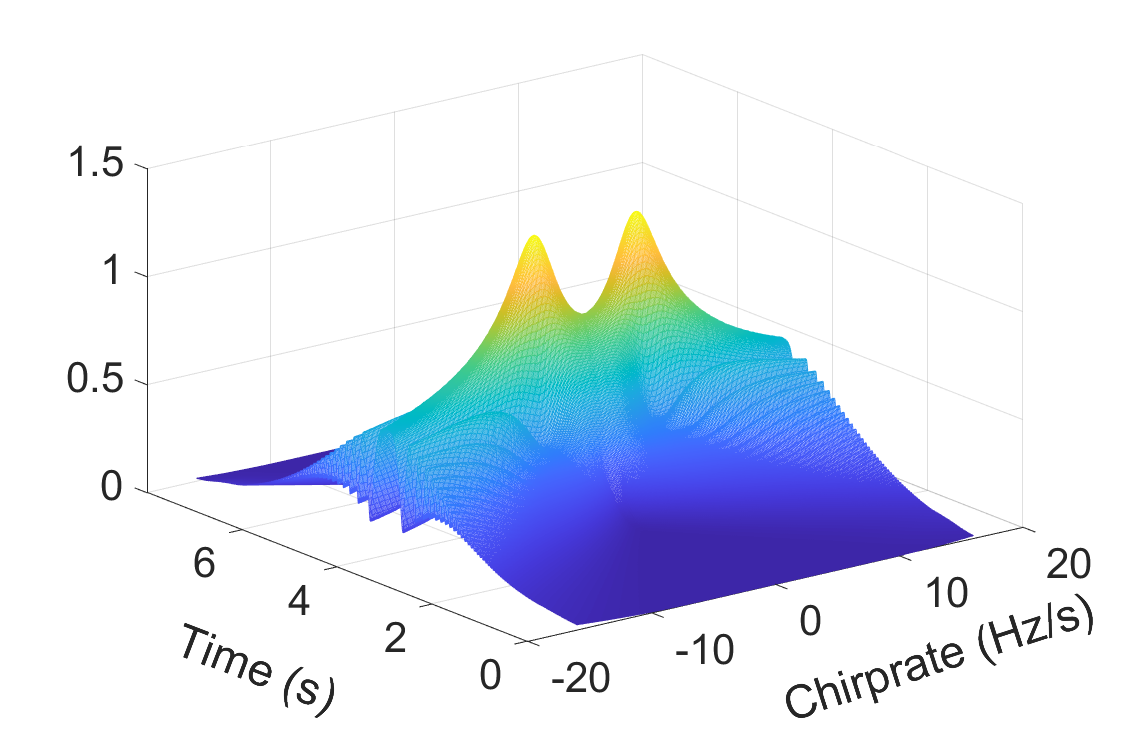}}
	\end{tabular}
	\caption{\small  Left: IFs of $x_1, x_2$. \quad Middle: WCT of $x$ with $a=\frac 1{26}$. \quad Right: 
Proposed  XWCT of $x$ with $a=\frac 1{26}$.}
	\label{figure:Example1_IFs}
\label{figure:Example1_WCT0}
\label{figure:Example1_WCT}
\end{figure}

{\bf Example 1.} Let $x(t)$ be the non-stationary signal given by  
\begin{equation}
\label{def_x}
x(t)=x_1(t)+x_2(t), \; x_1(t)=e^{i2\pi(42t-2t^2)}, \; x_2(t)=e^{i2\pi(10t+2t^2)}, \; t\in [0, 8).  
\end{equation}
With $\phi_1'(t)=42-4t, \phi_2'(t)=10+4t$, one observes that the IFs of $x_1$ and $x_2$ are cossover at $t=4$ with $\phi_1'(4)=\phi_2'(4)=26$. The IFs of $x_1, x_2$ are shown in the left panel of Fig.\ref{figure:Example1_IFs}.

%%%%%%%%%%%%%%%%%%%the beginning of figure 1 %%%%%%%%%%%%%%%
%%%%%%%%%%%%%%%%the end of figure 1  %%%%%%%%%%%%%%%%%%%%%

The ideal representation of $x(t)$ in 3D space is 
$$
{\rm IR}{}_x(a,b,\gl)=x_1(b)\gd\big(1/a-\phi_1'(b)\big)\gd\big(\gl-\phi_1''(b)\big)+x_2(b)\gd\big(1/a-\phi_2'(b)\big)\gd\big(\gl-\phi_2''(b)\big). 
$$ 
In particular, 
\begin{eqnarray}
\label{Example1_IR_b=4}
\big|{\rm IR}{}_x(1/{26},b,\gl)\big|\hskip -0.6cm &&=\gd\big(26-\phi_1'(b)\big)\gd(\gl+4)+\gd\big(26-\phi_2'(b)\big)\gd(\gl-4)\\
\nonumber &&=\gd(b-4)\gd(\gl+4)+\gd(b-4)\gd(\gl-4). 
\end{eqnarray}

The WCT (in moduls; the same applies below) of $x(t)$ with $a=\frac 1{26}$, that is 
$|U_x(\frac 1{26}, b, \gl)|$, is shown in  the middle panel of Fig.\ref{figure:Example1_WCT0}. Observe that along the time $b$ direction, $|U_x(\frac 1{26}, b, \gl)|$ approaches to 0 very fast as $b$ moves away from $b=4$. 
However $|U_x(\frac 1{26}, 4, \gl)|$ decays very slowly when $\gl$ moves away from $\gl=-4$ or from $\gl=4$.

%%%%%%%%%%%%%%%%%%%the beginning of figure 2 %%%%%%%%%%%%%%%
\begin{figure}[th]
	\centering
	\begin{tabular}{cccc}
	\resizebox {1.5in}{1.1in}{\includegraphics{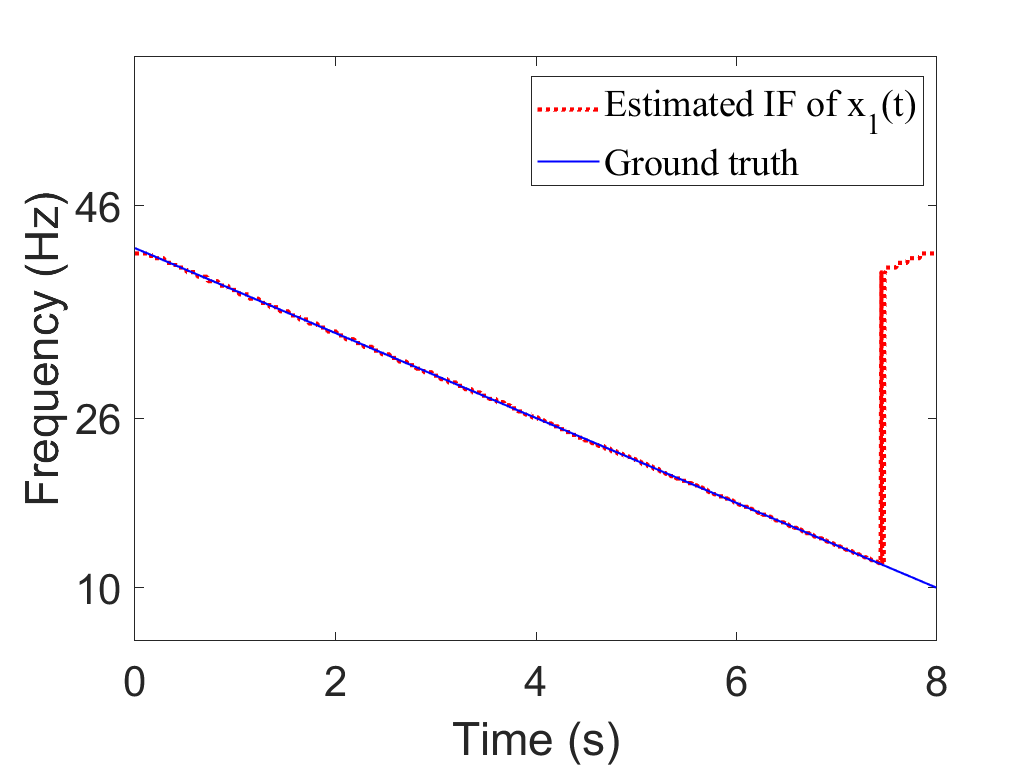}}&   \resizebox {1.5in}{1.1in} {\includegraphics{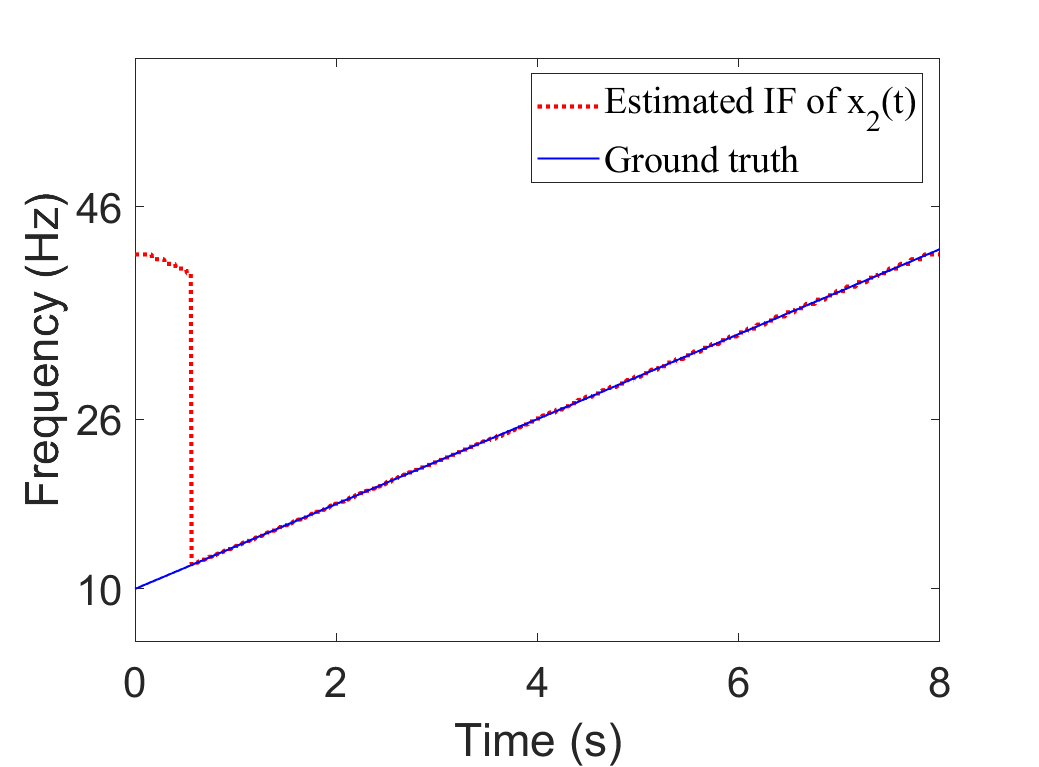}}
&\resizebox {1.5in}{1.1in} {\includegraphics{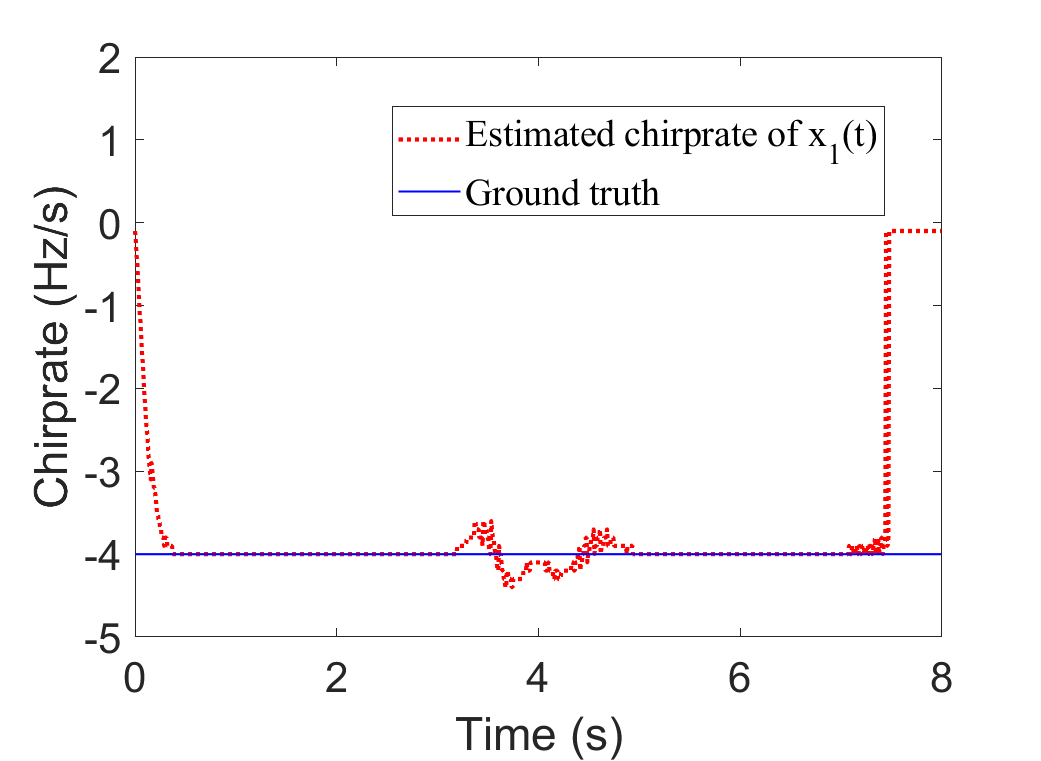}}
& \resizebox {1.5in}{1.1in} {\includegraphics{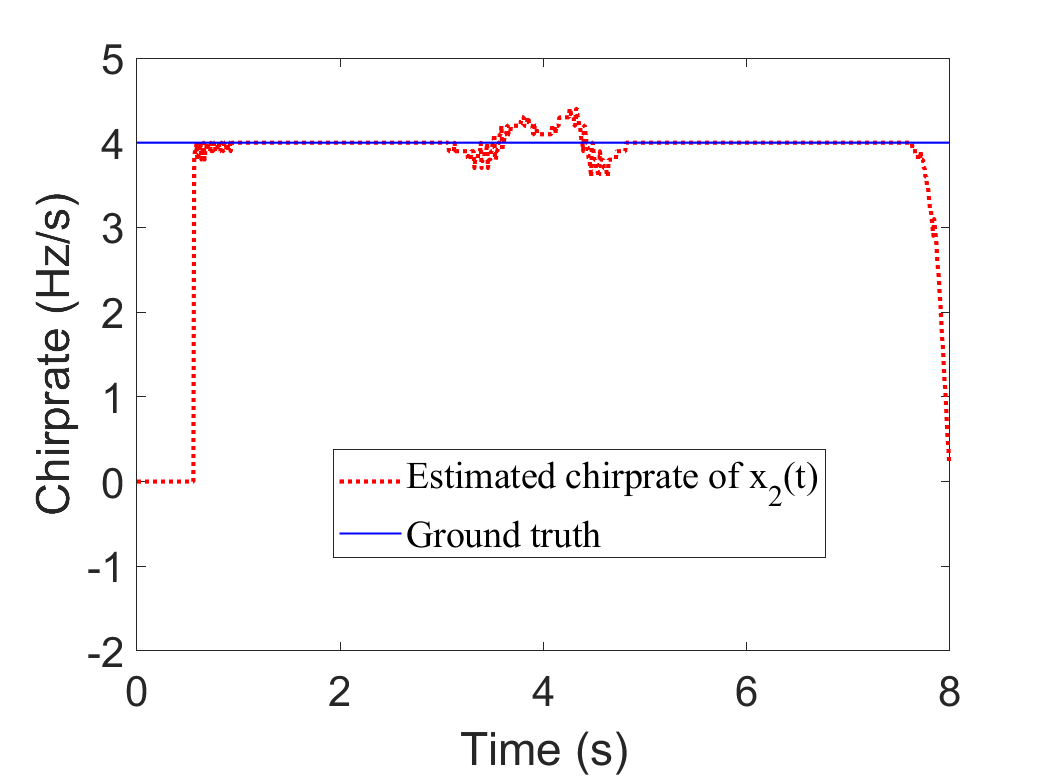}}
	\end{tabular}
	%\vskip -0.3cm 
\caption{\small Estimated IFs $\wt \phi'_\ell(b)=1/\check a_\ell(b)$ by \eqref{def_max_eta} (1st, 2nd panels). 
 Estimated chirprates $\wt \phi''_\ell(b)=\check \gl_\ell(b)$ by \eqref{def_max_eta}(3rd, 4th panels).}
	\label{figure:Example1_IFs_est0}
\end{figure}
%%%%%%%%%%%%%%%%the end of figure 2  %%%%%%%%%%%%%%%%%%%%%

The 1st and 2nd panels (from the left) of Fig.\ref{figure:Example1_IFs_est0} show the estimated IFs $\wt \phi'_\ell(b)=\mu/\check a_\ell(b)$ by \eqref{def_max_eta} with $\mu=1$. Overall the estimates are quite accurate except for $b$ near the end points, which is caused by the boundary issue. However for chirprate estimate, there are big errors near $b=4$  where the IFs of $x_1(t)$ and $x_2(t)$ are crossover as showns in 
the 3rd and 4th panels (from the left) of Fig.\ref{figure:Example1_IFs_est0}.  It is desriable to have a method which makes the chirprate  decays faster.  To this regard, the authors propose an X-ray transform-composed wavelet-based chirplet transform, which is defined in the next section. 

\section {X-ray wavelet-chirplet transform}
For a signal $x(t)=A(t)e^{i2\pi\phi(t)}$, when $A(t)$ changes slowly and $\phi'''(t)$ is small (in modulus), then 
for small $a$ (refer to \eqref{Ux_approx}), 
\begin{eqnarray*}
U^g_x(a, b, \gl)\approx x(b) \wb g\big(\mu- a\phi'(b), (\gl-\phi''(b))a^2\big), 
\end{eqnarray*}
where $\wb g$ is defined by \eqref{def_PFT}. 
Therefore, if ideally $\wb g(\eta, \gl)=\gd(\eta)\gd(\gl)$, then one has an ideal time-scale-chirprate representation for $x(t)$:
$$
U^g_x(a, b, \gl)\approx x(b)\gd\big(\mu- a\phi'(b)\big)\gd\big(\gl-\phi''(b)\big). 
$$
Of course, there does not exist a window function having such an ideal property. To get  a sharp time-scale-chirprate representation for $x(t)$, it is desirable that $|\wb g(\eta, \gl)|$ decays fast indepently for vaiables $\eta$ and $\gl$ near $0$. 

Next let us consider the case $g$ is the Gaussian window function defined by 
\begin{equation}
\label{def_g}
g_\gs(t)=\frac 1{\gs \sqrt {2\pi}} \; e^{-\frac {t^2}{2\gs^2}},   
\end{equation}
where $\gs>0$.  Then one has (refer to \cite{li2020adaptivestft}) 
\begin{equation}
\label{g_PFT} 
\wb g_\gs(\eta, \gl)=\frac 1{\sqrt{1+i2\pi\gs^2 \gl}} e^{-\frac{2\pi^2\gs^2 \eta ^2}{1+i2\pi \gs^2\gl}},   
\end{equation}
where $\sqrt{1+i2\pi\gs^2\gl}$ denotes the square root of $1+i2\pi\gs^2\gl$ with Re($\sqrt{1+i2\pi\gs^2\gl})>0$.  
%lying in the same quadrant as $1+i2\pi\gs^2\gl$.
Note that 
\begin{equation}
\label{g_abs}
|\wb g_\gs(\eta, \gl)|={(1+4\pi^2\gs^4 \gl^2)^{-\frac 14}} e^{-\frac{2\pi^2\gs^2 \eta ^2}{
1+4\pi^2\gs^4 \gl^2}},
\end{equation}
Thus with respect to the frequency variable $\eta$, $|\wb g_\gs(\eta, \gl)|$ decays very fast, while it does not with respect to the chirprate variable $\gl$.  Actually the decay of $|\wb g_\gs(\eta, \gl)|$ with respect to $\gl$ is very slow with a rate essential equivalent to the first factor of the product on the right-hand side of \eqref{g_abs}:
$$
f_\gs(\gl):={(1+4\pi^2\gs^4 \gl^2)^{-\frac 14}}. 
$$
Observe that $f_\gs(\gl)$ has its maximum value 1 at $\gl=0$ and decays with the order $|\gl|^{-\frac 12}$ as $\gl \to \infty$.

Let $C(f)$ be the chirp transform of $f(t)\in L_1(\RR)$ defined by 
\begin{equation}
\label{def_chirpT}
C(f) (\gl):=\int_{\RR} f(t) e^{-i\pi \gl t^2}dt. 
\end{equation} 
Note that the transform $\wb g (\eta, \gl)$ of $g$ defined by \eqref{def_PFT} can be written as 
$$
\wb g (\eta, \gl)=C(g_\eta)(\gl), \quad \hbox{where $g_\eta(t)=g(t) e^{-i2\pi\eta t}$}. 
$$
Thus the decay property of $C(g)(\gl)$ will provide useful information for the dacay of   $|\wb g (\eta, \gl)|$ along the direction $\gl$. 

The theoretical result on the chirp transform in \cite{chen2023disentangling} states that the chirp transform of $t^n g(t)$, where $n$ is a positive integer, decays faster than $g$.  Thus $\big(t^n g_\gs(t)\wb{\big)\;}(\eta, \gl)$ should decay faster than $\wb g(\eta, \gl)$ along the direction $\gl$.  
Here and below $\big(t^n g_\gs(t)\wb{\big)\;}(\eta, \gl)$ denotes the second-order  polynomial Fourier transform defined by \eqref{def_PFT} with $g(t)$ there replaced by $t^n g_\gs(t)$.  Indeed, one has by direct calculations, 
 \begin{eqnarray}
 &&\big(t^2 g_\gs(t)\wb{\big)\;}(\eta, \gl)
\label{tjg2} 
=\gs^2\Big(\frac 1{(1+i2\pi\gs^2 \gl)^{\frac 32}}-
\frac {(2\pi \gs \eta)^2}{(1+i2\pi\gs^2 \gl)^{\frac 52}}\Big) e^{-\frac{2\pi^2\gs^2 \eta ^2}{1+i2\pi \gs^2\gl}}, \\
&&\big(t^4 g_\gs(t)\wb{\big)\;}(\eta, \gl)
%=-\frac 1{\pi^2} \frac {\partial^2}{\partial^2 \gl}\wb g_\gs(\eta, \gl)\\
\label{tjg4} %&&\qquad  
=\gs^4\Big(\frac 3{(1+i2\pi\gs^2 \gl)^{\frac 52}}-
\frac {6(2\pi \gs \eta)^2}{(1+i2\pi\gs^2 \gl)^{\frac 72}}+
\frac {(2\pi \gs \eta)^4}{(1+i2\pi\gs^2 \gl)^{\frac 92}}
\Big) e^{-\frac{2\pi^2\gs^2 \eta ^2}{1+i2\pi \gs^2\gl}}. 
 \end{eqnarray}
Thus the decays of $|\big(t^2 g_\gs(t)\wb{\big)\;}(\eta, \gl)|$ and $|\big(t^4 g_\gs(t)\wb{\big)\;}(\eta, \gl)|$
with respect to $\gl$ are essentially controlled respectively by the factors 
$$
f_{\gs, 2}(\gl):={(1+4\pi^2\gs^4 \gl^2)^{-\frac 34}}, \; 
f_{\gs, 4}(\gl):={(1+4\pi^2\gs^4 \gl^2)^{-\frac 54}}. 
$$
Indeed, $f_{\gs, 2}(\gl)$ and  $f_{\gs, 4}(\gl)$ decay faster (have sharper graphs) near $\gl=0$ than $f_\gs$ 
and they have the decay order $|\gl|^{-\frac 32}$ and $|\gl|^{-\frac 52}$ resp. as $\gl \to \infty$.  
%See the pictures of $f_{\gs, 2}(\gl)$ and  $f_{\gs, 4}(\gl)$ with $\gs=1$ in Fig.\ref{figure:graph_f_gs}. 
Thus the authors of \cite{chen2023disentangling, chen2024multiple} use $t^2g_\gs(t)$ as the window functions. 

%%%%%%%%%%%%%%%%%%% figure 3 refer to chirp_transform.txt for following pictures %%%%%%%%%%%%%%%
\begin{figure}[th]
	\centering
	\begin{tabular}{cc}
	\resizebox {1.6in}{1.2in} {\includegraphics{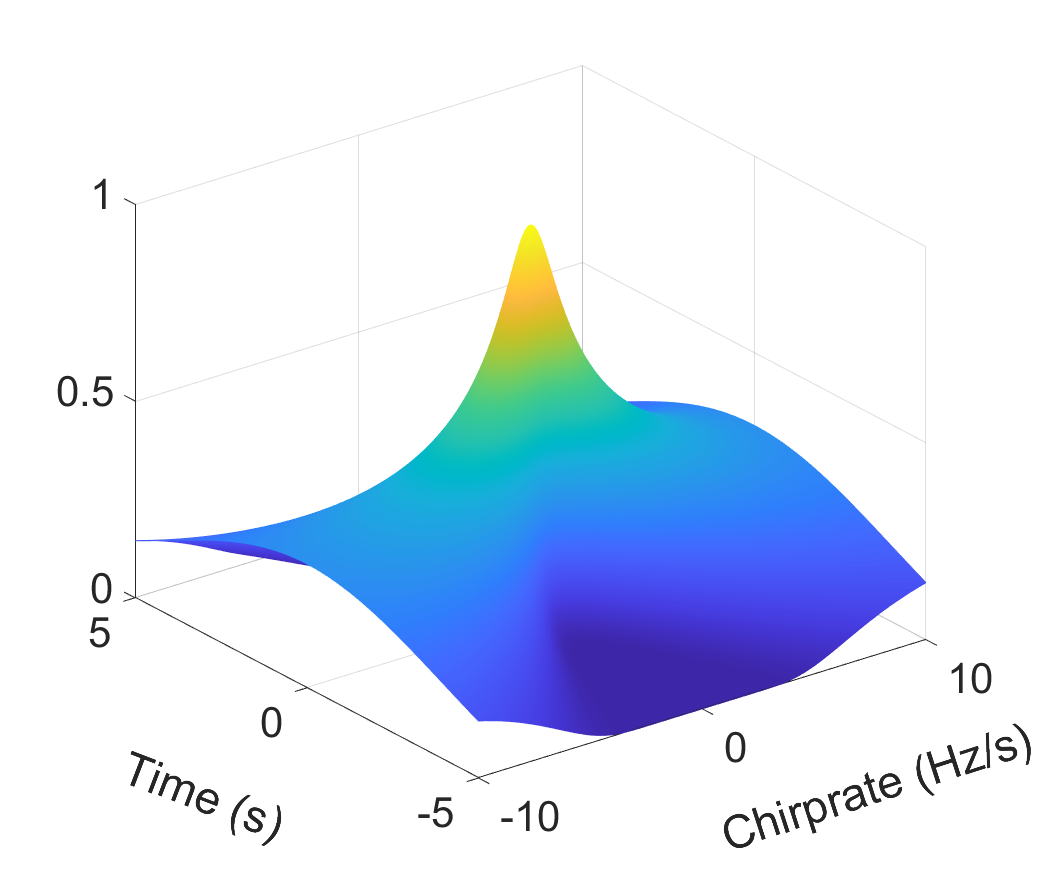}}	\quad & \quad  
\resizebox {1.6in}{1.2in} {\includegraphics{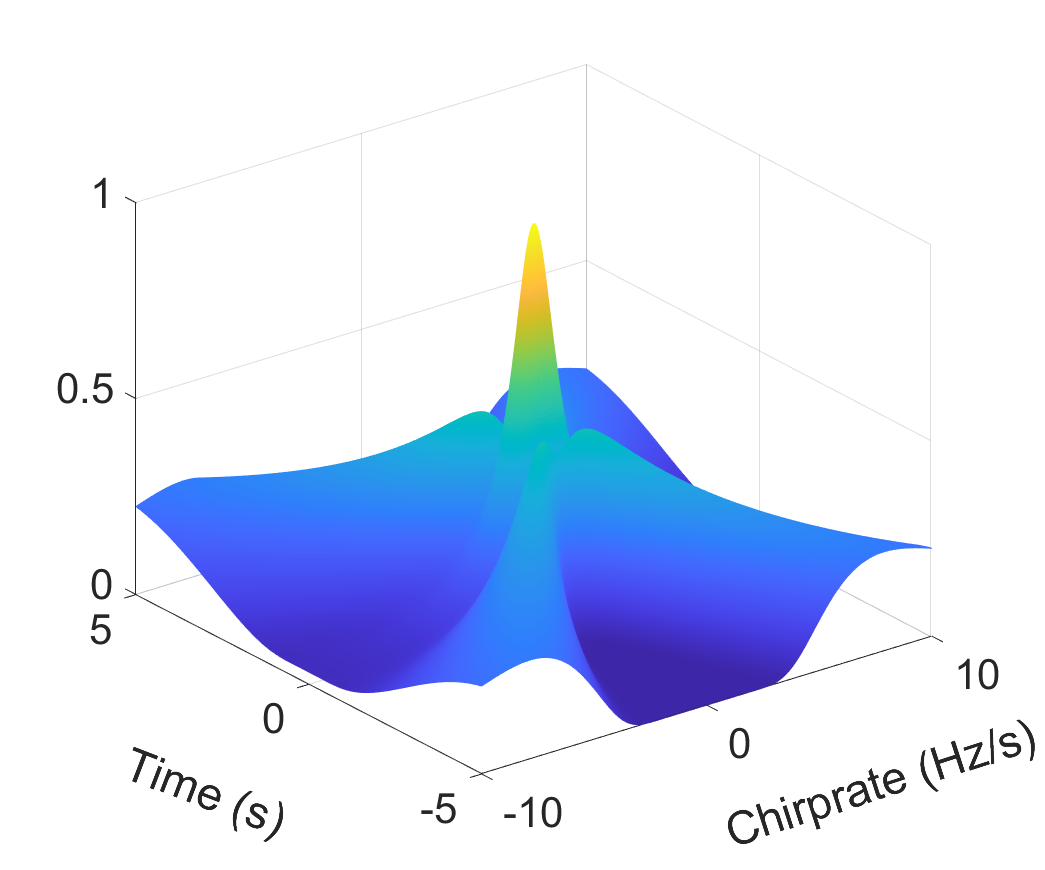}}
	\end{tabular}
	\caption{\small  2nd-order Fourier transform of $g_\gs(t)$ (Left) and   $t^2g_\gs(t)$ (Right)}% $\big|\big(t^2 g_\gs(t)\wb{\big)\;}(\eta, \gl)\big|$ (Right)}
	\label{figure:chirp_transform}
\end{figure}
%%%%%%%%%%%%%%%% end of figure 3  %%%%%%%%%%%%%%%%%%%%%

It seems the larger $j$, the faster decay of  $\big(t^j g_\gs(t)\wb{\big)\;}(\eta, \gl)$ along $\gl$, and hence the better $U^{t^j g}_x(a, b, \gl)$ in chirprate estimate. However it is not the case if one uses a window $t^j g_\gs(t)$ with a large $j$. The reason is probably that the other terms in $\big(t^j g_\gs(t)\wb{\big)\;}(\eta, \gl)$ (meaning,  the second term in \eqref{tjg2} for $\big(t^2 g_\gs(t)\wb{\big)\;}(\eta, \gl)$, 
the second and third terms in \eqref{tjg4} for $\big(t^4 g_\gs(t)\wb{\big)\;}(\eta, \gl)$) also play roles. Fig.\ref{figure:chirp_transform} shows  
 $|\wb g_\gs(\eta, \gl)|$ and  $|\big(t^2 g_\gs(t)\wb{\big)\;}(\eta, \gl)|$ (normazlied with its maximum to be 1), where $\gs=1/\sqrt{2\pi}$. Though  $|\big(t^2 g_\gs(t)\wb{\big)\;}(\eta, \gl)|$ decays faster than  $|\wb g_\gs(\eta, \gl)|$ along the chirprate direction, there are large oscillations in it.  The larger $j$ is, more oscillations  $|\big(t^j g_\gs(t)\wb{\big)\;}(\eta, \gl)|$ has. These large oscillations may cause the mixture of components in the 3D space.  
Thus using window functions $t^j g_\gs(t)$ cannot solve the fundmental issue of slow decay of the chirprate. In this paper the authors propose an X-ray transform composed wavelet-chirprate transform.

First, let us examine the core concept behind this novel transform.
Suppose $g(t)$ is the Gausssian window function $g_\gs(t)$, and $x(t)$ is a linear chirp or a linear frequency modulation signal:
\begin{equation}
\label{def_chirp0}
x(t)= Ae^{i2\pi \phi(t)}=A e^{i2\pi (ct +\frac 12 r t^2)},   
\end{equation}
where $A>0, c$ and $r$ are constants. With $\phi'(t)=c+rt, \phi''(t)=r$,  one has 
$$
x(b+at)=x(b)e^{i2\pi(\phi'(b)at +\frac 12 \phi''(b) (at)^2)},
$$
and hence,  
$$
U_x^{g_\gs}(a, b, \gl) = x(b) \wb g_\gs\big(\mu-a \phi'(b), a^2(\gl-\phi''(b))\big). 
$$
Following \eqref{g_abs}, one can conclude 
\begin{equation}
\label{Ux_abs}
|U_x^{g_\gs}(a, b, \gl)|=A \big(1+4\pi^2\gs^4 a^4(\gl-\phi''(b))^2\big)^{-\frac 14} 
e^{-\frac{2\pi^2\gs^2 \big(\mu-a\phi'(b)\big) ^2}{1+4\pi^2\gs^4 a^4(\gl-\phi''(b))^2}}. 
\end{equation}
The term $\big(\mu-a\phi'(b)\big)^2$ in the left-hand side of \eqref{Ux_abs} in the exponent of 
$e^{-1}$ plays a crtical role in the fast decay of 
$|U_x^{g_\gs}(a, b, \gl)|$ along the IF ridge $a= \mu/{\phi'(b)}$. One hopes that the term $(\gl-\phi''(b))$
also in the exponent of $e^{-1}$ so that $|U_x^{g_\gs}(a, b, \gl)|$ decays fast along the chirprate ridge $\gl= \phi''(b)$. After rigorous derivation and calculations, the quantity $\big|U^{g_\gs}_x(\frac {a\mu}{\mu+v a\gl}, b+v, \gl)\big|$ has such a property. 

Indeed, for $\big|U^{g_\gs}_x(\frac {a\mu}{\mu+v a\gl}, b+v, \gl)\big|$, the term corresponding to $\big(\mu-a\phi'(b)\big)^2$ for $|U_x^{g_\gs}(a, b, \gl)|$ is, with $a$ and $b$ replaced respectively by $\frac {a\mu}{\mu+v a\gl}$ and $b+v$, 
\begin{eqnarray}
 \label{vconstraint}&&\big(\mu-\frac {a\mu}{\mu+v a\gl} \phi'(b+v)\big)^2=\big(\mu-\frac {a\mu}{\mu+v a\gl} (\phi'(b)+v\phi''(b))\big)^2\\
\nonumber&&\qquad =\frac {\mu^2}{(\mu+v a\gl)^2} \big(\mu-a\phi'(b)+av (\gl-\phi''(b))\big)^2, 
\end{eqnarray}
which leads to 
\begin{eqnarray}
\nonumber &&\big|U^{g_\gs}_x(\frac {a\mu}{\mu+v a\gl}, b+v, \gl)\big|=A \Big(1+4\pi^2\gs^4 
\big(\frac {a\mu}{\mu+v a\gl})^4(\gl-\phi''(b))^2\Big)^{-\frac 14} \times \\ 
\label{XRUx_abs} && \hskip 5cm  {\rm exp}\Big(-\frac{2\pi^2\gs^2 \mu^4a^4 \big(\mu-a\phi'(b)+av (\gl-\phi''(b)) \big) ^2}{(\mu+v a \gl)^4+4\pi^2\gs^4 \mu ^4 a^4(\gl-\phi''(b))^2}\Big). 
\end{eqnarray}
If $v\not =0$, with $a>0$, the term $\mu-a\phi'(b)+av (\gl-\phi''(b))$ in the exponent of 
$e^{-1}$ will result fast decay of $\big|U^{g_\gs}_x(\frac {a\mu}{\mu+v a\gl}, b+v, \gl)\big|$ along the 3D ridge 
$$
a=\mu/{\phi'(b)}, \; \gl=\phi''(b). 
$$
This is exactly what one needs to well separate a multicomponent signal in the 3D space. Motivated by this observation, the following transform is proposed.

 \begin{mdef} 
{\bf (X-ray wavelet-chirplet transform (XWCT))}  
Let 
$U^g_x(a, b, \gl)$ be the WCT of $x(t)$ defined by \eqref{def_TSC0} with a window function $g(t)$ and $\mu>0$.
 Define the XWCT of $x(t)$ by 
\begin{equation}
\label{def_XRWCT}
\cU_x^g(a, b, \gl):=\int_{-\infty}^\infty \big|U^g_x(\frac {a\mu}{\mu+va\gl}, b+v, \gl)\big| h(v) dv, 
\end{equation} 
where $h(t)$ is a nonnegative window function decaying fast near $0$ and $\int_{-\infty}^\infty h(t)dt=1$. 
\end{mdef}

Notice that the XWCT defined in \eqref{def_XRWCT} involves an integral with a window function \(h(v)\). The rationale for applying this averaging lies in the constraint imposed on $v$. On  the one hand, from \eqref{XRUx_abs}, one can see the larger $v$ (in modulus) is, the sharper  $\big|U^{g_\gs}_x(\frac {a\mu}{\mu+v a\gl}, b+v, \gl)\big|$ will be in the chirprate $\gl$ direction. 
On the other hand, one has used $\phi'(b+v)=\phi'(b)+v\phi''(b)$ in the derivation of \eqref{vconstraint}. This relation holds for the phase function $\phi(t)$ of a linear chirp signal. For non-linear chirp signals, one may use the relation 
  $$
\phi'(b+v)\approx \phi'(b)+v\phi''(b). 
$$
The smaller $v$ (in modulus) is, the accurate the above approximation.  One may compromise these two ``conflict'' ideal conditions by taking the average of $\big|U^{g}_x(\frac {a \mu}{\mu+v a\gl}, b+v, \gl)\big|$ with $v$ ranging over a neighborhood of $0$. This leads to the defintion of X-ray wavelet-chirplet transform in \eqref{def_XRWCT} below, where the integral  involving a window function \(h(v)\) acts as a (weighted) average.

Next, let us look at how $\cU_x^g(a, b, \gl)$ is related to the X-ray transform of $|U^g_x(a, b, \gl)|$.  
The continuous X-ray transform of a 3D function $f(z_1, z_2, z_3), (z_1, z_2, z_3)\in \RR^3$, denoted by $Pf$, is defined by the set of
all line integrals of $f$ \cite{averbuch20043d}. For a line $L$, defined by a unit vector $\gth$ and a point $z=(z_1, z_2, z_3)$ on $L$, one expresses $L$ as 
$$
L(v) = z + v\gth, \; v\in \RR 
$$
The X-ray transform of $f$ on $L$, written as $Pf(z, \gth)$ or $P_\gth f(z)$,  is defined by
$$
Pf (z, \gth):=\int_{-\infty}^\infty f (z +v\gth) dv.
$$
Let $h(v)$ be a window function with $\int_{-\infty}^\infty h(v) dv=1$. One may call 
$$
{\mathcal P} f (z, \gth):=\int_{-\infty}^\infty f (z +v\gth) h(v) dv
$$
a localized X-ray transform of $f$.  This localized version integrates the function along the line, but weights the contribution of each point on the line by the window function $h(v)$. This localization can be useful for various applications, as it focuses the transform on a specific region of the line.

For a scale variabe $a$, one knows $\frac \mu a$ is the corresponding frequency of $U_x^g(a, b, \gl)$. Denote 
$$
\xi=\frac \mu a, \;  \hbox{and} \; \mathbb{U} _x^g(\xi, b, \gl)=U_x^g(a, b, \gl)=U_x^g(\frac \mu \xi, b, \gl). 
$$
Then 
$$
\mathbb{U}_x^g(\xi+\gl v, b+v, \gl)=U_x^g(\frac \mu {\xi+\gl v}, b+v, \gl)=U_x^g(\frac \mu {\frac \mu a+\gl v}, b+v, \gl)
$$
Thus 
\begin{equation}
\label{XCT}
\int_{-\infty}^\infty \big|U^g_x(\frac {a\mu}{\mu+va\gl}, b+v, \gl)\big| h(v) dv=
\int_{-\infty}^\infty \big|\mathbb{U}^g_x(\xi+\gl v, b+v, \gl)\big| h(v) dv,
\end{equation}
which is ${\mathcal P}(|\mathbb{U}^g_x|)\big((\xi, b, \gl), \gth\big)$ with $\gth=(\gl, 1, 0)$. That is 
XWCT $\cU_x^g(a, b, \gl)$ defined as the quantity on the left-hand side of the above equation is, in the time-frequency-chirprate space, the localized X-ray transform 
of $|\mathbb{U}^g_x|$ at $(\frac \mu a, b, \gl)$ with (unormalized) direction  $(\gl, 1, 0)$. That is why  
$\cU_x^g(a, b, \gl)$ is called the {X-ray transform-composed wavelet-based chirplet transform}, the {X-ray wavelet-chirplet transform (XWCT)} for short. 

The preceding discussion has elaborated on why \(\cU_x^g(a, b, \gl)\) exhibits faster decay along the \(\gl\) direction than \(U^g_x(a, b, \gl)\) when $g$ is the Gaussian window function \(g_\gs\) defined in \eqref{def_g}. This assertion will now be demonstrated through the following three illustrative examples. For the purposes of this paper, we set \(\mu = 1\) and fix the window function as \(g = g_\gs\). The first consideration involves selecting the parameter \(\gs\), which in this work is determined based on the 
R${\rm \acute e}$nyi entropy of the WCT. More precisely, let 
\begin{equation}
 \label{def_renyi_entropy_spec}
E_{\gs}: =
 \frac{1}{{1 - \ell }}\Big(\log _2 
  {\iint_{\RR^2}\int_0^\infty \left|U^{g_\gs}_x(a, b, \gl) \right|^{2\ell}\frac {da}a db d\gl}
- \log _2 {\Big(\iint_{\RR^2}\int_0^\infty \left |U^{g_\gs}_x(a, b, \gl)\right|^2\frac {da}a db d\gl \Big)^\ell}\Big),
\end{equation}
where $\ell$ is usually greater than 2.  In this paper, the authors choose $\ell=2.5$, a common value used in other papers, see for examples \cite{ sheu2017entropy,stankovic2001measure}. The R${\rm \acute e}$nyi entropy is %a commonly used 
a measurement of the concentration of a TF representation such as STFT, SST, etc. of a signal $x(t)$, see \cite{ stankovic2001measure}. 
In this paper,  $E_\gs$ defined above measures the concentration of the time-scale-chirprate representation of $x(t)$.  The smaller the R${\rm \acute e}$nyi entropy, the more 
concentrating the time-scale-chirprate representation. $\gs$ will be selected to be the one with smallest R${\rm \acute e}$nyi entropy:
\begin{equation}
\label{entropy_gs0}
\gs_0=\min_{\gs>0} E_\gs. 
\end{equation}

In the following examples, for simplicity, $U_x(a, b, \gl)$ and  $\cU_x(a, b, \gl)$ are used to denote 
$U^{g_\gs}_x(a, b, \gl)$ and  $\cU^{g_\gs}_x(a, b, \gl)$ respectively with $\gs=\gs_0$ defined by \eqref{entropy_gs0}. About the window $h(v)$ for XWCT, $h(v)=g_{\gga}(v), -1\le v\le 1$ with $\gga=0.25$ is chosen. 

\bigskip 

{\bf Example 1 (Continued).} Let $x(t)$ be the non-stationary signal given by \eqref{def_x}  
with $t$ discretized as $t_n=n \Delta t, 0\le n<8\times 128$ with sampling step $\Delta=\frac 1{128}$.

\begin{figure}[th]
	\centering	\begin{tabular}{cc}
	\resizebox {2.0in}{1.5in} {\includegraphics{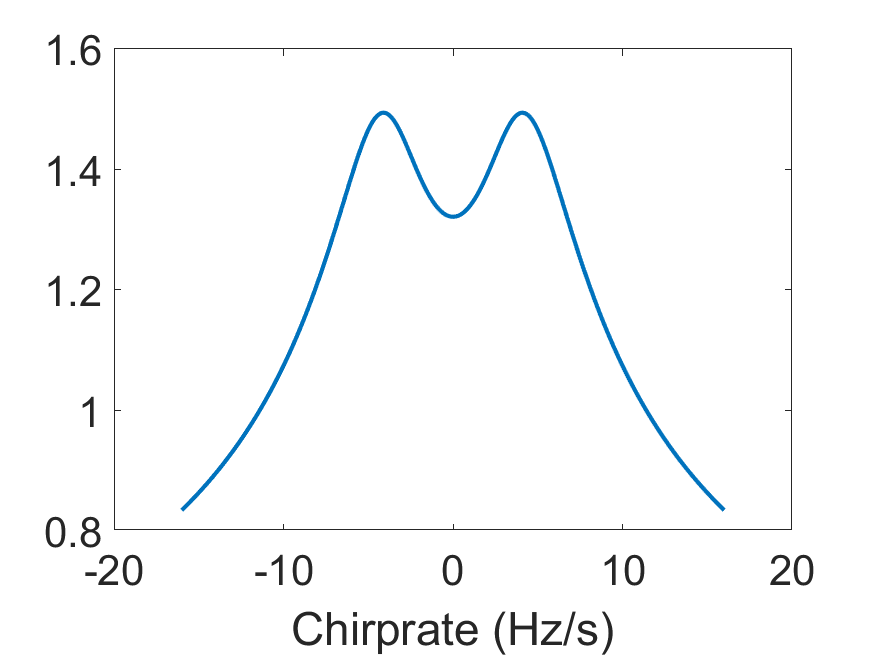}}	\quad & \quad  \resizebox {2.0in}{1.5in} {\includegraphics{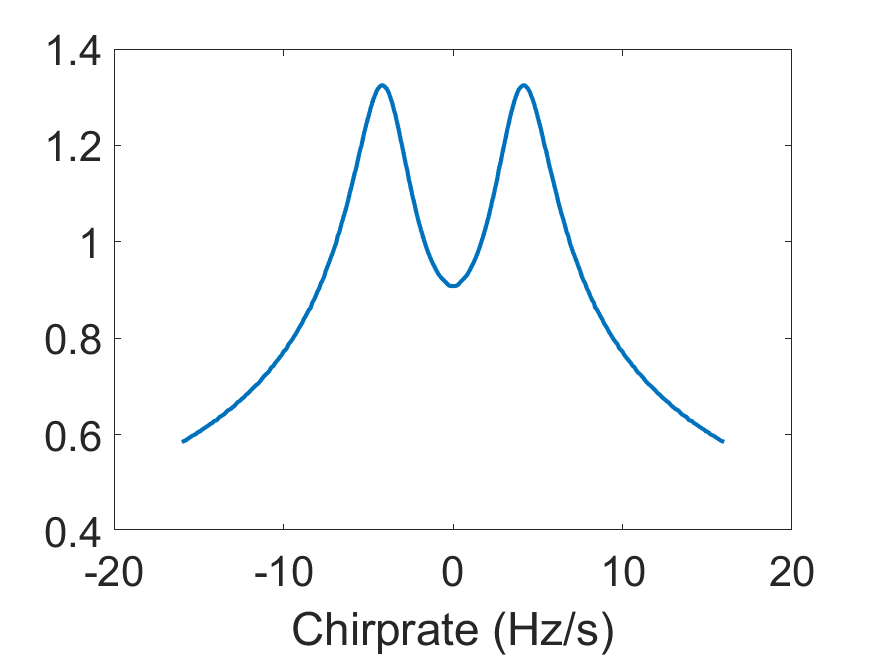}}
	\end{tabular}
%	%\vskip -0.3cm 
	\caption{\small Slice of WCT $|U_x(\frac 1{26}, 4, \gl)|$ (Left) and slice of XWCT $|\cU_x(\frac 1{26}, 4, \gl)|$ (Right)}	
\label{figure:Example1_WCT_slice}
\end{figure}

For $x(t)$ given by \eqref{def_x}, the correspoding parameter $\gs_0$ determined by \eqref{entropy_gs0} is  6.32. The XWCT of $x(t)$ with $a=\frac 1{26}$ (e.g. $|\cU_x(\frac 1{26}, b, \gl)|$) is shown in the right panel of Fig.\ref{figure:Example1_WCT}. Observe that along the time $b$ direction, both $|U_x(\frac 1{26}, b, \gl)|$ (see Fig.\ref{figure:Example1_WCT0}) and $|\cU_x(\frac 1{26}, b, \gl)|$ approach to 0 fast as $b$ moves away from $b=4$. In constrast to $|U_x(\frac 1{26}, 4, \gl)|$,  Fig.\ref{figure:Example1_WCT} shows $|\cU_x(\frac 1{26}, 4, \gl)|$ decays fairly fast, much faster than $|U_x(\frac 1{26}, 4, \gl)|$  when $\gl$ moves away from $\gl=-4$ or from $\gl=4$.  Fig.\ref{figure:Example1_WCT_slice} shows the slice of WCT and XWCT:  $|U_x(\frac 1{26}, 4, \gl)|$ (left panel) and  $|\cU_x(\frac 1{26}, 4, \gl)|$ (right panel). Clearly $|\cU_x(\frac 1{26}, 4, \gl)|$  decays much faster than $|U_x(\frac 1{26}, 4, \gl)|$  when $\gl$ moves away from $\gl=-4$ or from $\gl=4$.

%%%%%%%%%%%%%%%%%%%the beginning of figure 4 %%%%%%%%%%%%%%%
%%%%%%%%%%%%%%%%the end of figure 4  %%%%%%%%%%%%%%%%%%%%%

\begin{figure}[th]
	\centering
	\begin{tabular}{cccc}
		\resizebox {1.4in}{1.1in} {\includegraphics{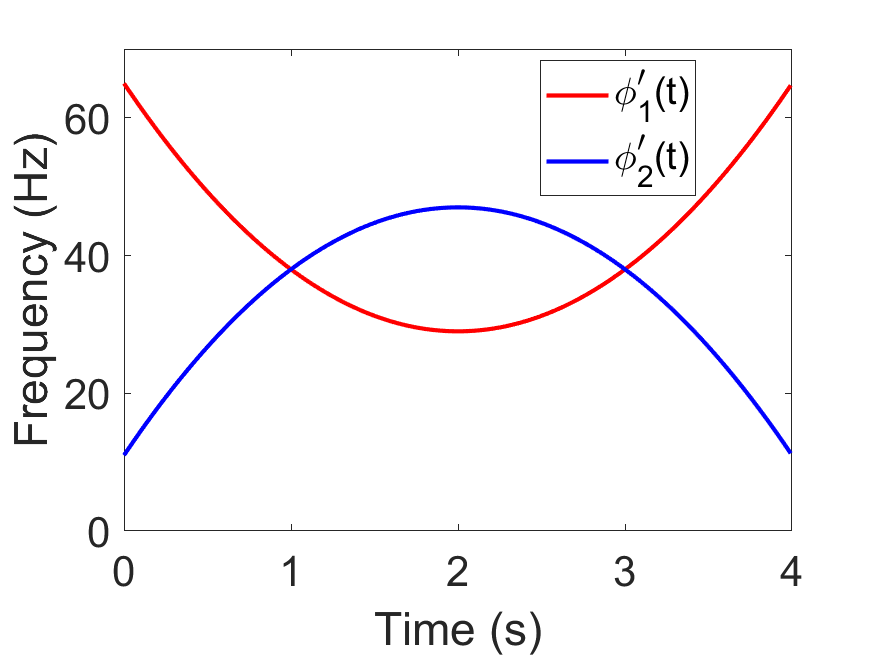}}
		& \resizebox {1.4in}{1.1in} {\includegraphics{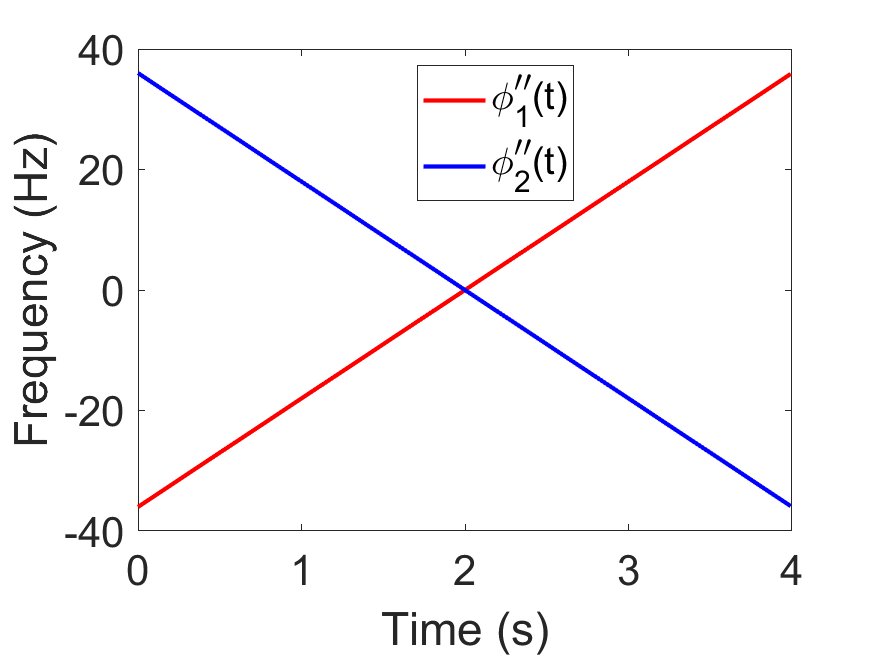}}
&\resizebox {1.5in}{1.2in} {\includegraphics{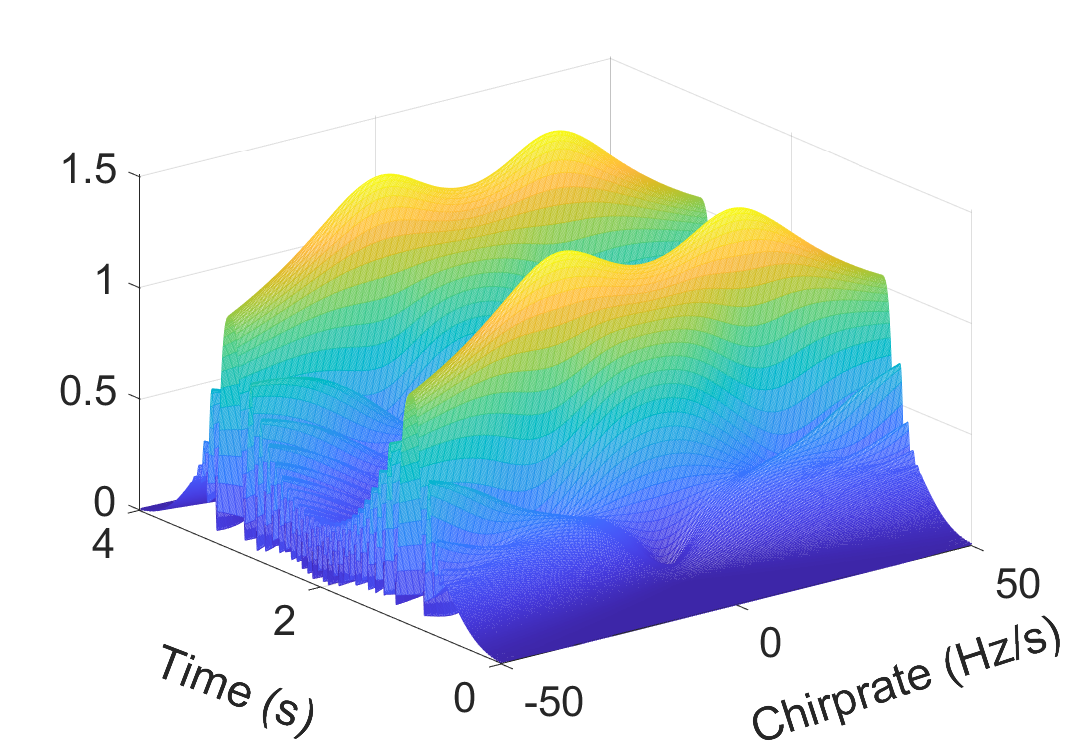}}	
&  \resizebox {1.5in}{1.2in} {\includegraphics{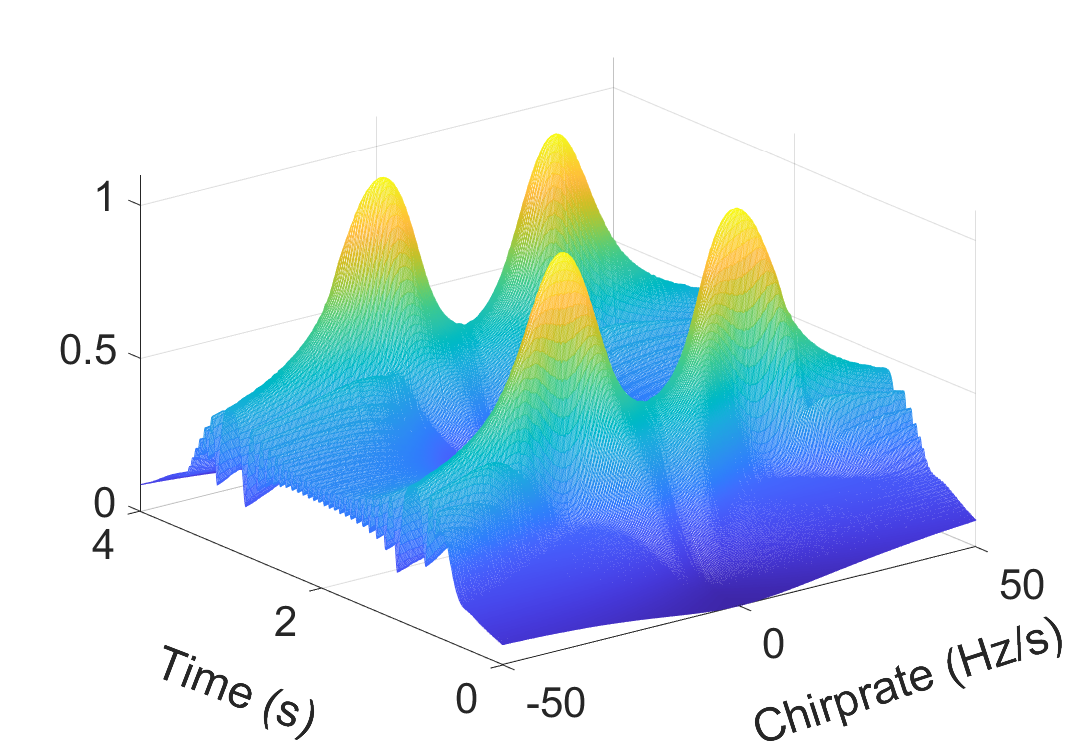}}
	\end{tabular}
	\caption{\small From left to right:  IFs of $y_1, y_2$ (1st); Chirprates of $y_1, y_2$ (2nd); WCT of $y$  with $a=\frac 1{38}$ (third); XWCT of $y$ with $a=\frac 1{38}$ (4th)}
	\label{figure:Example2_WCT}
	\label{figure:Example2_IFs}
\end{figure}

\bigskip 
{\bf Example 2.} Let 
\begin{equation}
\label{def_y}
y(t)=y_1(t)+y_2(t), \; y_1(t)=e^{i2\pi(3(t-2)^3+29t)}, y_2(t)=e^{i2\pi(-3(t-2)^3+47t)}, \; t\in [0, 4),  
\end{equation} 
with $t$ discretized as $t_n=n \Delta t, 0\le n<4\times 128$ with sampling step $\Delta=\frac 1{128}$. With $\phi_1'(t)=9(t-2)^2+29, \phi_2'(t)=-9(t-2)^2+47$, one observes that the IFs of $y_1$ and $y_2$ are crossover at $t=1$ and $t=3$ with 
$
\phi_1'(1)=\phi_1'(3)=\phi_2'(1)=\phi_2'(3)=38. 
$
In addition,  $\phi_1''(t)=18(t-2)$ and $\phi_2''(t)=-18(t-2)$ are crossover at $t=2$. 
The IFs and chirprates of $x_1, x_2$ are shown in the 1st and 2nd panels respectively (from the left) of  Fig.\ref{figure:Example2_IFs}.

The ideal representation of $y(t)$ in 3D space is 
$$
{\rm IR}{}_y(a,b,\gl)=y_1(b)\gd\big(1/a-\phi_1'(b)\big)\gd\big(\gl-\phi_1''(b)\big)+y_2(b)\gd\big(1/a-\phi_2'(b)\big)\gd\big(\gl-\phi_2''(b)\big). 
$$ 
Next let us look at the WCT and XWCT of $y$ when $a=\frac 1{38}$, which correspoding to $\phi_k'(1), \phi_k'(3), k=1, 2$. First one has 
\begin{equation}
\label{Example2_IR_b=4}
\big|{\rm IR}{}_y(1/{38},b,\gl)\big|
=\gd(b-1)\gd(\gl+18)+\gd(b-1) \gd(\gl-18)+\gd(b-3)\gd(\gl+18)+\gd(b-3)\gd(\gl-18). 
\end{equation}

%%%%%%%%%%%%%%%%%%%the beginning of figure 5 %%%%%%%%%%%%%%%
%%%%%%%%%%%%%%%%the end of figure 5  %%%%%%%%%%%%%%%%%%%%%

For this signal $y(t)$, the correspoding parameter $\gs_0$ determined by \eqref{entropy_gs0} is  4.21. 
$|U_y(\frac 1{38}, b, \gl)|$ (WCT with $a=\frac 1{38}$) and $|\cU_y(\frac 1{38}, b, \gl)|$ (XWCT with $a=\frac 1{38}$) are present on the 3rd panel and the 4th panel respectively of  Fig.\ref{figure:Example2_WCT}. One can see along the time $b$ direction, both $|U_y(\frac 1{38}, b, \gl)|$ and $|\cU_y(\frac 1{38}, b, \gl)|$ approach to 0 fast as $b$ moves away from $b=1$ or $b=3$.  This is 
exactly the property of the ideal represestation  $\big|{\rm IR}{}_y(\frac 1{38}, b, \gl)\big|$ has: 
$\big|{\rm IR}{}_y(\frac 1{38},b,\gl)\big|=0$ for $b\not=1, 3$.

\begin{figure}[th]	\centering
	\begin{tabular}{cc}
	\resizebox {2.0in}{1.5in} {\includegraphics{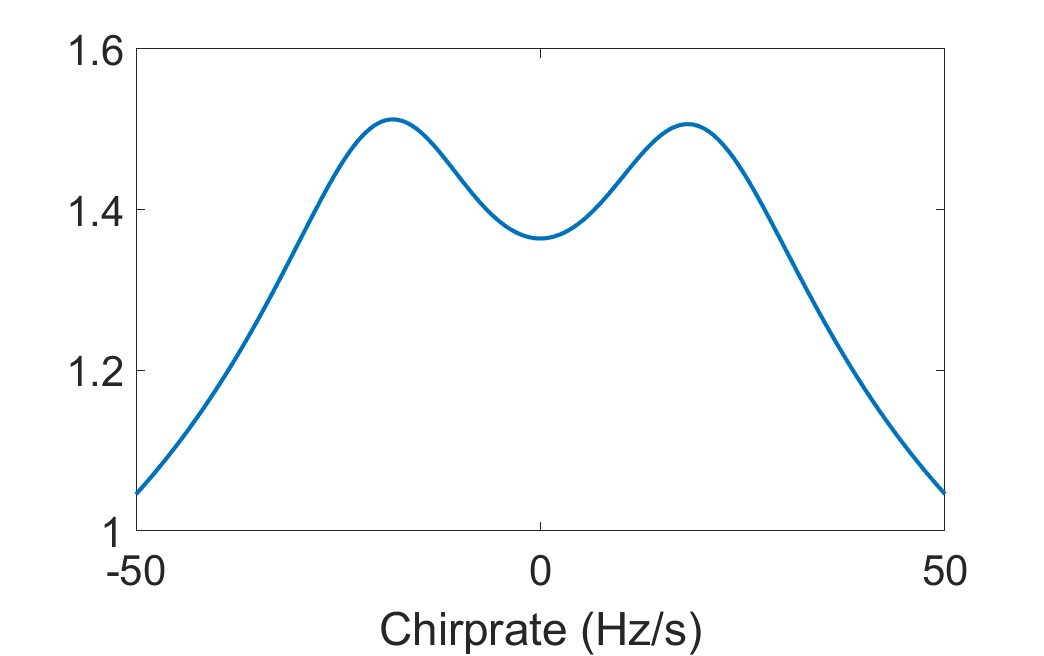}}	\quad & \quad  \resizebox {2.0in}{1.5in} {\includegraphics{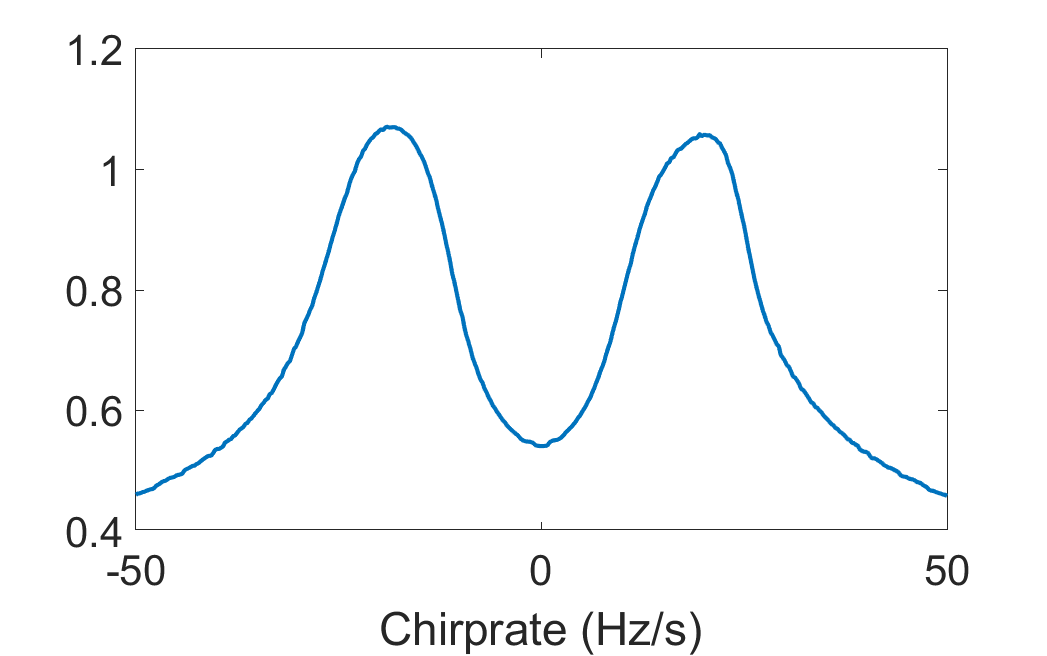}}
	\end{tabular}
	\caption{\small Slice of WCT $|U_y(\frac 1{38}, 1, \gl)|$ (Left) and slice of XWCT $|\cU_y(\frac 1{38}, 1, \gl)|$ (Right)}	\label{figure:Example2_WCT_slice}
\end{figure}

Along the chirprate $\gl$ direction,  the ideal represestation  $\big|{\rm IR}{}_y(\frac 1{38},b,\gl)\big|=0$ if $b\not=1, 3$, and  
$$
\big|{\rm IR}{}_y(1/{38},1,\gl)\big|=\big|{\rm IR}{}_y(1/{38},3,\gl)\big|=
\gd(\gl+18)+\gd(\gl-18). 
$$
From the 3rd panel of Fig.\ref{figure:Example2_WCT}, one can see 
both $|U_y(\frac 1{38}, 1, \gl)|$ and $|U_y(\frac 1{38}, 3, \gl)|$
have two peaks along the direction $\gl$, which correspod to $\gd(\gl+18)$ and $\gd(\gl-18)$ respectively. Observe that $|U_y(\frac 1{38}, 1, \gl)|$ and $|U_y(\frac 1{38}, 3, \gl)|$ decay very slowly when $\gl$ moves away from $\gl=-18$ or from $\gl=18$. However, as shown on the right panel of Fig.\ref{figure:Example2_WCT}, $|\cU_y(\frac 1{38}, 1, \gl)|$  ($|\cU_y(\frac 1{38}, 3, \gl)|$) decays much faster than $|U_y(\frac 1{38}, 1, \gl)|$ ($|U_y(\frac 1{38}, 3, \gl)|$ respectively)   when $\gl$ moves away from $\gl=-18$ or from $\gl=18$.  Fig.\ref{figure:Example2_WCT_slice} shows the slice of  WCT and XWCT:  $|U_y(\frac 1{38}, 1, \gl)|$ (on the left) and  $|\cU_y(\frac 1{38}, 1, \gl)|$ (on the right). Clearly $|\cU_y(\frac 1{38}, 1, \gl)|$  decays fairly fast when $\gl$ moves away from $\gl=-18$ or from $\gl=18$.

%%%%%%%%%%%%%%%%%%%the beginning of figure 6 %%%%%%%%%%%%%%%
%%%%%%%%%%%%%%%%the end of figure 6  %%%%%%%%%%%%%%%%%%%%%

\begin{figure}[th]
	\centering
	\begin{tabular}{cccc}
		\resizebox {1.4in}{1.1in} {\includegraphics{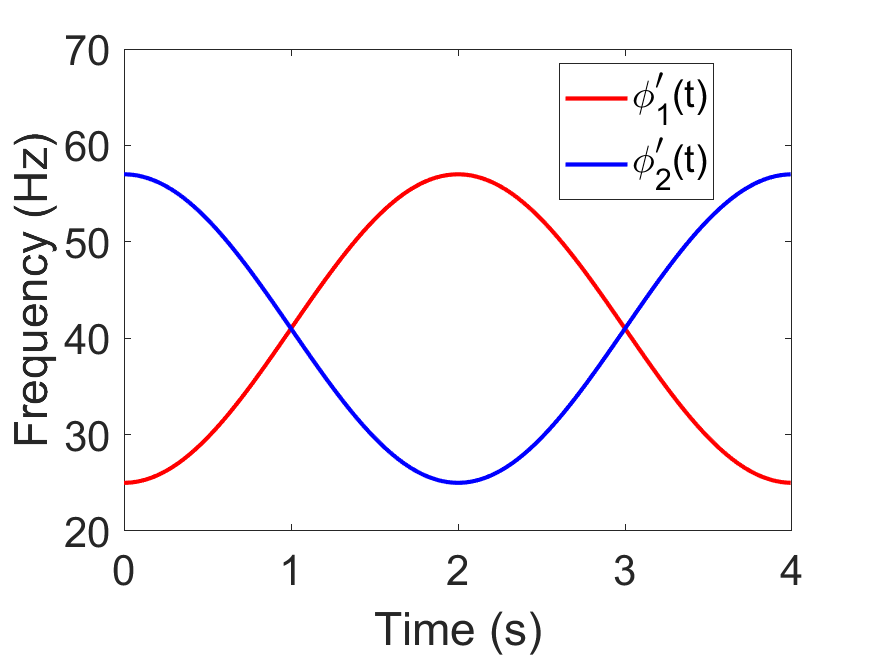}}
		&  \resizebox {1.4in}{1.1in} {\includegraphics{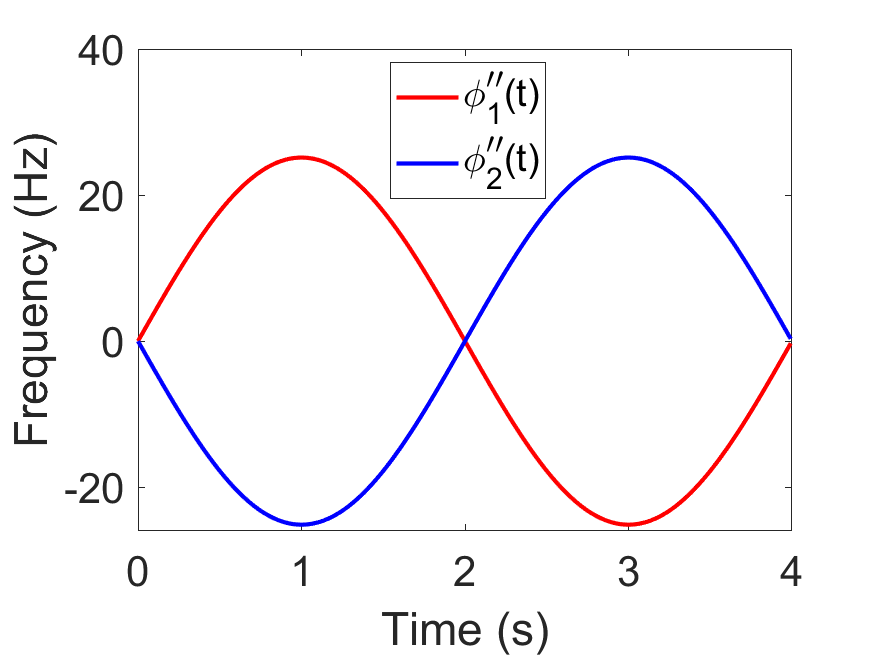}}
\resizebox {1.5in}{1.2in} {\includegraphics{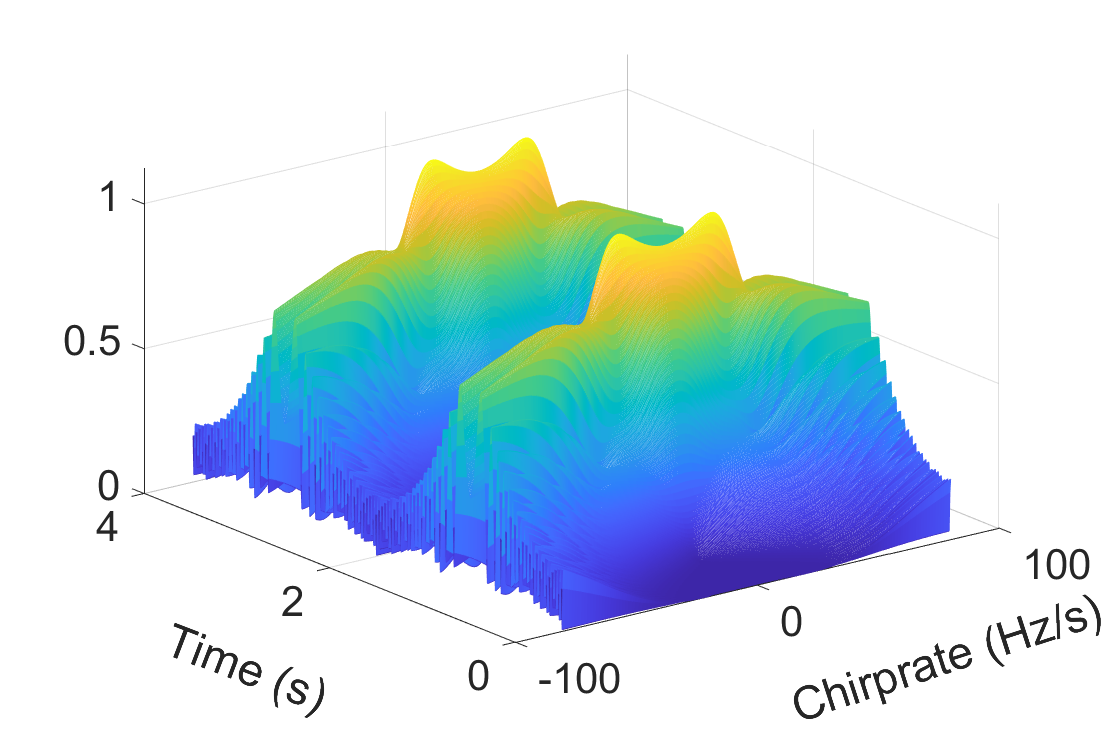}}
& \resizebox {1.5in}{1.2in} {\includegraphics{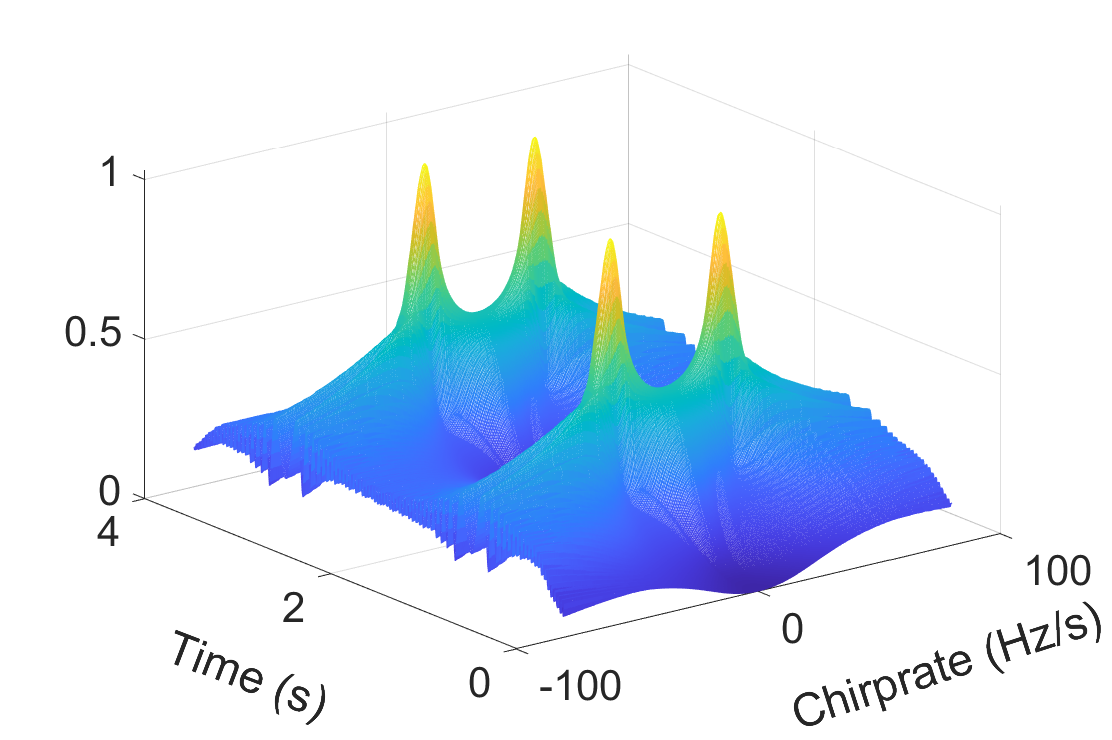}}
	\end{tabular}
	%\vskip -0.3cm 
\caption{\small From left to right:  IFs of $z_1, z_2$ (1st); Chirprates of $z_1, z_2$ (2nd); WCT of $z$  with $a=\frac 1{41}$ (third); XWCT of $z$ with $a=\frac 1{41}$ (4th)}
	\label{figure:Example3_IFs}
\label{figure:Example3_WCT}
\end{figure}

\bigskip 
{\bf Example 3.} Let 
\begin{equation}
\label{def_z}
z(t)=z_1(t)+z_2(t), \; z_1(t)=e^{i2\pi\phi_1(t)}, z_2(t)=e^{i2\pi \phi_2(t)}, \; t\in [0, 4),  
\end{equation}
where 
$
\phi_1(t)=41t-\frac{32}\pi \sin(\frac \pi 2t), \; \phi_2(t)=41t+\frac{32}\pi \sin(\frac \pi 2t),  
$
and $t$ is also discretized as $t_n=n \Delta t, 0\le n<4\times 128$ with sampling step $\Delta=\frac 1{128}$. 
With 
\begin{eqnarray*}
&&\phi_1'(t)=41-16\cos(\frac \pi 2t), \; \phi_2'(t)=41+16\cos(\frac \pi 2t), \phi_1''(t)=8\pi \sin(\frac \pi 2t), \; \phi_2''(t)=-8\pi \sin(\frac \pi 2t), 
\end{eqnarray*}
one knows $\phi_1'(t)$ and $\phi_2'(t)$ are crossover at $t=1, t=3$, while $\phi_1''(t)$ and $\phi_2''(t)$ are crossover at $t=2$. The IFs and chirprates of $z_1, z_2$ are shown in the 1st and 2nd panels (from the left) of  Fig.\ref{figure:Example3_IFs}.
 
%%%%%%%%%%%%%%%%%%%the beginning of figure 7 %%%%%%%%%%%%%%%
%%%%%%%%%%%%%%%%the end of figure 7  %%%%%%%%%%%%%%%%%%%%%

The ideal representation of $z(t)$ in 3D space is 
$$
{\rm IR}{}_z(a,b,\gl)=z_1(b)\gd\big(1/a-\phi_1'(b)\big)\gd\big(\gl-\phi_1''(b)\big)+z_2(b)\gd\big(1/a-\phi_2'(b)\big)\gd\big(\gl-\phi_2''(b)\big). 
$$ 
 Note that $\phi_k'(1)=\phi_k'(3)=41, k=1, 2$.  Let us consider the case $a=1/41$. One can obtain 
\begin{equation}
\label{Example3_IR_b=4}
\big|{\rm IR}{}_z(1/{41},b,\gl)\big|
=\gd(b-1)\gd(\gl+8\pi)+\gd(b-1)\gd(\gl-8\pi)+\gd(b-3)\gd(\gl+8\pi)+\gd(b-3)\gd(\gl-8\pi). 
\end{equation}

\begin{figure}[H]
	\centering
	%\begin{tabular}{ccc}
	\begin{tabular}{cc}
	\resizebox {2.0in}{1.5in} {\includegraphics{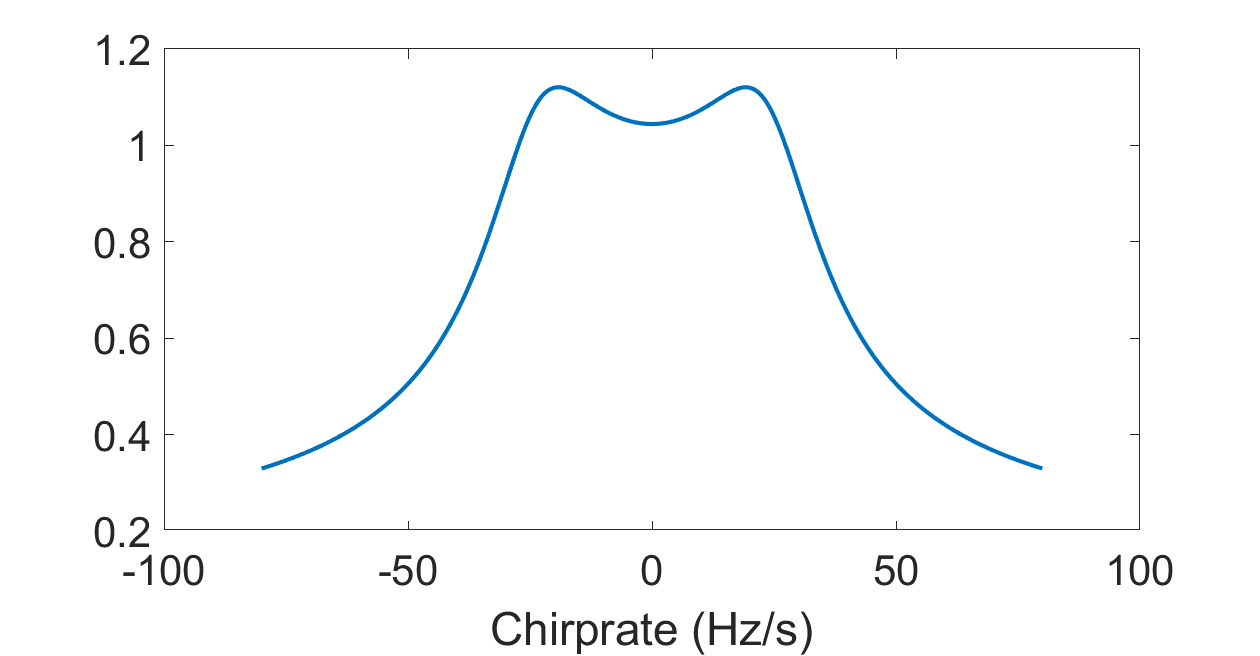}}	\quad & \quad  \resizebox {2.0in}{1.5in} {\includegraphics{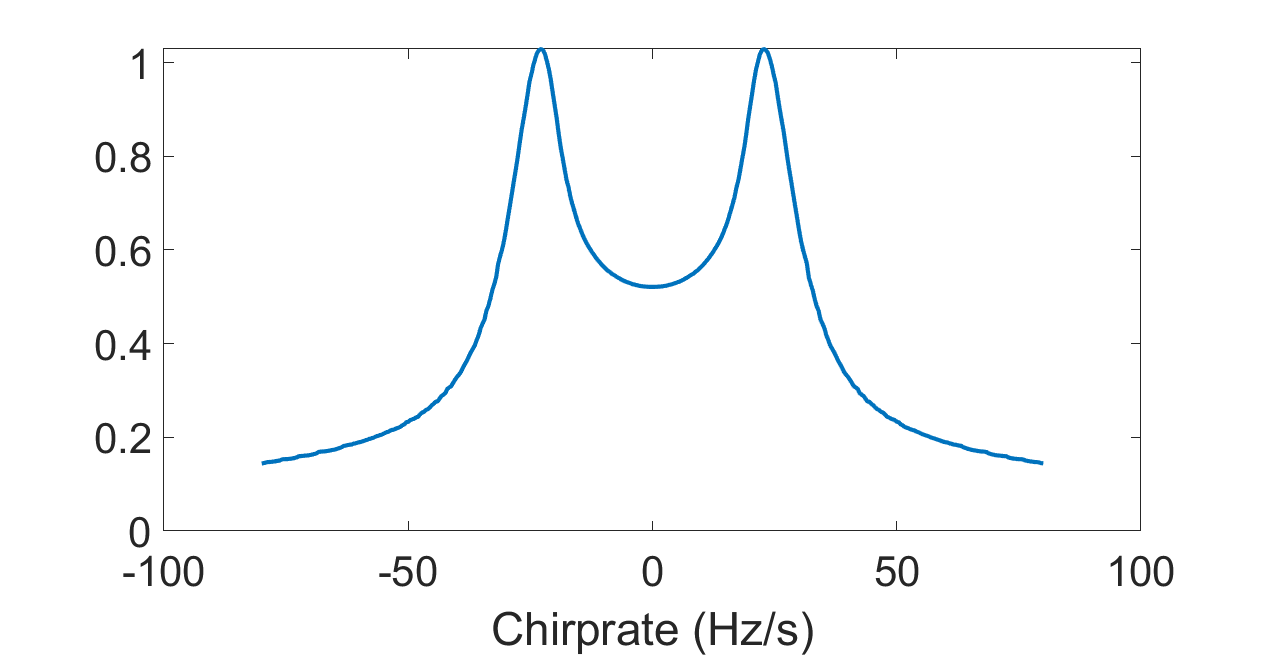}}
	\end{tabular}
	%\vskip -0.3cm 
	\caption{\small Slice of WCT $|U_z(\frac 1{41}, 1, \gl)|$ (Left) and slice of XWCT $|\cU_z(\frac 1{41}, 1, \gl)|$ (Right)}
	\label{figure:Example3_WCT_slice}
\end{figure}

The parameter $\gs_0$ determined by \eqref{entropy_gs0} for $z(t)$ is  5.02.  
$|U_z(\frac 1{41}, b, \gl)|$ (WCT with $a=\frac 1{41}$) and $|\cU_z(\frac 1{41}, b, \gl)|$ (XWCT with $a=\frac 1{41}$) are provided on the 3rd panel and the 4th panel respectively of  Fig.\ref{figure:Example3_WCT}. Along the chirprate $\gl$ direction,  the ideal represestation  $\big|{\rm IR}{}_z(\frac 1{41},b,\gl)\big|=0$ if $b\not=1, 3$, and  
$$
\big|{\rm IR}{}_z(1/{41},1,\gl)\big|=\big|{\rm IR}{}_z(1/{41},3,\gl)\big|=
\gd(\gl+8\pi)+\gd(\gl-8\pi). 
$$
The 3rd panel of Fig.\ref{figure:Example3_WCT} shows that
both $|U_z(\frac 1{41}, 1, \gl)|$ and $|U_z(\frac 1{41}, 3, \gl)|$
have two peaks along the direction $\gl$, which are associated with $\gd(\gl+8\pi)$ and $\gd(\gl-8\pi)$ respectively. Observe that $|U_z(\frac 1{41}, 1, \gl)|$ and $|U_z(\frac 1{41}, 1, \gl)|$ decay very slowly when $\gl$ moves away from $\gl=-8\pi$ or from $\gl=8\pi$. However, as shown on the 4th panel of Fig.\ref{figure:Example3_WCT}, $|\cU_y(\frac 1{41}, 1, \gl)|$  ($|\cU_z(\frac 1{41}, 3, \gl)|$) decays much faster than $|U_z(\frac 1{41}, 1, \gl)|$ ($|U_z(\frac 1{41}, 3, \gl)|$ resp.)   when $\gl$ moves away from $\gl=-8\pi$ or from $\gl=8\pi$.   Fig.\ref{figure:Example3_WCT_slice} shows the slice of 
WCT and XWCT:  $|U_z(\frac 1{41}, 1, \gl)|$ (on the left) and  $|\cU_z(\frac 1{41}, 1, \gl)|$ (on the right). Clearly $|\cU_z(\frac 1{41}, 1, \gl)|$  decays fairly fast when $\gl$ moves away from $\gl=-8\pi$ or from $\gl=8\pi$.

%%%%%%%%%%%%%%%%%%%the beginning of figure 8 %%%%%%%%%%%%%%%
%%%%%%%%%%%%%%%%the end of figure 8  %%%%%%%%%%%%%%%%%%%%%

From the above examples, one can see XWCT, the WCT incorporated with a special X-ray transform, indeed has much fast decay than WCT. In particular the XWCT of $z(t)$ on the 4th panel of  Fig.\ref{figure:Example3_WCT} looks close to 
$$
\gd(b-1)\gd(\gl+8\pi)+\gd(b-1)\gd(\gl-8\pi)+\gd(b-3)\gd(\gl+8\pi)+\gd(b-3)\gd(\gl-8\pi),
$$
the ideal representation of the signal $z(t)$ when $a=\frac 1{41}$.

\section{Third-order synchrosqueezed  X-ray wavelet-chirplet transform } 
	
To consider a synchrosqueezing transform, first one needs to define the IF  reference function and the chirprate reference function (also called the IF reassignment operator and the chirprate reassignment operator respectively). An IF reference function $\gO^g_x(a, b, \gl)$ and chirprate reference function $\gL^g_x(a, b, \gl)$ are said to be of order 2 (order 3 resp.) if they are respectively $\phi'(t)$ and $\phi''(t)$ for any linear (quadratic resp.) chirps $x(t)=Ae^{i2\pi \phi(t)}$, and the corresponding synchrosqueezed WCT transform ({SWCT}) is called the 2nd-order SWCT (3rd-order SWCT resp.).  In the literature, the SWCT or SCT is referred to the 2nd-order SWCT or SCT. 
Thus the 3rd-order SWCT/SCT is a higher-order squeezed transform.

Though this paper focuses on the 3-order SWCT and the 3-order XSWCT this paper, for self-containing purpose and for comparison, the 2nd-order $\gO^g_x(a, b, \gl)$ and $\gL^g_x(a, b, \gl)$ will be provided. 

\subsection{Synchrosqueezed wavelet-chirplet transform} 

 $x(t)$ is said to be a (generalized) linear chirp if
\begin{equation}
\label{def_chirp}
x(t)=A(t) e^{i2\pi \phi(t)}=A e^{pt+\frac 12 qt^2} e^{i2\pi (ct +\frac 12 r t^2)},   
\end{equation}
where $A, p, q, c$ and $r$ are real constants with $p, q$ small (in modulus). 

In the following $U^{g'}_x(a, b, \gl), U^{g''}_x(a, b, \gl), U^{b^j g(b)}_x(a, b, \gl)$ will be used to denote the WCTs of $x(t)$  define by \eqref{def_TSC} with the window function $g(b)$ replaced by $g'(b), g''(b)$ and $b^j g(b)$ resp., where $j\ge 1$. 

First one has for a signal $x(t)$,  
\begin{eqnarray}
\nonumber {\partial _b} U^g_x(a, b, \gl)\hskip -0.6cm &&
= \int_{\RR} x'(b+at)  g(t) e^{-i2\pi \mu  t -i\pi \gl a^2 t^2}dt= \int_{\RR} \frac 1a \frac {\partial}{\partial t} \big( x(b+at) \big) g(t) e^{-i2\pi \mu  t -i\pi \gl a^2 t^2}dt\\
\nonumber&&= -\frac 1a \int_{\RR} x(b+at) \frac {\partial}{\partial t} \big(  g(t) e^{-i2\pi \mu  t -i\pi \gl a^2 t^2} \big)dt
\\ \nonumber &&
= -\frac 1a \int_{\RR} x(b+at) \big (g'(t)+g(t)(-i2\pi \mu   -i2\pi \gl a^2 t) \big)e^{-i2\pi \mu  t -i\pi \gl a^2 t^2} dt. 
\end{eqnarray}
Thus 
\begin{equation}
\label{derU1}
{\partial _b} U^g_x(a, b, \gl) =-\frac 1a  U^{g'}_x(a, b, \gl)+\frac{i 2\pi \mu }a  U^g_x(a, b, \gl) +i2\pi \gl a  U^{bg(b)}_x(a, b, \gl).   
\end{equation}

On the other hand, for $x(t)$ given by \eqref{def_chirp}, one has 
$
x'(t)=\big(p+qt +i2\pi(c+rt)\big)x(t). 
$
Thus
\begin{eqnarray}
\nonumber {\partial _b} U^g_x(a, b, \gl)\hskip -0.6cm &&= \int_{\RR} x'(b+at)  g(t) e^{-i2\pi \mu  t -i\pi \gl a^2 t^2}dt\\
\nonumber&&= \int_{\RR} \big(p+q(b+at)+i2\pi(c+r(b+at))\big)x(b+at)  g(t) e^{-i2\pi \mu  t -i\pi \gl a^2 t^2}dt\\
\label{derU2}&&= \big(p+qb+i2\pi(c+rb)\big) U^g_x(a, b, \gl)+ (q+i2\pi r) a U^{bg(b)}_x(a, b, \gl). 
\end{eqnarray}
Hence, when \(U^g_x(a, b, \gl)\neq 0\), one has 
\begin{equation}
  \label{eqution1stdiff}  {\partial_b U^g_x(a, b, \gl)}/{U^g_x(a, b, \gl)}=\big(p+qb+i2\pi(c+rb)\big) + (q+i2\pi r) a {U^{bg(b)}_x(a, b, \gl)}/{U^{g}_x(a, b, \gl)}. 
\end{equation}
Differentiating both sides of \eqref{eqution1stdiff} with respect to \( b \), one has 
\begin{equation*}
    \partial_b\Big({\partial_b U^g_x}/{U^g_x}\Big)  = (q + i 2 \pi r) + (q + i 2 \pi r) a \partial_b\Big({U^{bg(b)}_x}/{U^g_x}\Big),  
 \end{equation*}
which leads to  
\begin{eqnarray}   
  \label{lambda}  q + i 2 \pi r \hskip -0.6cm&&= \frac{\partial_b\left({\partial_b U^g_x}/{U^g_x}\right)}{1 + a \partial_b\left({U^{bg(b)}_x}/{U^g_x}\right)}. 
\end{eqnarray}  
Plugging $q + i 2 \pi r $ in \eqref{lambda} into \eqref{eqution1stdiff} results in 
\begin{eqnarray}  
  %\nonumber 
p+qb+i2\pi(c+rb)%\hskip -0.6cm&&
= \frac{\partial_b U^g_x}{U^g_x}
- (q+i2\pi r) a \frac{U^{bg(b)}_x}{U^{g}_x}
\label{omega}%\hskip -0.6cm&&
=\frac{\partial_b U^g_x} {U^g_x}
-\frac{a U^{bg(b)}_x\partial_b\left(\frac{\partial_b U^g_x }{U^g_x}\right)}{ U^g_x \big(1 + a \partial_b\left(\frac{U^{bg(b)}_x}{U^g_x}\right)\big)}. 
\end{eqnarray}

Denote 
\begin{equation}
\label{def_D}
D_x^g(a,b,\gl):= {U^{g'}_xU_x^{bg(b)}-U^{g}_xU_x^{bg'(b)}+i 2\pi \lambda a^2(U_x^{b^2g(b)}U_x^{g}-U_x^{bg(b)}U_x^{bg(b)})}. 
\end{equation}
Applying \eqref{derU1} and  \eqref{derU1} with $g(b)$ replaced by $g'(b)$ and $bg(b)$ respectively    
to the right-hand sides of \eqref{lambda} and \eqref{omega}, one has for $a, b, \gl$ with $D_x^g(a,b,\gl)\not=0$, 
\begin{eqnarray*}
 &&\frac q{i2\pi}+r=\gl+\frac 1{i a^2 2\pi} \frac {{U_x^{g}U_x^{g''}-U_x^{g'}U_x^{g'}+i 2\pi \lambda a^2(U_x^{bg(b)}U_x^{g'}-U_x^{bg'(b)}U_x^{g}-U_x^{g}U_x^{g})}}{(U^{g'}_xU_x^{bg(b)}-U^{g}_xU_x^{bg'(b)})+i 2\pi \lambda a^2(U_x^{b^2g(b)}U_x^{g}-U_x^{bg(b)}U_x^{bg(b)})},\\
&&\frac {p+qb}{i2\pi}+c+rb\\
&&=\frac \mu a-\frac 1{i a 2\pi} \frac {U_x^{bg(b)}U_x^{g''}-U_x^{bg'(b)}U_x^{g'}+i 2\pi \lambda a^2(U_x^{b^2g(b)}U_x^{g'}-U_x^{bg'(b)}U_x^{bg(b)}-U_x^{bg(b)}U_x^g)}{(U^{g'}_xU_x^{bg(b)}-U^{g}_xU_x^{bg'(b)})+i 2\pi \lambda a^2(U_x^{b^2g(b)}U_x^{g}-U_x^{bg(b)}U_x^{bg(b)})}. 
 \end{eqnarray*}
Therefore the IF $\phi'(b)$ and chirprate $\phi''(b)$ of $x(t)$ can be obtained by 
\begin{equation*}
\phi'(b)=c+rb=\gO^g_x(a, b, \gl), \; \phi''(b)=r =\gL^g_x(a, b, \gl), 
\end{equation*}
where 
\begin{eqnarray}
\label{def_gO}&&\gO^g_x(a, b, \gl):= \\
&& \nonumber \quad \frac \mu a -\frac 1a {\rm Re}\Big(\frac 1{i2\pi} \frac {U_x^{bg(b)}U_x^{g''}-U_x^{bg'(b)}U_x^{g'}+i 2\pi \lambda a^2(U_x^{b^2g(b)}U_x^{g'}-U_x^{bg'(b)}U_x^{bg(b)}-U_x^{bg(b)}U_x^g)}{ U^{g'}_xU_x^{bg(b)}-U^{g}_xU_x^{bg'(b)}+i 2\pi \lambda a^2(U_x^{b^2g(b)}U_x^{g}-U_x^{bg(b)}U_x^{bg(b)})}\Big),\\
\label{def_gL} &&\gL^g_x(a, b, \gl) :=
%\\ && \nonumber \quad 
\gl+\frac 1{a^2}{\rm Re} \Big(\frac 1{i2\pi} \frac {{U_x^{g}U_x^{g''}-U_x^{g'}U_x^{g'}+i 2\pi \lambda a^2(U_x^{bg(b)}U_x^{g'}-U_x^{bg'(b)}U_x^{g}-U_x^{g}U_x^{g})}}{ U^{g'}_xU_x^{bg(b)}-U^{g}_xU_x^{bg'(b)}+i 2\pi \lambda a^2(U_x^{b^2g(b)}U_x^{g}-U_x^{bg(b)}U_x^{bg(b)})}\Big). 
\end{eqnarray}
Hence for a general signal $x(t)=A(t)e^{i2\pi \phi(t)}$, the quantities on the right-hand sides of 
%the above equations 
\eqref{def_gO} \eqref{def_gL}
are good candidates for IF $\phi'(t)$ and chirprate $\phi''(t)$. To summarize, one has the following proposition. 
\begin{pro}
 For a signl $x(t)=A(t)e^{i2\pi \phi(t)}$, define its IF reference function $\gO^g_x(a, b, \gl)$ and chirprate reference function $\gL^g_x(a, b, \gl)$ by \eqref{def_gO} and \eqref{def_gL} respectively. 
If $x(t)$ is a linear chirp given in \eqref{def_chirp}, then $\gO^g_x(a, b, \gl)=\phi'(b), \gL^g_x(a, b, \gl)=\phi''(b)$. 
\end{pro}	

The 2nd-order synchrosqueezed ({SWCT}) is defined by 
\begin{equation}
	\label{def_STSC}
	\begin{split}
	S_x(\xi, b, \gga) &:= \iint_{\{(a,\gl):U^g_x(a,b,\gl)\not=0, \quad  D^g_x(a,b,\gl)\not=0\}} \hskip -1.6cm U^g_x(a,b,\gl) 
%\\ &\quad \times 
 \gd\big(\xi-\gO^g_x(a, b, \gl)\big)\gd\big(\gga-\gL^g_x(a, b, \gl)\big)\frac{da}{a} d\gl, 
	\end{split}
\end{equation}
where $D^g_x(a,b,\gl)$ is defined by \eqref{def_D}. 

Next the multiple SWCT ({MSWCT}), which is further synchrosequeezed transform of SWCT, is defined. Notice that SWCT is a frequency-time-chirprate representation of a signal, while one may use 
$\gO^g_x(a, b, \gl)$ and $\gL^g_x(a, b, \gl)$, which are functions of scale, time and chirprate, to  define MSWCT. Thus one needs to use the relation  between frequency and scale: 
$$
\xi=\mu/a.  
$$
Denote $S^{[1]}_x(\xi, b, \gga):=S_x(\xi, b, \gga)$. The MSWCT is defined iteratively by 
\begin{equation}
	\label{def_MSTSC}
	\begin{split}
	S^{[n]}_x(\xi, b, \gga) &:= \iint_{\big\{(\eta,\gl): U^g_x(\frac{u}{\eta}, b, \gl)\not=0, D^g_x(\frac{u}{\eta}, b, \gl)\not=0 \big\}}  \hskip -1.6cm S^{[n-1]}_x(\eta, b, \gl) 
%\\ &\quad \times 
\gd\big(\xi-\gO^g_x(\frac{\mu}{\eta}, b, \gl)\big)\gd\big(\gga-\gL^g_x(\frac{\mu}{\eta}, b, \gl)\big)\frac{d\eta}{\eta} d\gl,
	\end{split}
\end{equation}
where $n= 2,3 \cdots$. Note that $\frac {d\eta}\eta$ in \eqref{def_MSTSC}, instead of $d\eta$, is used. This is due to the fact that in the implementation of WCT and SWCT, the scale variable $a$ is discretized as $2^{vj} \gD t, j\in \ZZ$, where  $v$ is a number and $\gD t$ is the sample rate of the time variable $b$. Considering the relationship between $a$ and $\eta$, it will be better to discretize $\eta$ in the same way as $a$ when one implements MSWCT. Oberve that \eqref{def_MSTSC} can be written as the following equation. 
\begin{equation}
	\label{def_MSTSC1}
	\begin{split}
	S^{[n]}_x(\xi, b, \gga) &= \iint_{\big\{(a,\gl):U^g_x(a, b, \gl)\not=0, D^g_x(a, b, \gl)\not=0\big\}}  \hskip -1.6cm S^{[n-1]}_x\left({\mu}/{a}, b, \gl\right) %\\	&\quad \times 
\gd\big(\xi-\gO^g_x(a, b, \gl)\big)\gd\big(\gga-\gL^g_x(a, b, \gl)\big)\frac{da}{a} d\gl.
	\end{split}
\end{equation}
Furthermore, $S^{[n]}_x(\xi, b, \gga)$ can be written as 
\begin{equation}
\label{def_MSTSC2}
\begin{split}
S^{[n]}_x(\xi, b, \gga) &= \iint_{\big\{(a,\gl):U^g_x(a, b, \gl)\not=0, D^g_x(a, b, \gl)\not=0\big\}} \hskip -1.6cm U_x(a, b, \gl) 
%\\&\quad \times 
\gd\big(\xi-\gO^{g, n-1}_x(a, b, \gl)\big)\gd\big(\gga-\gL^{g, n-1}_x(a, b, \gl)\big)\frac {da}a  d\gl,
\end{split}
\end{equation}
where 
\begin{equation}
\label{def_iterativeO}
\begin{array}{l}
\gO^{g, 0}_x(a, b, \gl):=\gO^{g}_x(a, b, \gl),  \gL^{g, 0}_x(a, b, \gl):=\gL^{g}_x(a, b, \gl),  
\\
\gO^{g, j}_x(a, b, \gl):=\gO^{g}_x\Big(\frac \mu{\gO^{g, j-1}_x(a, b, \gl)}, b, \gL^{g, j-1}_x(a, b, \gl)\Big),
\\
\gL^{g, j}_x(a, b, \gl):=\gL^{g}_x\Big(\frac \mu{\gO^{g, j-1}_x(a, b, \gl)}, b, \gL^{g, j-1}_x(a, b, \gl)\Big), \quad j=1,2 \cdots.
\end{array}
\end{equation}

\subsection{3rd-order synchrosqueezed wavelet-chirplet transform} 
In this subsection the 3rd-order SWCT and the 3rd-order SXWCT are considered.  Let $x(t)$ be a (generalized) quadratic chirp:
\begin{equation}
\label{def_chirp_cubic}
x(t)=A(t) e^{i2\pi \phi(t)}=e^{d(t)} e^{i2\pi \phi(t)},   \;  d(t)=d_0+d_1t+d_2 t^2 +d_3t^3, 
\phi(t)=c_0+c_1t+c_2 t^2 +c_3t^3,
\end{equation}
where $d_j, c_j\in \RR$. With  $x'(t)=\big(d'(t)+i2\pi \phi'(t)\big)x(t)$, one has  
\begin{eqnarray*}
&& \nonumber {\partial_b} U^g_x(a, b, \gl)= \int_{\RR} x'(b+at)  g(t) e^{-i2\pi \mu  t -i\pi \gl a^2 t^2}dt\\
\nonumber&&= \int_{\RR} \Big(d'(b)+i2\pi \phi'(b)+ (d''(b)+i2\pi \phi''(b))at+ 
\frac 12 (d^{(3)}(b)+i2\pi \phi^{(3)}(b))(at)^2\Big)\times \\
&& \qquad \qquad x(b+at)  g(t) e^{-i2\pi \mu  t -i\pi \gl a^2 t^2}dt
\\
\nonumber &&= \big(d'(b)+i2\pi \phi'(b)
\big) U^g_x(a, b, \gl)+ a (d''(b)+i2\pi \phi''(b)) U^{bg(b)}_x(a, b, \gl) 
\\\nonumber && \qquad  
+\frac {a^2}2 
(d^{(3)}(b)+i2\pi \phi^{(3)}(b)) U^{b^2g(b)}_x(a, b, \gl). 
\end{eqnarray*}
Thus one has 
\begin{equation}
\label{3rd_eq1}
\frac {\partial_bU^g_x}{i2\pi U^g_x}= \frac {d'(b)}{i2\pi}+\phi'(b)
+ a \big(\frac {d''(b)}{i2\pi} + \phi''(b)\big) \frac {U^{bg(b)}_x}{U^g_x}+\frac {a^2}2 
\big(\frac {d^{(3)}(b)}{i2\pi}+ \phi^{(3)}(b)\big) \frac {U^{b^2g(b)}_x}{U^g_x}. 
\end{equation}
Taking $\partial _\gl$ and $\partial_b$ respectively to the both side of \eqref{3rd_eq1}, one reaches 
\begin{eqnarray*}
&& \frac 1{i2\pi} \partial_\gl\Big( \frac {\partial_bU^g_x}{U^g_x}\Big)=
a \big(\frac {d''(b)}{i2\pi} + \phi''(b)\big)\partial_\gl\Big(\frac {U^{bg(b)}_x}{U^g_x}\Big)+\frac {a^2}2 
\big(\frac {d^{(3)}(b)}{i2\pi}+ \phi^{(3)}(b)\big) \partial_\gl\Big(\frac {U^{b^2g(b)}_x}{U^g_x}\Big)\\
&&\frac 1{i2\pi} \partial_b\Big( \frac {\partial_bU^g_x}{U^g_x}\Big)=
\big(\frac {d''(b)}{i2\pi} + \phi''(b)\big)\Big(a\partial_b\Big(\frac {U^{bg(b)}_x}{U^g_x}\Big)+1\Big)+
\\ && \qquad \qquad \qquad \qquad \qquad 
\big(\frac {d^{(3)}(b)}{i2\pi}+ \phi^{(3)}(b)\big)\Big( 
\frac {a^2}2  \partial_b\Big(\frac {U^{b^2g(b)}_x}{U^g_x}\Big)+a \frac {U^{bg(b)}_x}{U^g_x}\Big). 
\end{eqnarray*}
Thus 
$$
\left[
\begin{array}{cc}
a\partial_\gl\Big(\frac {U^{bg(b)}_x}{U^g_x}\Big)&\frac {a^2}2 
\partial_\gl\Big(\frac {U^{b^2g(b)}_x}{U^g_x}\Big)\\
a \partial_b\Big(\frac {U^{bg(b)}_x}{U^g_x}\Big)+1
&\frac {a^2}2  \partial_b\Big(\frac {U^{b^2 g(b)}_x}{U^g_x}\Big)+a \frac {U^{bg(b)}_x}{U^g_x}
\end{array}
\right]
\left[
\begin{array}{l}
 \frac {d''(b)}{i2\pi} + \phi''(b)\\
\frac {d^{(3)}(b)}{i2\pi}+ \phi^{(3)}(b)
\end{array}
\right]
=
 \frac 1{i2\pi} 
\left[
\begin{array}{l}
\partial_\gl\Big( \frac {\partial_bU^g_x}{U^g_x}\Big)\\
\partial_b\Big( \frac {\partial_bU^g_x}{U^g_x}\Big)
\end{array}
\right]
$$

Denote 
\begin{equation}
\label{def_D0}
D^g_{x,0}:=\partial_\gl\Big(\frac {U^{bg(b)}_x}{U^g_x}\Big)\Big( a\partial_b\Big(\frac {U^{b^2 g(b)}_x}{U^g_x}\Big)+2 \frac {U^{bg(b)}_x}{U^g_x}\Big)-
\partial_\gl\Big(\frac {U^{b^2g(b)}_x}{U^g_x}\Big)\Big(
a \partial_b\Big(\frac {U^{bg(b)}_x}{U^g_x}\Big)+1\Big). 
\end{equation}
Then, if $D^g_{x, 0}\not=0$, one has 
$$
\left[
\begin{array}{l}
 \frac {d''(b)}{i2\pi} + \phi''(b)\\
\frac {d^{(3)}(b)}{i2\pi}+ \phi^{(3)}(b)
\end{array}
\right]
=
 \frac 1{i\pi a^2 D^g_{ x,0}} 
\left[
\begin{array}{cc}
\frac {a^2}2  \partial_b\Big(\frac {U^{b^2 g(b)}_x}{U^g_x}\Big)+a \frac {U^{bg(b)}_x}{U^g_x}
&-\frac {a^2}2 
\partial_\gl\Big(\frac {U^{b^2g(b)}_x}{U^g_x}\Big)\\
-a \partial_b\Big(\frac {U^{bg(b)}_x}{U^g_x}\Big)-1
&a\partial_\gl\Big(\frac {U^{bg(b)}_x}{U^g_x}\Big)
\end{array}
\right]
\left[
\begin{array}{l}
\partial_\gl\Big( \frac {\partial_bU^g_x}{U^g_x}\Big)\\
\partial_b\Big( \frac {\partial_bU^g_x}{U^g_x}\Big)
\end{array}
\right], 
$$
which leads to
\begin{eqnarray}
\label{dd_phi_2}&&\frac {d''(b)}{i2\pi} + \phi''(b)=\frac 1{i\pi a} \frac{D^g_{x,1}}{D^g_{x,0}}, 
\label{ddd_phi_3}\frac {d^{(3)}(b)}{i2\pi}+ \phi^{(3)}(b)=\frac 1{i\pi a^2} \frac{D^g_{x,2}}{D^g_{x,0}}, 
\end{eqnarray}
where
\begin{eqnarray}
\label{def_D1}
&&D^g_{x,1}:=\Big( \frac {a}2  \partial_b\Big(\frac {U^{b^2 g(b)}_x}{U^g_x}\Big)+\frac {U^{bg(b)}_x}{U^g_x}\Big)\partial_\gl\Big( \frac {\partial_bU^g_x}{U^g_x}\Big)-\frac {a}2 
\partial_\gl\Big(\frac {U^{b^2g(b)}_x}{U^g_x}\Big)\partial_b\Big( \frac {\partial_bU^g_x}{U^g_x}\Big), 
\\
\label{def_D2} &&D^g_{x,2}:=-\Big(a \partial_b\Big(\frac {U^{bg(b)}_x}{U^g_x}\Big)+1\Big)\partial_\gl\Big( \frac {\partial_bU^g_x}{U^g_x}\Big)
+a\partial_\gl\Big(\frac {U^{bg(b)}_x}{U^g_x}\Big)\partial_b\Big( \frac {\partial_bU^g_x}{U^g_x}\Big). 
\end{eqnarray}
Plugging the expressions of $\frac {d''(b)}{i2\pi} + \phi''(b)$ and 
$\frac {d^{(3)}(b)}{i2\pi}+ \phi^{(3)}(b)$ in \eqref{dd_phi_2} %and \eqref{ddd_phi_3} 
into \eqref{3rd_eq1}, one has 
\begin{equation*}
 \frac {d'(b)}{i2\pi}+\phi'(b)=\frac {\partial_bU^g_x}{i2\pi U^g_x}-
 \frac 1{i\pi} \frac{D^g_{x,1}}{D^g_{x,0}}\frac {U^{bg(b)}_x}{U^g_x}-
\frac 1{i2\pi} \frac{D^g_{x,2}}{D^g_{x,0}}
 \frac {U^{b^2g(b)}_x}{U^g_x}. 
\end{equation*}
Define 
\begin{eqnarray}
\label{3rd_IF}&&\gT_x^g(a, b, \gl):={\rm Re} \Big(\frac 1{i2\pi U^g_x}\big( \partial_bU^g_x-
 2\frac{D^g_{x,1}}{D^g_{x,0}}\; U^{bg(b)}_x-
\frac{D^g_{x,2}}{D^g_{x,0}}\;  U^{b^2g(b)}_x\big)\Big),\\
\label{3rd_chirprate}&&\Xi^g_x(a, b, \gl):= {\rm Re} \Big(\frac 1{i\pi a} \frac{D^g_{x,1}}{D^g_{x,0}}\Big). 
\end{eqnarray}
By the above derivation, one has the following proposition.
 
\begin{pro}
\label{pro2}
 For a signl $x(t)=A(t)e^{i2\pi \phi(t)}$, define its 3rd-order IF reference function $\gT^g_x(a, b, \gl)$ and chirprate reference function $\Xi^g_x(a, b, \gl)$ by \eqref{3rd_IF} and \eqref{3rd_chirprate} respectively with $D_{x, 0}, D^g_{x,1}$ and $D^g_{x,2}$ given by   \eqref{def_D0}, \eqref{def_D1} and \eqref{def_D2} respectively. 
Then $\gT^g_x(a, b, \gl)=\phi'(b), \Xi^g_x(a, b, \gl)=\phi''(b)$ provided that $x(t)$ is a quadratic chirp given in  \eqref{def_chirp_cubic}. 
\end{pro}	
 
The 3rd-order synchrosqueezed WCT ({SWCT}), denoted as $T_x(\xi, b, \gga)$ here, is a synchrosqueezed WCT with  $\gT_x^g(a, b, \gl), \Xi^g_x(a, b, \gl)$ used as the IF and chirprate reference functions: % and the chirprate reference function: 
\begin{equation}
	\label{def_SWCT}
	\begin{split}
	T_x(\xi, b, \gga) &:= \iint_{\{(a,\gl):U^g_{x}(a,b,\gl)\not=0, \quad D^g_{x, 0}(a,b,\gl)\not=0\}}\hskip -1.6cm  U^g_x(a,b,\gl) 
\gd\big(\xi-\gT^g_x(a, b, \gl)\big)\gd\big(\gga-\Xi^g_x(a, b, \gl)\big)\frac{da}{a} d\gl,
	\end{split}
\end{equation}
where $D^g_{x, 0}(a,b,\gl)$ is defined by \eqref{def_D0}. 
The IF reference function $\gT^g_x(a, b, \gl)$ and the chirprate reference function $\Xi^g_x(a, b, \gl)$ are also used to define the 3rd-order synchrosqueezed XWCT ({SXWCT}), denoted as ${\mathcal T}_x(\xi, b, \gga)$: 
\begin{equation}
	\label{def_SXWCT}
	\begin{split}
	\cT_x(\xi, b, \gga) &:= \iint_{\{(a,\gl):U^g_{x}(a,b,\gl)\not=0 ,D^g_{x, 0}(a,b,\gl)\not=0\}} \hskip -1.6cm \cU^g_x(a,b,\gl) 
	%\\ &\quad \times 
\gd\big(\xi-\gT^g_x(a, b, \gl)\big)\gd\big(\gga-\Xi^g_x(a, b, \gl)\big)\frac{da}{a} d\gl,
	\end{split}
\end{equation}
where $\cU^g_x(a,b,\gl)$ is the XWCT of $x(t)$ defined by \eqref{def_XRWCT}. 

The multiple 3rd-order SWCT  can be defined iteratively as \eqref{def_MSTSC}  and it can written as \eqref{def_MSTSC2} merely with $\gO^g_x(a, b, \gl)$ and $\gL^g_x(a, b, \gl)$ replaced by $\gT^g_x(a, b, \gl)$ and $\Xi^g_x(a, b, \gl)$ respectively. 
Additionally, the multiple 3rd-order SXWCT is defined in a corresponding manner.
%iteratively as \eqref{def_MSTSC} and they can written as \eqref{def_MSTSC2}.

To implement the 3rd-order SWCT and SXWCT, one needs to compute  $\gT^g_x(a, b, \gl)$, $\Xi^g_x(a, b, \gl)$. Their expressions, which are given in term of $U_x^{b^\ell g(b)}$,  $U_x^{b^\ell g'(b)}$ and $U_x^{g''(b)}$,  are provided in the appendix. 

\section{Implementation and mode retrieval experimental results}

In this section,  the implementations of WCT, SWCT and multiple SWCT (MSWCT)  
are provided.   Experimenetal results of IF and chirprate estimation and mode retrieval by SWCT, MSWCT and SXWCT are present. 

\subsection{Implementation}
	
In this paper the Fourier transform of a signal $x(t)$ is defined by 
$$
\wh x(\xi):=\int_\RR x(t) e^{-i2\pi \xi t}dt. 
$$	
Then one has  
$$
\int_\RR x(t)\overline {y(t)}dt =\int_\RR \wh x(\eta)\overline {\wh y(\eta)}d\eta, 
$$
or 
$$
\int_\RR x(t){y(t)}dt =\int_\RR \wh x(\eta) \wh y(-\eta)d\eta.  
$$

%Note that $\wb g (0, 0)=1$ since $\int_\R g(t) dt=1$. 

Next we rewrite WCT of a signal $x(t)$ defined by \eqref{def_TSC} as follows. 
\begin{eqnarray}
\nonumber U^g_x(a, b, \gl) \hskip -0.6cm  &&= \int_{\RR} x(b+at)  g(t) e^{-i2\pi \mu  t -i\pi \gl a^2 t^2}dt\\
\nonumber  &&=\int_{\RR} \big(x(b+at)\big)^{\wedge}(\eta) \big( g(t) e^{-i2\pi \mu  t -i\pi \gl a^2 t^2}\big)^{\wedge}(-\eta)d\eta\\
\nonumber &&=\int_{\RR} \frac 1a e^{\frac{i2\pi \eta b}a} \wh x(\frac\eta a) \wb g(\mu-\eta, a^2\gl) d\eta\\
\label{TSC_freq}&&=\int_{\RR} \Big( \wh x(\eta) \wb g(\mu-a\eta, a^2\gl)\Big) e^{i2\pi \eta b} d\eta,  
\end{eqnarray}
where in the above $\big(x(b+at)\big)^{\wedge}(\eta)$ denotes the Fourier transform of $x(b+at)$ with respect to variable $t$.  Thus one can use FFT (the fast Fourier transform, for $x(t)$) and IFFT (the inverse fast Fourier transform, for  $\wh x(\eta) \wb g(\mu-a\eta, a^2\gl)$ with variable $\eta \in \R$) to implement WCT.

Next we consider calculating the WCT $U^g_x(a, b, \gl)$ of a signal with a window function $g(t)$ by applying 
\eqref{TSC_freq}.  In practice, for a particular signal $x(t)$, its WCT $U^g_x(a, b, \gl)$ 
lies in a region of the scale-time-chirprate plane: 
$$
\{(a, b, \gl): \; A_0 \le a\le A_1, 0\le b \le B_1, -R_0\le \gl \le R_0\}
$$ 
for some $0<A_0, A_1, B_1, R_0<\infty$. That is $U^g_x(a, b, \gl)$ 
is negligible for  $(a, b, \gl)$ outside this region. 

Suppose the input signal $x(t)$ is discretized uniformly at points
$$
t_n=t_0+n\gD t, \; n=0, 1, \cdots, N-1, t_0=0, 
$$
where $\gD t$ is the time-variable sample step.
%Denote $x_n=x(t_n).$
Let ${\bf x}\in \CC^N$ denote the discretization of $x(t)$ with sampling points $t_n$:
$$
{\bf x}:=\big[x(t_n)\big]_{n=N-1}^0.
$$

Let
\begin{equation}
\label{def_etak}
\eta_k:=\left\{ \begin{array}{ll}
k\gD\eta, \; & \hbox{for $0\le k\le \lfloor\frac N2\rfloor$},\\
(k-N)\gD\eta, \;  & \hbox{for $\lfloor\frac N2\rfloor<k<N$},
\end{array}
\right.
\end{equation}
where $\gD\eta:=\frac 1{N\gD t}$,
be the sampling points for the frequency variable $\eta$. Then $\wh x(\eta_k), 0\le k\le N-1$,
the discretization of the Fourier transform $\wh {x}(\eta)$ of $x(t)$ can be approximated by 
\begin{equation}
\label{approx_FT_x1}
 \big[\wh x(\eta_k)\big]_{k=N-1}^0  \approx \gD t \;   \wh {\bf x}=\gD t \; {\rm F}_N{\bf x},
\end{equation}
%\todo[inline,color=yellow!40]{I'm not sure whether it should be \( k = N-1 \) to \( 0 \) or \( k = 0 \) to \( N-1 \). This issue seems to appear later as well.}
where $F_N$ is the discrete Fourier transformation matrix of dimension $N$. % defined in Section 2.4.  
$\wh {\bf x}:=F_N{\bf x}$ can be obtained by FFT.

As in \cite{thakur2013synchrosqueezing}, the scale variable is discretized as
$$
a_j=2^{j \gD\wt a} \gD t,  \; j=1, 2, \cdots, J_0,
$$
where
$$
J_0:=\Big\lceil \frac {\log_2 N-1}{\gD \wt a}\Big\rceil,
$$
and $\gD \wt a$ is a sample step of $\ln a$ which the user can choose. One may choose $\gD \wt a$ to be a number such as 
$\frac 1{32}$, $\frac 1{64}$, or $\frac 1{128}$.

The time variable $b$ and the chirprate variable $\gl$ in the WCT are discretized respectively by
$$
b_m=m \gD t, \; m=0, 1, \cdots, N-1; \; \gl_\ell=-R_0+\ell \gD\gl, \; \ell=0, 1, \cdots, L-1, 
$$
where the user can choose the sampling step $\gD\gl$ and $L=\Big\lfloor \frac {2R_0}{\gD \gl}\Big\rfloor$. 

Observe that for $0\le m\le N-1$ and $0\le k\le N/2$,
$$
b_m \eta_k=\frac {m k}N, \; a_j \eta_k=\frac {2^{j \gD \wt a} \; k}N.
$$
For each $j$ and $\ell$, let $\wb {\bf G}_{j, \ell}$ denote a column vector of dimension $N$ with the $(k+1)$th-component, 
denoted by $\big(\wb {\bf G}_{j, \ell}\big)_k$, 
 given by
$$
\big(\wb {\bf G}_{j, \ell}\big)_k:=\wb g(\mu-a_j \eta_k, a_j^2\gl_\ell), \; k=0, 1, \cdots, N-1.
$$
Then one has 
\begin{eqnarray*}
&&{ U^g_x(a_j, b_m, \gl_\ell)}=\int_{-\infty}^\infty \wh x(\eta) \wb g(\mu-a_j \eta, a_j^2\gl_\ell) e^{i2\pi b_m \eta} d\eta
\\
&& \approx  \gD \eta \sum_{k=0}^{N-1} \wh{ x}(\eta_k)\wb g(\mu-a_j\eta_k, a_j^2\gl_\ell) e^{i2\pi b_n\eta_k} \\
&&\approx \frac{1}{N \gD t}\sum_{k=0}^{N} \gD t  (F_N {\bf x})_k 
\big(\wb {\bf G}_{j, \ell}\big)_k e^{i2\pi m k/N}\\
&&=\frac{1}{N}\sum_{k=0}^{N-1}  \Big(F_N {\bf x}\odot 
\wb {\bf G}_{j, \ell}\Big)_k
e^{i2\pi m k/N}\\
&&=\Big(F_N^{-1}\big(F_N{\bf x}\odot \wb {\bf G}_{j, \ell}\big)\Big)_m,
\end{eqnarray*}
where $F_N^{-1}$ is the inverse DFT of dimension $N$, $F_N \bf x\odot \wb {\bf G}_{j, \ell}$ denotes the element-wise product of vectors $F_N \bf x$ and $\wb {\bf G}_{j, \ell}$ obtained by multiplying these two vectors element by element.
Thus ${ U^g_x(a_j, b_m, \gl_\ell)}$ can be approximated by $\wt U^g_x(j, m, \ell)$ given by
\begin{equation}
\label{TSC_discret0}
\Big[\wt U^g_x(j, m, \ell)\Big]_{m=N-1}^0=F_N^{-1}\big(F_N {\bf x}\odot \wb {\bf G}_{j, \ell}\big),
\end{equation}
where $1\le j\le J_0, 0\le \ell \le L-1$. Note that $F_N^{-1}$ can be calculated by ifft. Thus ${ \wt U^g_x(j, m, \ell)}$ can be obtained by 
\begin{equation}
\label{TSC_discret}
\Big[\wt U^g_x(j, m, \ell)\Big]_{m=N-1}^0=\text{ifft}\big(\text{fft}({\bf x})\odot \wb {\bf G}_{j, \ell}\big). 
\end{equation}

Next we consider the implementation of SWCT and MSWCT. For the chirprate variable $\gga$ of SWCT $S_x(\xi, b, \gga)$, we may 
discretize it as $\gga_p=-R_0+p \gD \gga, p=0, 1, 2, \cdots$ for some sampling step $\gD \gga$ (one may just set  $\gD \gga =\gD \gl$). For the frequency variable $\xi$,  if we just consider SWCT, we may simply discretize $\xi$ uniformly: $\xi_k=\xi_0+k \gD \xi, k=0, 1, \cdots$ for some sampling step $\gD \xi$. 
Let  $a_j, b_m, \gl_\ell,  j, m, \ell=0, 1, \cdots, $ be the the sampling points of $a, b, \gl$ respectively didcussed above and let ${ U^g_x(a_j, b_m, \gl_\ell)}$ (actually $\wt U^g_x(j, m, \ell)$) be the WCT of $x(t)$ calculated above. Then the SWCT of $x(t)$ is given  by
\begin{equation}
\label{def_STSC_dis}
S_x(\xi_k, b_m, \gga_p)= 
\mathop{\sum}_{(j, \ell)\in O_{k, m, p} }
 U^g_x(a_j, b_m, \gl_\ell)a_j^{-1} (\gD a)_j \gD\gl,
\end{equation}
where $(\gD a)_j:=a_{j+1}-a_j$, and 
$$
 O_{k, m, p}:=\left\{ \begin{array}{l}
(j, \ell): \;  |\gO^g_x(a_j, b_m, \gl_\ell)-\xi_k|\le \frac 12 \gD\xi, 
 |\gL^h_x(a_j, b_m, \gl_\ell)-\gga_p|\le \frac 12 \gD\gga\\
\qquad \quad  |U^g_x(a_j, b_m, \gl_\ell)|>0, |D^g_{x,0}(a_j, b_m, \gl_\ell)|>0
 \end{array}
 \right\}. 
 $$
 
\bigskip 
When we consider MSWCT, as we mentioned above, we should discretize the frequency varilable $\xi$ in the same way as the scale variable $a$. Recall $a$ is discretized as $a_j=2^{j \gD\wt a} \gD t,  \; 
j=1, 2, \cdots, J_0$. We now discretzie $\xi$ as  
\begin{equation}
\label{def_xik}
\xi_k=\frac \mu{a_{{}_{J_0-k+1}}}, \; k=1, 2, \cdots, J_0. 
\end{equation}
Denote 
$$
(\gD \xi)_k:=\xi_{k+1}-\xi_k. 
$$

With $\xi$ discretized by \eqref{def_xik},  SWCT defined by \eqref{def_STSC} 
is \begin{equation}
\label{def_STSC_dis_power}
S_x(\xi_k, b_m, \gga_p)= 
\mathop{\sum}_{(j, \ell)\in \wt O_{k, m, p} }
 U^g_x(a_j, b_m, \gl_\ell)a_j^{-1} (\gD a)_j \gD\gl,
\end{equation}
where 
\begin{equation}
\label{def_wt_O}
\wt  O_{k, m, p}:=\left\{ \begin{array}{l}
(j, \ell): \;  -\frac 12 (\gD\xi)_{k-1}<\gO^g_x(a_j, b_m, \gl_\ell)-\xi_k\le \frac 12 (\gD\xi)_k, 
 |\gL^h_x(a_j, b_m, \gl_\ell)-\gga_p|\le \frac 12 \gD\gga\\
\qquad \quad  |U^g_x(a_j, b_m, \gl_\ell)|>0, |D^g_{x,0}(a_j, b_m, \gl_\ell)|>0
 \end{array}
 \right\}. 
 \end{equation}

With $S^{[1]}_x(\xi_k, b_m, \gga_p):=S_x(\xi_k, b_m, \gga_p)$ defined by \eqref{def_STSC_dis_power}, the MSWCT after $n$ iterations ($n$ times of synchrosqueezing) is given by  
\begin{equation}
\label{def_MSTSC_dis}
S^{[n]}_x(\xi_k, b_m, \gga_p)= \mathop{\sum}_{ 
(j, \ell) \in \wt O_{k, m, p}}  S^{[n-1]}_x(\frac{\mu}{a_{{}_{J_0-j+1}}}, b_m, \gl_\ell)a_j^{-1} (\gD a)_j \gD\gl, 
\end{equation}
where $\wt O_{k, m, p}$ is defined by \eqref{def_wt_O}. 

To implement SWCT and MSWCT, we need to calculate $\gO^{g}_x$ and $\gL^g_x$. Next we consider the case $g(t)=g_\gs(t)$.
Notice that 
$$
g'_\gs(t)=-\frac 1{\gs^2}tg_\gs(t), \; g''_\gs(t)=-\frac 1{\gs^2}g_\gs(t)+\frac 1{\gs^4}t^2g_\gs(t).  
$$
Then one can obtain when the window functions $g=g_\gs$, % and $h=g_{\wt \gs}$, 
then  $\gO^{g}_x$ and $\gL^g_x$ defined by  \eqref{def_gO} and 
 \eqref{def_gL} respectively are reduced to 
\begin{eqnarray}
\label{def_gO_simple1}\gO^{g_\gs}_x(a, b, \gl) \hskip -0.4cm &&=\frac \mu a -\frac 1a {\rm Re}\Big(\frac 1{i2\pi}  \frac {U^{g_\gs}_x U_x^{bg_\gs(b)}}{(U_x^{bg_\gs(b)})^2-U^{g_\gs}_x U_x^{b^2g_\gs(b)}}\Big),\\
\label{def_gL_simple1} 
\gL^{g_{\gs}}_x(a, b, \gl) \hskip -0.4cm &&=\gl+\frac 1{a^2}{\rm Re} \Big(\frac 1{i2\pi} \frac {(U^{g_{\gs}}_x)^2}{(U_x^{bg_{\gs}(b)})^2-U^{g_{\gs}}_x U_x^{b^2g_{\gs}(b)}}\Big).   
\end{eqnarray}

To implement the 3rd-order SWCT and SXWCT, we will use the expressions of $\gT^{g}_x$ and $\Xi^g_x$ provided in the appendix. In particular, 
when $g(t)=g_\gs(t)$, $\gT^{g_\gs}_x$ and $\Xi^{g_\gs}_x$ are given in \eqref{def_gT_simple},  
%and \eqref{def_Xi_simple} respectively, 
both in the terms of $U_x^{b^jg_\gs(b)}, 1\le j\le 4$. 	Thus, to calculate the IF and chirprate reference functions, we merely calculate $U_x^{b^jg_\gs(b)}, 1\le j\le 4$. 
$U_x^{b^jg_\gs(b)}$ can be computed as  $U_x^{g_\gs}$ by \eqref{TSC_discret} with $\wb g_\gs(\eta, \gl)$ replaced by 
$\big(t^j g_\gs(t)\wb{\big)\;}(\eta, \gl)$.  
$\big(t^2 g_\gs(t)\wb{\big)\;}(\eta, \gl)$ and 
 $\big(t^4 g_\gs(t)\wb{\big)\;}(\eta, \gl)$ have been provided above; and one can obtain   
 \begin{eqnarray*}
 &&\big(t g_\gs(t)\wb{\big)\;}(\eta, \gl)=-\frac 1{i2\pi} \frac {\partial}{\partial \eta}\wb g_\gs(\eta, \gl)
 =\frac {-i2\pi \gs^2\eta}{(1+i2\pi\gs^2 \gl)^{\frac 32}} e^{-\frac{2\pi^2\gs^2 \eta ^2}{1+i2\pi \gs^2\gl}}, \\ 
 &&\big(t^3 g_\gs(t)\wb{\big)\;}(\eta, \gl)=-\frac 1{2\pi^2} \frac {\partial^2}{\partial \gl \partial \eta}\wb g_\gs(\eta, \gl)\\
&& \qquad \qquad =-i2\pi \gs^4 \eta \Big(\frac 3{(1+i2\pi\gs^2 \gl)^{\frac 52}}-
\frac {(2\pi \gs \eta)^2}{(1+i2\pi\gs^2 \gl)^{\frac 72}}\Big) e^{-\frac{2\pi^2\gs^2 \eta ^2}{1+i2\pi \gs^2\gl}}. 
 \end{eqnarray*}

\subsection{Experimenets and results}
In this subsection experimenetal results of IF and chirprate estimation and mode retrieval by SWCT/MSWCT and SXWCT are provided. 

As in Section 3, the Guassian window, namely, $g(t)=g_\gs(t)$ defined by \eqref{def_g}, will be used. 
It's crucial to note that experimental results are significantly influenced by signal sampling rate and the window function's width, namely $\gs$. For fair comparisons between different methods, signals must possess similar sampling rates and utilize identical $\gs$ values. Unfortunately, some of recent research on CT-based and WCT-based IF and chirprate estimation do not specify $\gs$. Consequently, direct comparisons using their examples with the proposed method may be unreliable.
 In this paper the R${\rm \acute e}$nyi entropy of the WCT to determine $\gs$ given by \eqref{entropy_gs0} will be used. 

This paper mainly focuses on the 3rd-order SXWCT and compare its performance with that of the 3rd-order SWCT to show how much the X-ray transform-composed WCT improves the performance.  In the following the authors provide experiment results with the signals $x(t), y(t), z(t)$ considered in Examples 1-3 above. The procedures are as follows. For a given multicomponent signal $x(t)$, first one determines $\gs$ by the entropy of the WCT of the signal given by \eqref{entropy_gs0}. Next the authors calculate the 3rd-order IF and chirprate reference functions $\gT^g_x(a, b, \gl)$ and $\Xi^g_x(a, b, \gl)$ by \eqref{3rd_IF} and \eqref{3rd_chirprate} respectively, more precisely, by \eqref{def_gT_simple}.  
%and \eqref{def_Xi_simple} respectively. 
After that one obtains SWCT $T_x(\xi, b, \gga)$ by \eqref{def_SWCT} and SXWCT $\cT_x(\xi, b, \gga)$ by \eqref{def_SXWCT}. Then a 3D ridge extrctor to SWCT and SXWCT respectively is applied to obtain IF estimates $\check \eta_\ell(b)$ and chirprate estimates $\check \gl_\ell(b)$ of different components with these two methods.  

A classic (2D) time-frequency ridge detection technique was first introduced in \cite{carmona1997characterization} and has gained popularity in recent years due to its effectiveness and high efficiency  \cite{yu2018multisynchrosqueezing,yu2016general,chen2017separation,he2021local}. The reader is referred to \cite{laurent2021novel,meignen2023new,su2024ridge} and references therein for recent new development of 2D  ridge detection techniques. 
In this paper, the authors use a 3D extended version of the 2D ridge extrctor in  \cite{carmona1997characterization} to obtain the IF and chirprate estimation from SWST and SXWCT.  Interested readers can find more details on the specific algorithm implementation for this 3D ridge extractor in \cite{chen2024multiple, zhang2022two}.

 After obtaining the estimated IFs and chirprates, one then uses \eqref{approx_group} in Algorithm 2 (with \( \check a_\ell(b) = 1/\check{\eta}_\ell(b) \)) to retrieve the modes \( \wt{x}_k(b) \).
Our experiments show that for the signals considered in Examples 1-3, the multiple 3rd-order SWCT provides better IF and chirprate estimates than the 3rd-order SWCT, while the 3rd-order SXWCT already yields accurate estimates and the multiple 3rd-order SXWCT does not improve the accurancy. 
In the following the authors present the results with the 3rd-order SWCT, the multiple 3rd-order SWCT (with 5 times iterations) and the 3rd-order SXWCT. 

To measure the error of a retrieved mode, the root mean square error (RMSE) is used. More precisely, let $f=(f_1, f_2, \cdots, f_n)$ be a (discrete) signal and $\wt f$ be its estimate. Then the RMSE is defined by 
$
E_f:=\frac 1{\sqrt n}\|f-\wt f\|_2. 
$
In addition, to avoide the boundary issue, 
the error with the portion of $f$:
$(f_{[\frac n8]}, f_{[\frac n8]+1},\cdots, f_{[\frac {7n}8]})$ will be calculated. 

%%%%%%%%%%%%%%%%%%%the beginning of figure 9 %%%%%%%%%%%%%%%
\begin{figure}[th]
	\centering
	\begin{tabular}{cccc}
	\resizebox{1.5in}{1.0in}{\includegraphics{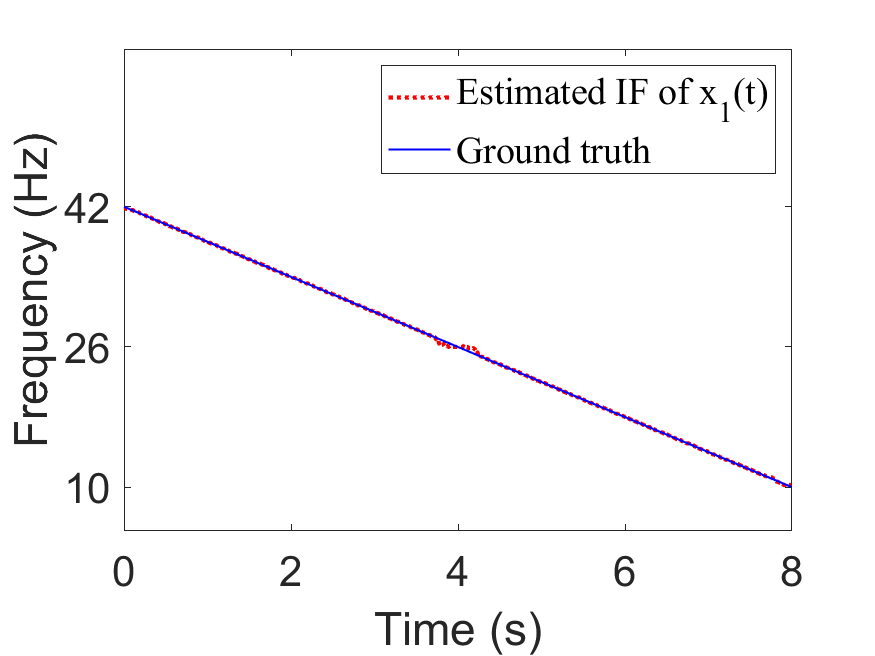}}	
& \resizebox{1.5in}{1.0in} {\includegraphics{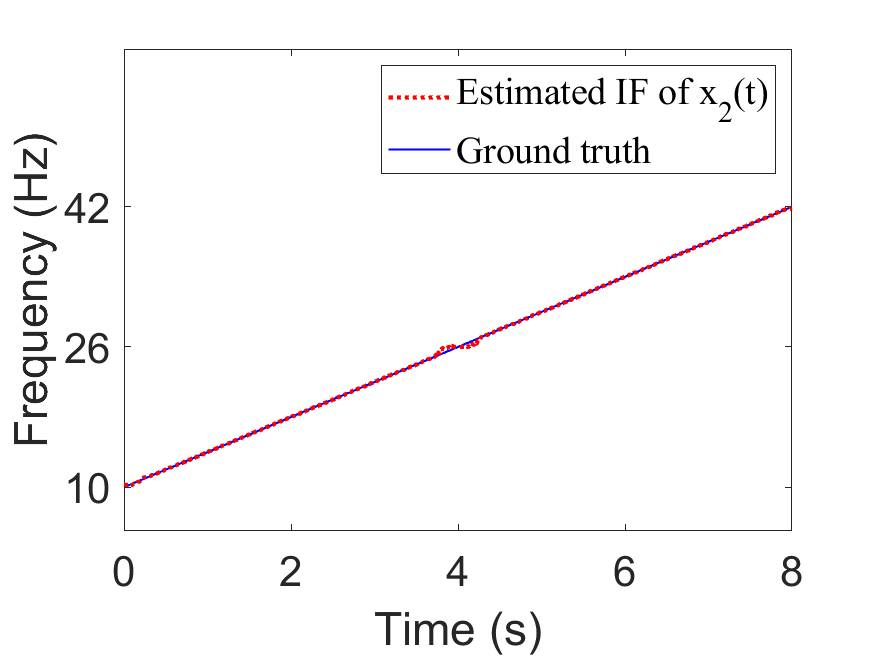}}
&
\resizebox {1.5in}{1.0in} {\includegraphics{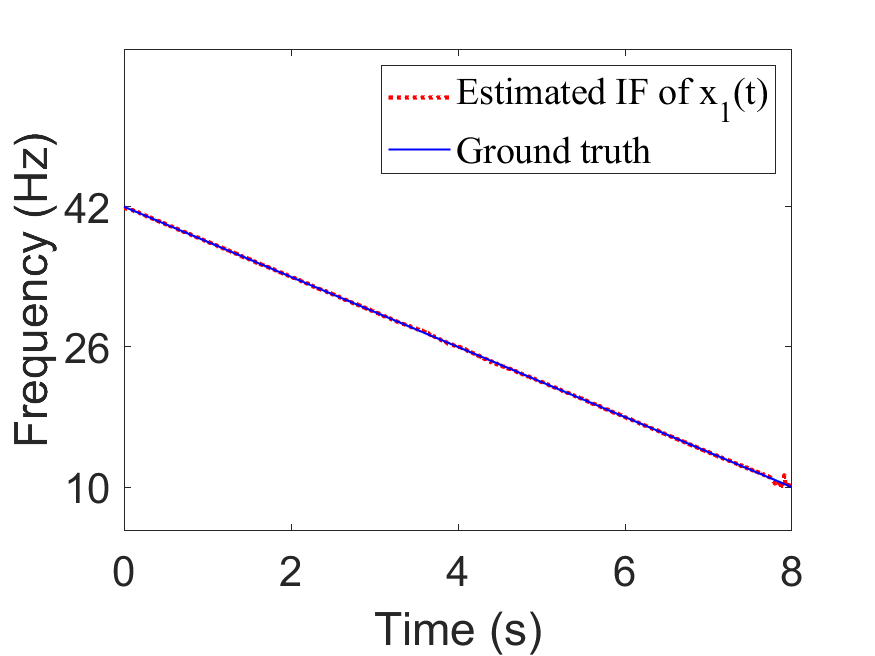}}
& \resizebox{1.5in}{1.0in} {\includegraphics{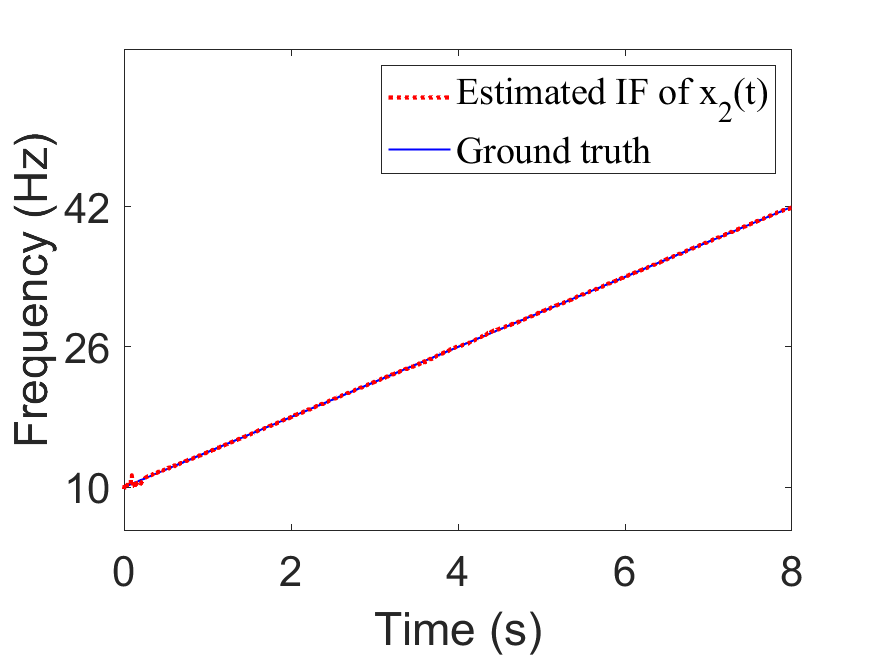}}\\
\resizebox{1.5in}{1.0in} {\includegraphics{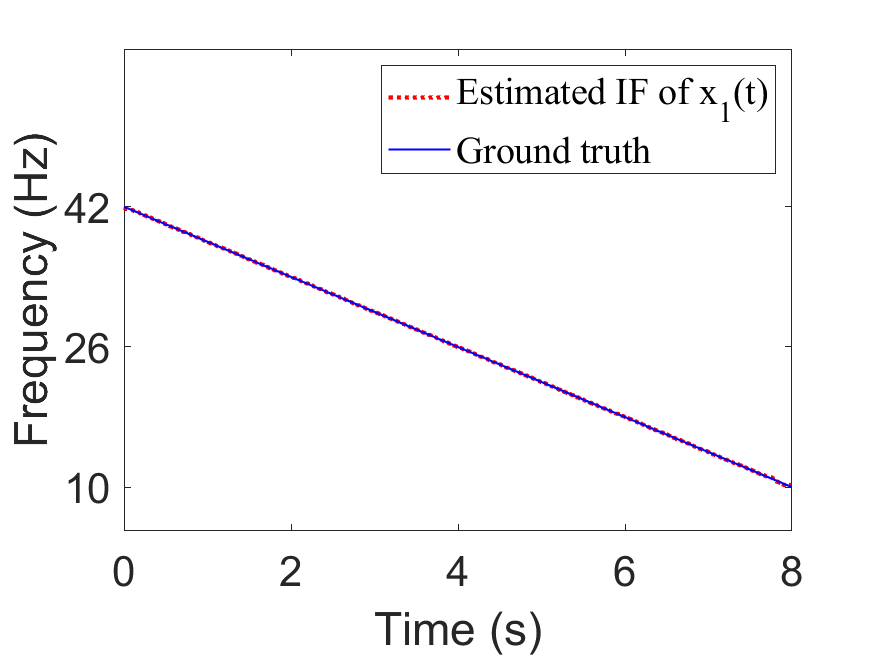}}	
&  \resizebox {1.5in}{1.0in} {\includegraphics{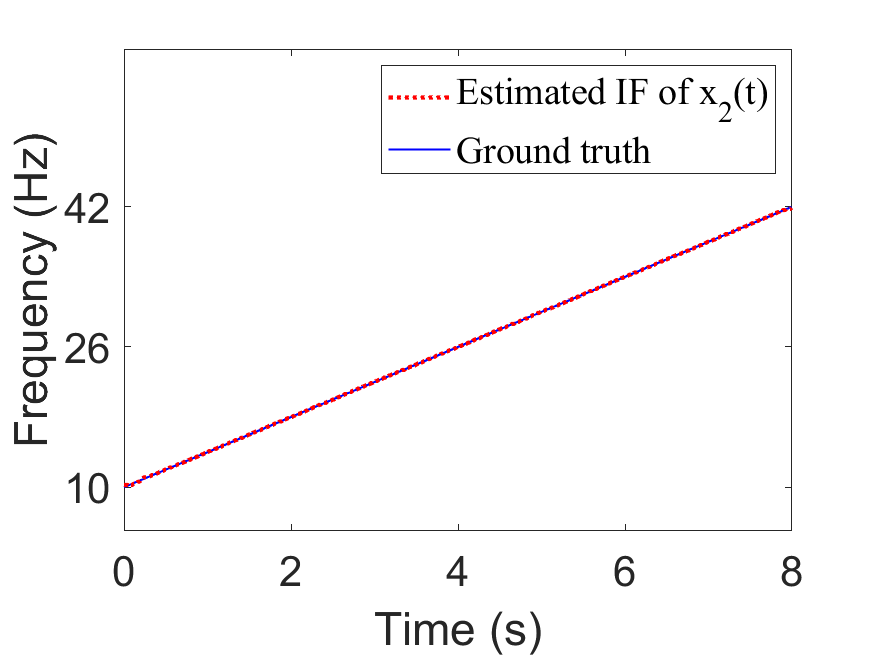}}
&&
\end{tabular}
	%\vskip -0.3cm 
\caption{\small IF estimation of $x_k(t)$. Top row (from left to right): by SWCT (1st and 2nd panels); by  MSWCT (3rd and 4th panels). Bottom row:  by SXWCT.}
	\label{figure:Example1_IFs_est}
\end{figure}
%%%%%%%%%%%%%%%%the end of figure 9  %%%%%%%%%%%%%%%%%%%%%

\bigskip 
{\bf Example 1 (Continued).} Let $x(t)$ be the signal given by \eqref{def_x}. Fig.\ref{figure:Example1_IFs_est} provides IF estimates by SWCT  (in 1st and 2nd panels in the top row), MSWCT  (3rd and 4th panels in top row), and SXWCT (in the bottom row). 
One observes  that overall the SWCT offers a good estimate of the IF, with the exception of the point at $t=4$, where there are some minor estimation errors. Recall the IFs of $x_1(t)$ and $x_2(t)$ is crossover at $t=4$. As shown in the 3rd and 4th panels in the top row and  the two panels in the bottom row of Fig.\ref{figure:Example1_IFs_est}, both the MSWCT and SXWCT provide very accurate IF estimates of both $x_1$ and $x_2$.

%%%%%%%%%%%%%%%%%%%the beginning of figure 10 %%%%%%%%%%%%%%%
\begin{figure}[H]
	\centering
	\begin{tabular}{cccc}
	\resizebox {1.5in}{1.0in}{\includegraphics{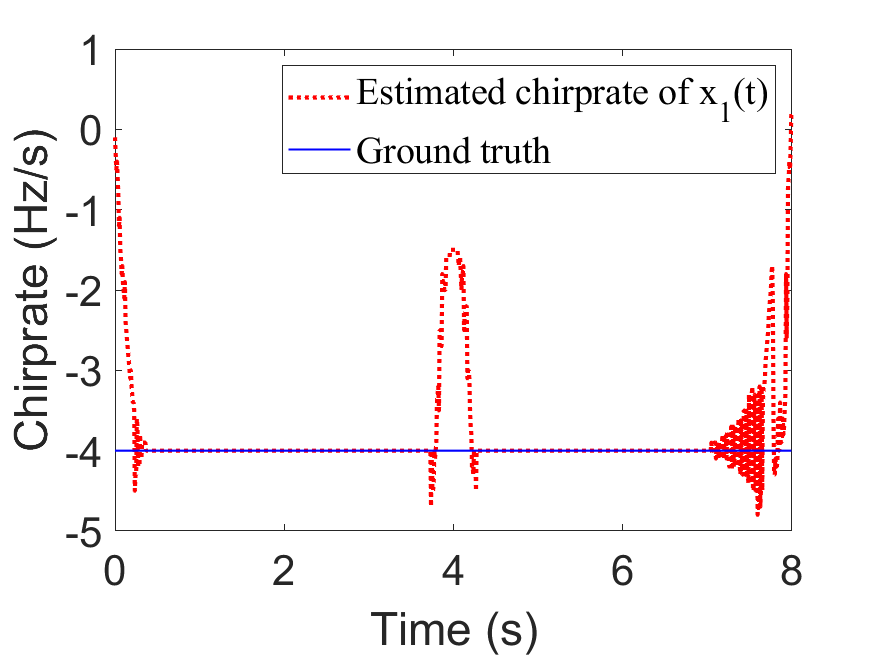}}	
&   
\resizebox {1.5in}{1.0in} {\includegraphics{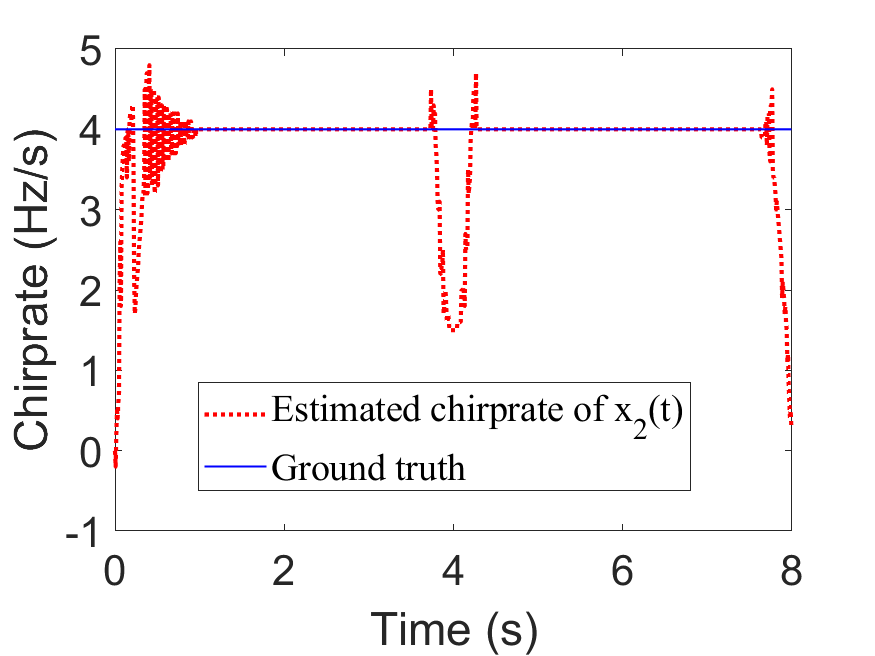}}
&\resizebox {1.5in}{1.0in}{\includegraphics{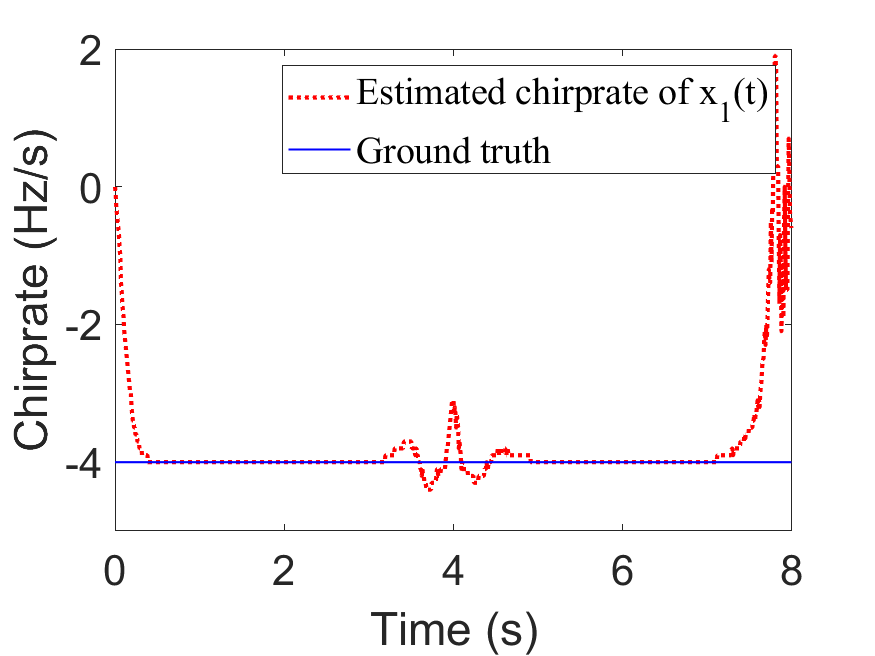}}
&  
\resizebox {1.5in}{1.0in}{\includegraphics{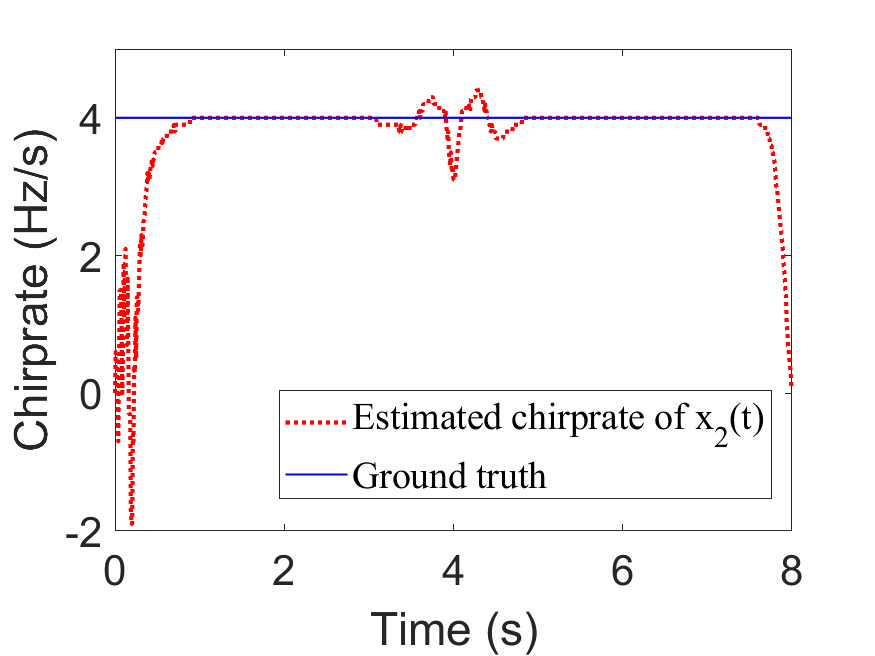}}\\
\resizebox {1.5in}{1.0in}{\includegraphics{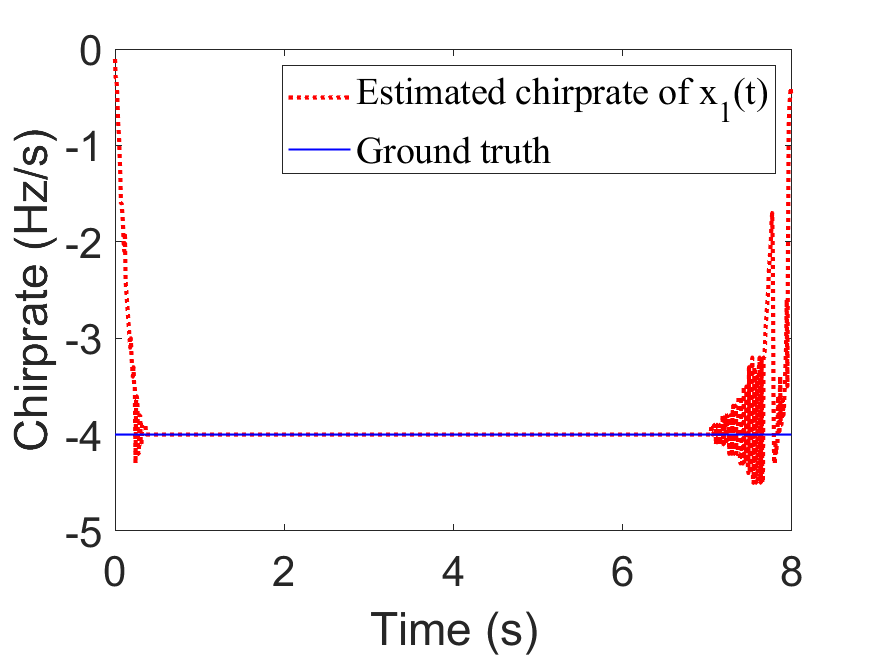}}	
&\resizebox {1.5in}{1.0in} {\includegraphics{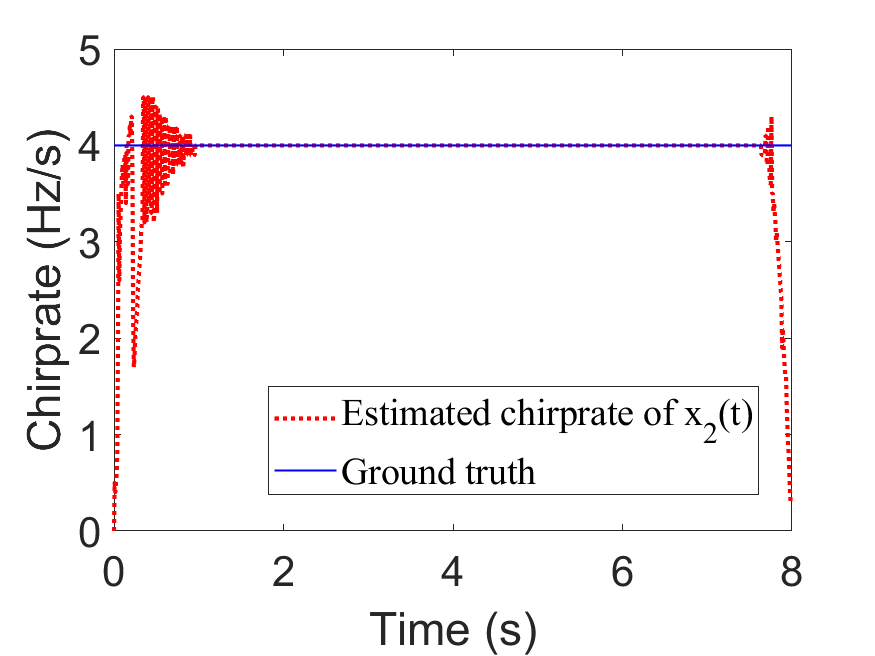}}
&&
	\end{tabular}
	%\vskip -0.3cm 
\caption{\small Chirprate estimation of $x_k(t)$. Top row (from left to right): by SWCT (1st and 2nd panels); by MSWCT (3rd and 4th panels). Bottom row:  by SXWCT.}
	\label{figure:Example1_chirprates_est}
\end{figure}
%%%%%%%%%%%%%%%%the end of figure 10  %%%%%%%%%%%%%%%%%%%%%

Fig.\ref{figure:Example1_chirprates_est} shows chirprate estimates by SWCT  (in 1st and 2nd panels in the top row), MSWCT  (3rd and 4th panels in top row), and SXWCT (in the bottom row). There are big chirprate estimate errors near $t=4$ with SWCT as shown in the top row in  Fig.\ref{figure:Example1_chirprates_est}. Though the MSWCT improves the accuracy of chirprate estimate as shown  in the 3rd and 4th panels in top row of  Fig.\ref{figure:Example1_chirprates_est}, there are still big errors near $t=4$.  There is essentially no chirprate estimate error with SXWCT except for the place when $t$ is near the end points.  Clearly SXWCT performs much better than SWCT and MSWCT in chirprate estimation. 

%%%%%%%%%%%%%%%%%%%the beginning of figure 11 %%%%%%%%%%%%%%%
\begin{figure}[th]
	\centering
	%\begin{tabular}{ccc}
	\begin{tabular}{cccc}
	\resizebox {1.5in}{1.0in} {\includegraphics{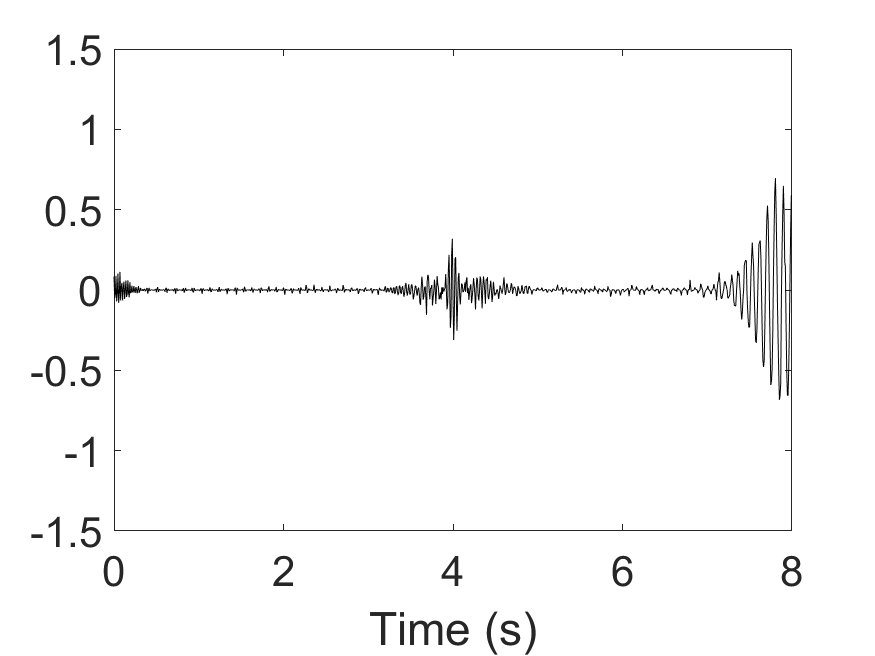}}	
&  \resizebox {1.5in}{1.0in} {\includegraphics{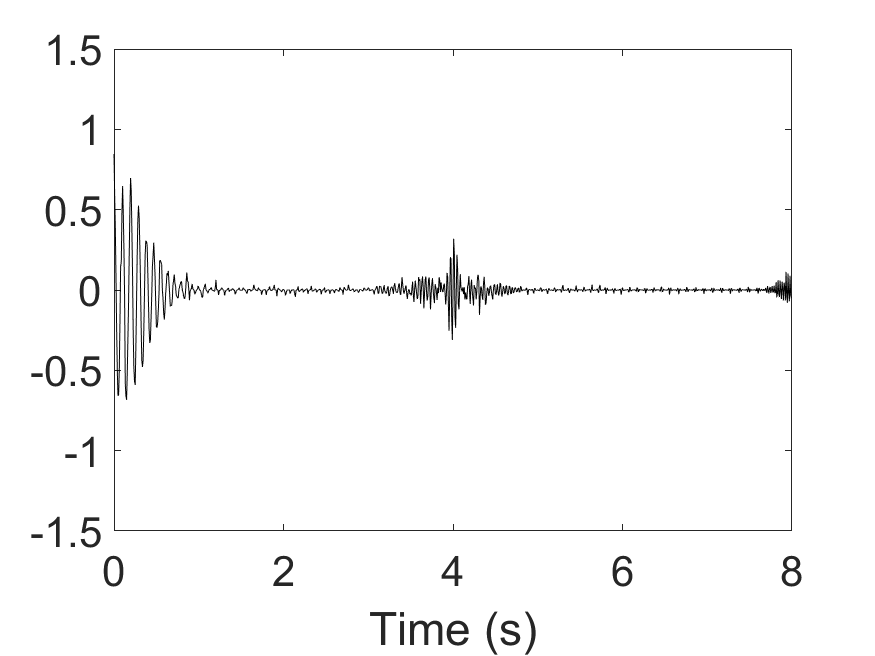}}&
\resizebox {1.5in}{1.0in} {\includegraphics{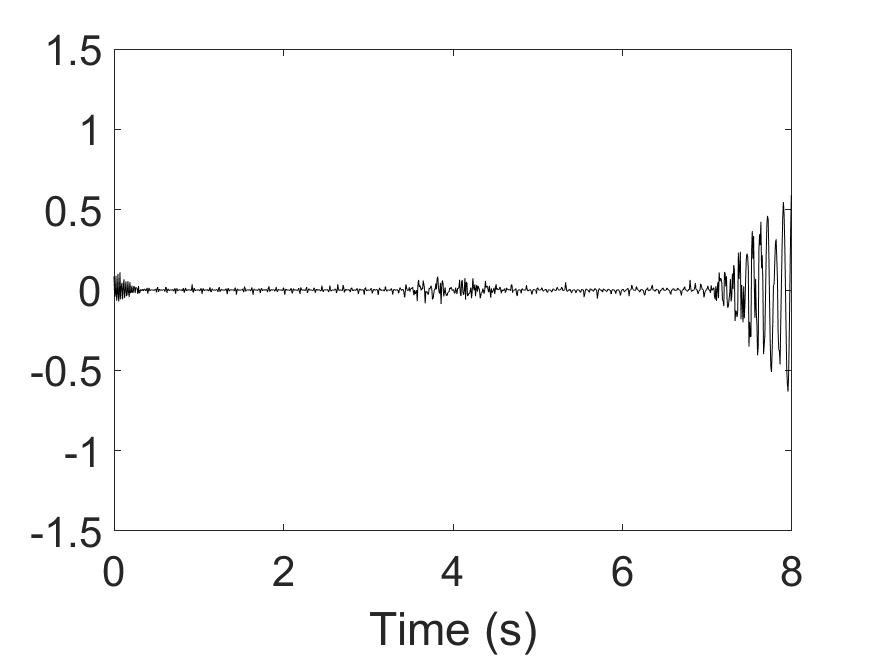}}
& \resizebox {1.5in}{1.0in} {\includegraphics{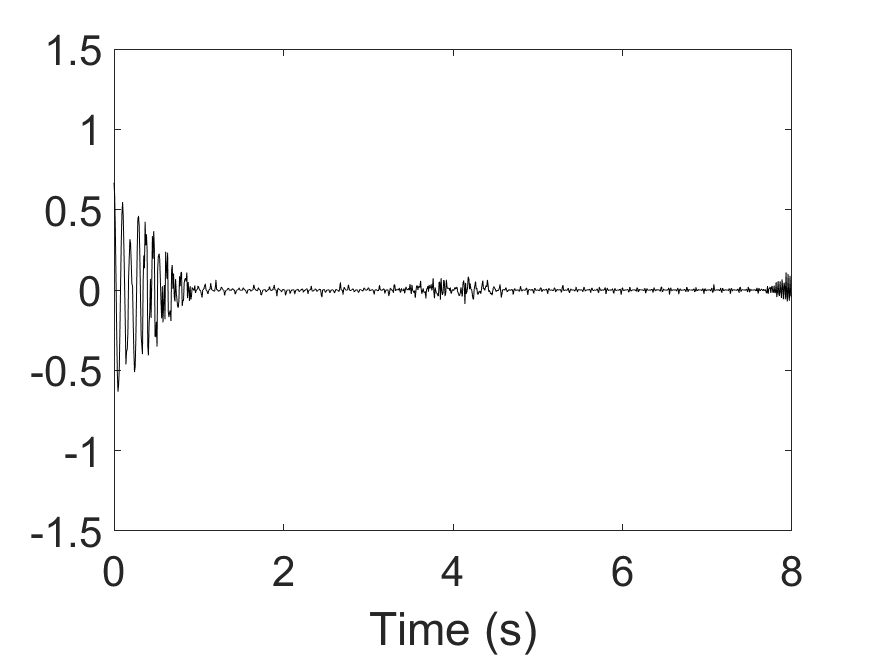}}
	\end{tabular}
\caption{\small From left to right: recover residual by MSWCT 
real$(x_1(t)-\wt x_1(t))$ (1st planel) and  real$(x_2(t)-\wt x_2(t))$ (2nd panel); 
recover residual by SXWCT real$(x_1(t)-\wt x_1(t))$ (3rd panel) and  real$(x_2(t)-\wt x_2(t))$ (4th panel).}
	\label{figure:Example1_recover}
\end{figure}
%%%%%%%%%%%%%%%%the end of figure 11  %%%%%%%%%%%%%%%%%%%%%

Fig.\ref{figure:Example1_recover} shows the real part of the mode retrieval residual 
by the 3rd-order MSWCT: real$(x_1(t)-\wt x_1(t))$ (1st panel) and  real$(x_2(t)-\wt x_2(t))$ (2nd panel) and that by the 3rd-order SXWCT:  real$(x_1(t)-\wt x_1(t))$ (3rd panel) and  real$(x_2(t)-\wt x_2(t))$ (4th panel). One can see that SXWCT performs better than %SWCT and 
MSWCT in %chirprate estimation and 
mode retrieval. The RMSEs for mode retrieval with the MSWCT for $x_1(t)$ and $x_2(t)$ are 
respectively 
$$
E_{x_1}=0.0526, \; E_{x_2}=0.0526, 
$$
while the errors with SXWCT are 
$$
\wt E_{x_1} =0.0237, \; \wt E_{x_2}=0.0232. 
$$ 

%%%%%%%%%%%%%%%%%%%the beginning of figure 12 %%%%%%%%%%%%%%%
\begin{figure}[H]
	\centering
	%\begin{tabular}{ccc}
	\begin{tabular}{cccc}
	\resizebox {1.5in}{1.0in} {\includegraphics{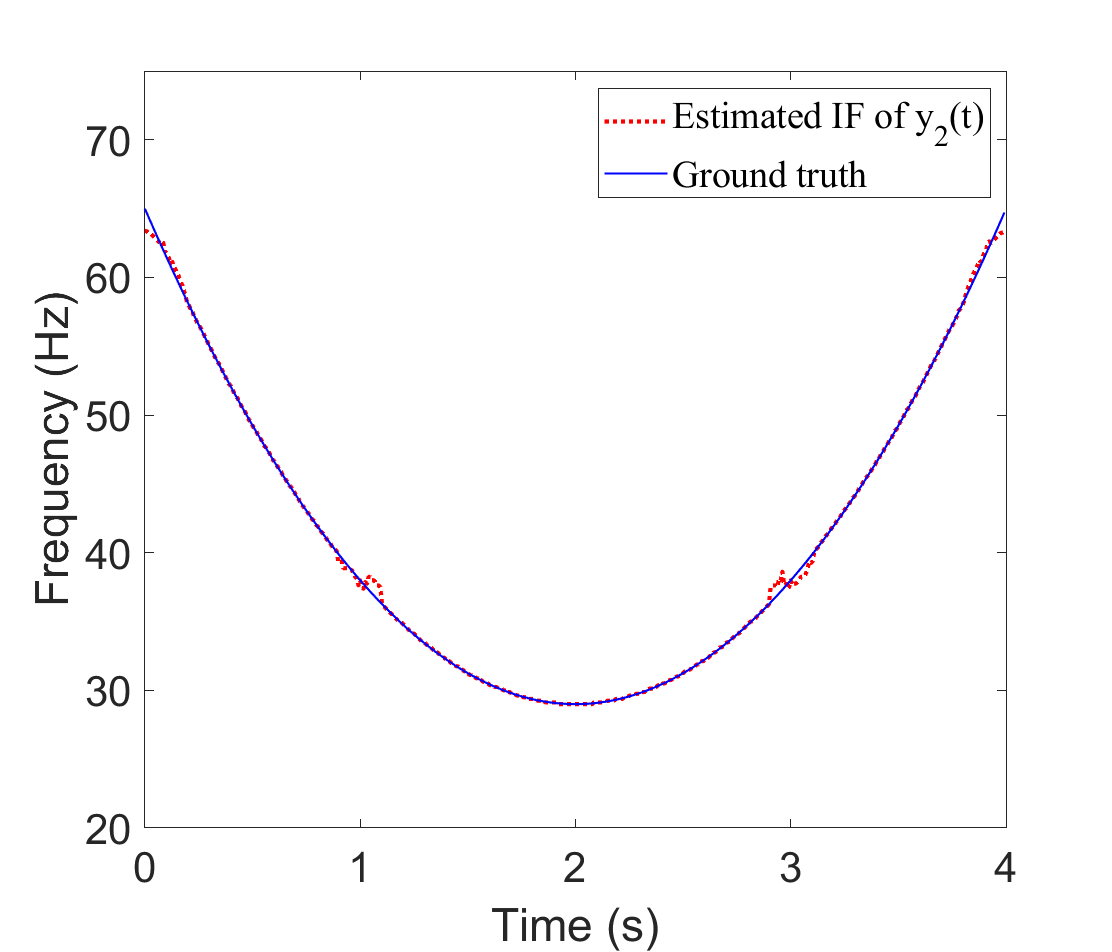}}	& \resizebox {1.5in}{1.0in} {\includegraphics{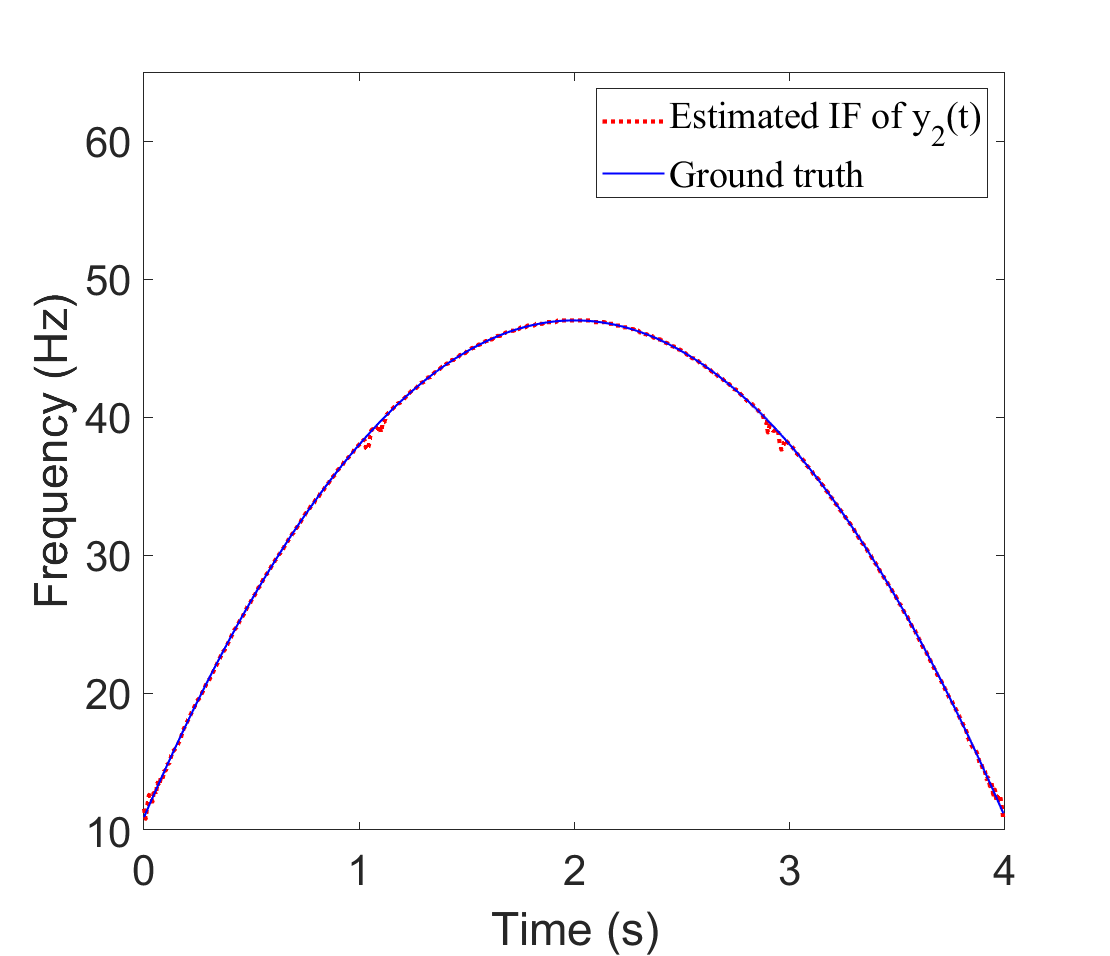}}
&\resizebox {1.5in}{1.0in} {\includegraphics{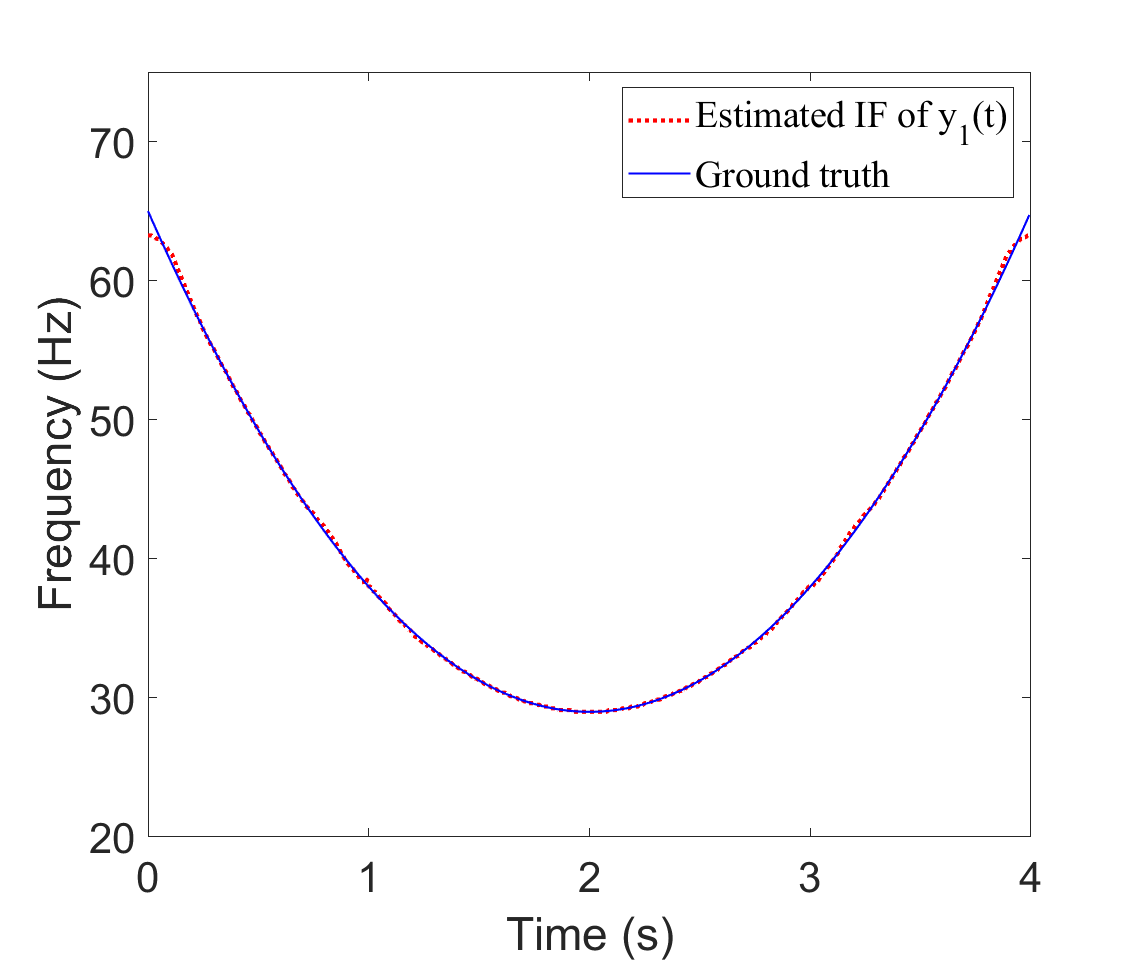}} & \resizebox {1.5in}{1.0in} {\includegraphics{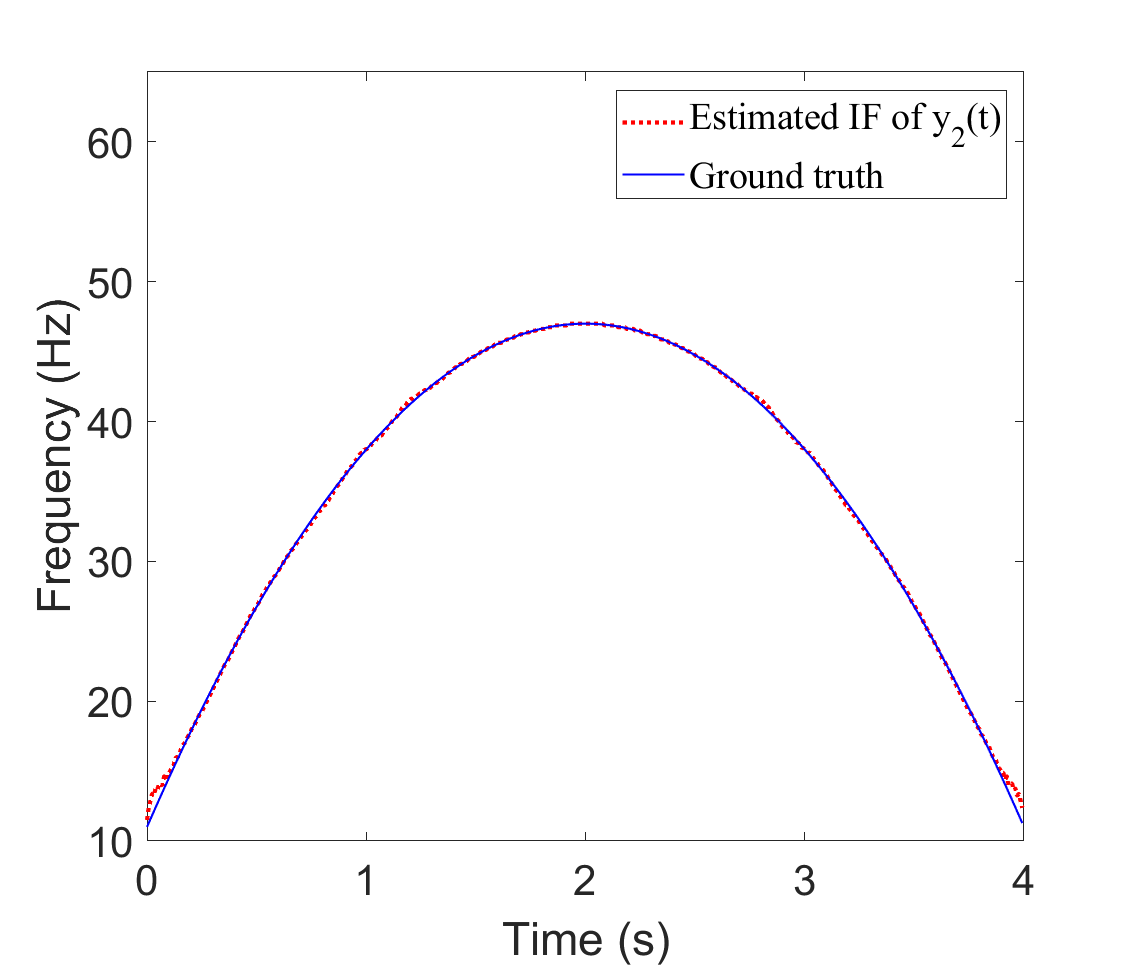}}\\
\resizebox {1.5in}{1.0in} {\includegraphics{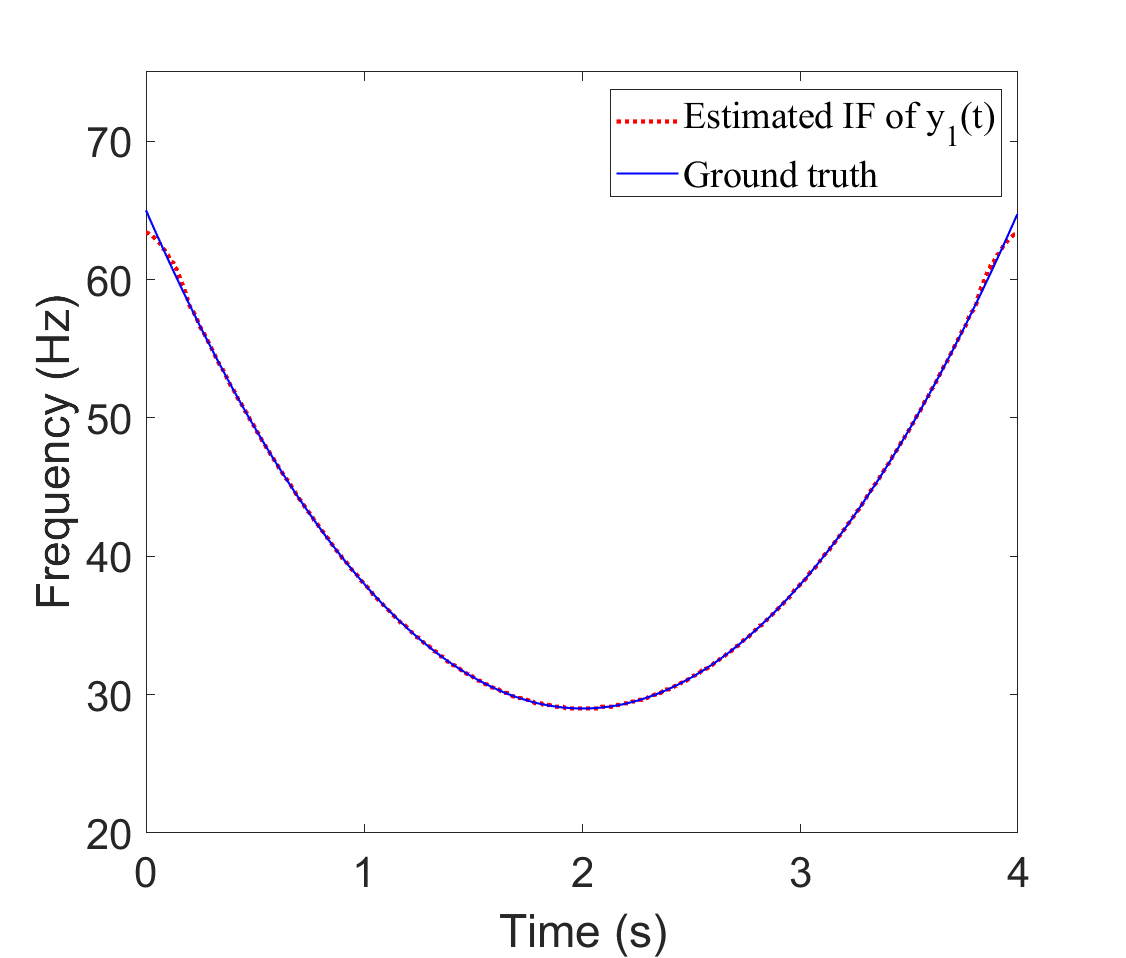}}	&\resizebox {1.5in}{1.0in} {\includegraphics{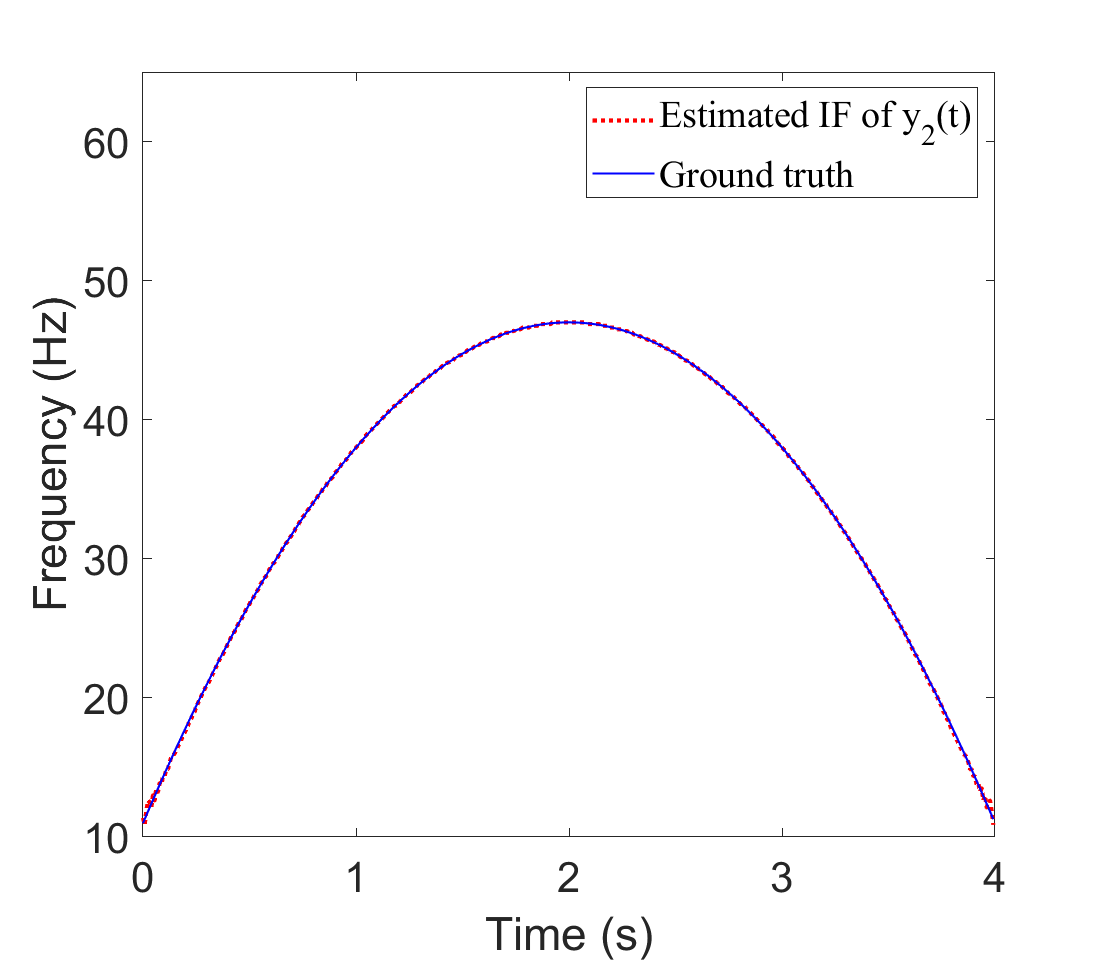}}
&
\resizebox {1.5in}{1.0in} {\includegraphics{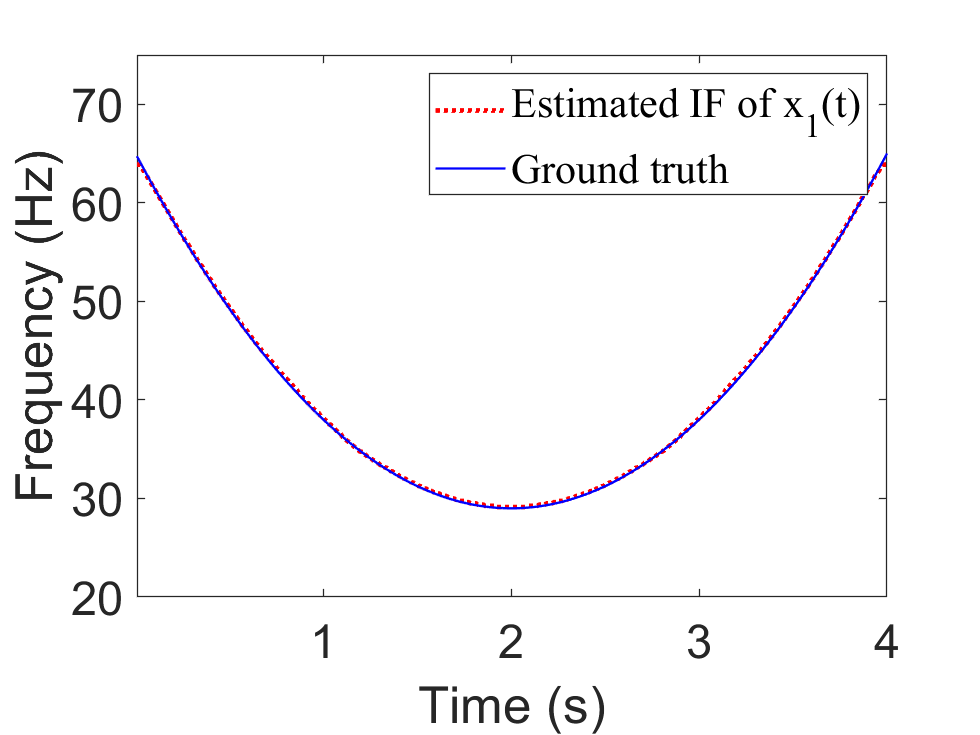}}	&\resizebox {1.5in}{1.0in} {\includegraphics{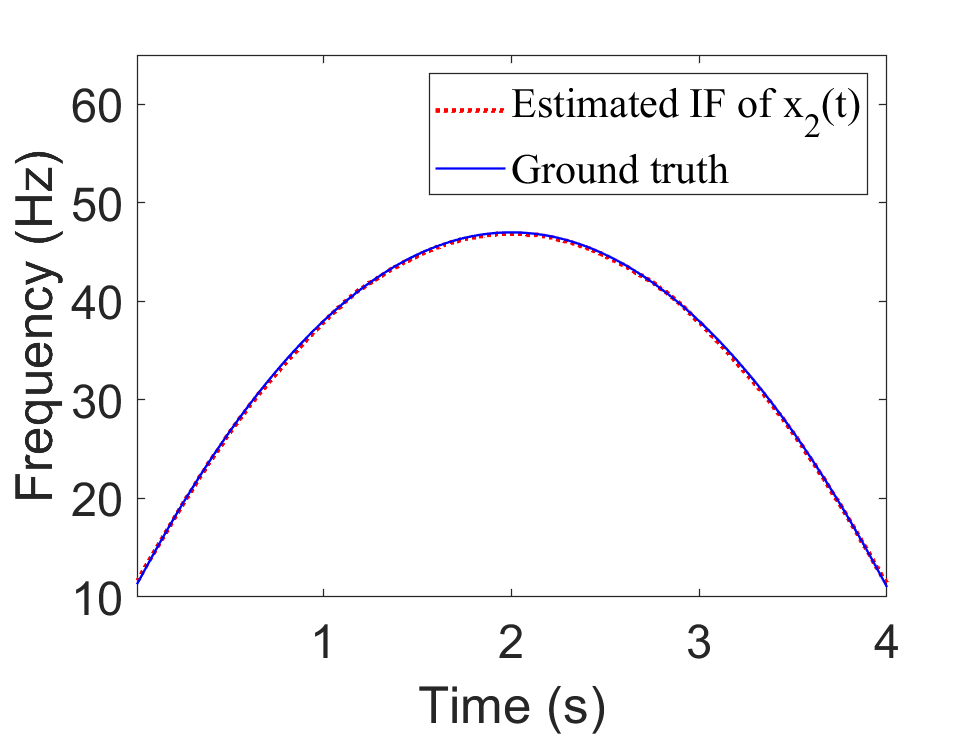}}
\end{tabular}
\caption{\small IF estimation of $y_k(t)$. Top row (from left to right): by SWCT (1st and 2nd panels); by MSWCT (3rd and 4th panels).
Bottom row (from left to right): by SXWCT (1st and 2nd panels); 
by MSCT (3rd and 4th panels).}
	\label{figure:example2_IFs_est}
\end{figure}
%%%%%%%%%%%%%%%%the end of figure 12  %%%%%%%%%%%%%%%%%%%%%

{\bf Example 2 (Continued).} Let $y(t)$ be the signal given by \eqref{def_y}. For this signal, the performance of IF and chirprate estimation and mode retrieval by multiple synchrosqueezed chirplet transform (MSCT) is also provided. Fig.\ref{figure:example2_IFs_est} provides IF estimates by SWCT  (1st and 2nd panels in the top row), MSWCT  (3rd and 4th panels in top row), SXWCT (1st and 2nd panels in the bottom row) and MSCT (3rd and 4th panels in the bottom row). From the top row of Fig.\ref{figure:example2_IFs_est}, one can see the IF estimate of $y_1(t)$ and $y_2(t)$ by SWCT is overall accurate except for at $t=1, 3$, where the IFs of $y_1(t)$ and $y_2(t)$ is crossover. 
As shown in the 3rd and 4th panels in the top row and  the panels in the bottom row of Fig.\ref{figure:example2_IFs_est}, MSWCT, SXWCT and MSCT all provide very accurate IF estimates of both $y_1$ and $y_2$.   

\begin{figure}[H]
	\centering
	\begin{tabular}{cccc}
	\resizebox {1.5in}{1.0in} {\includegraphics{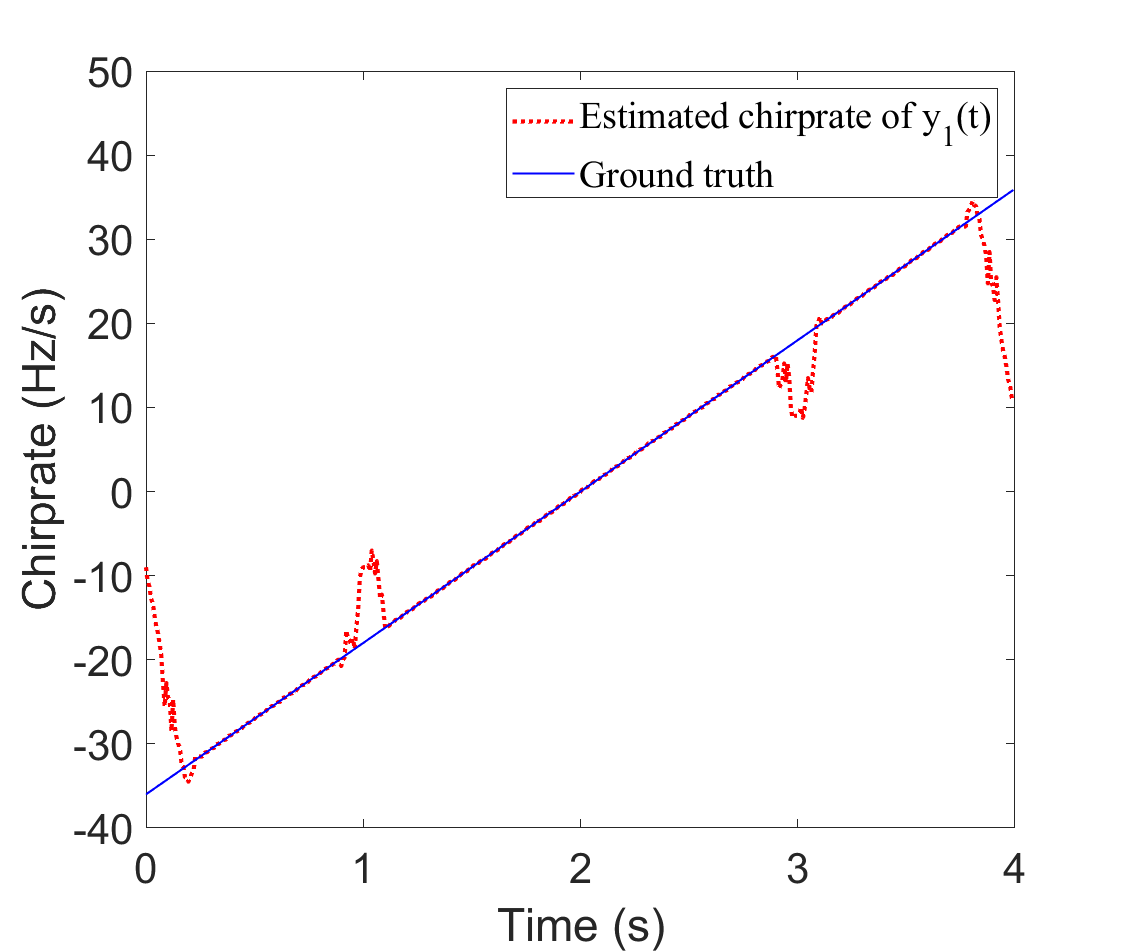}}	
&\resizebox {1.5in}{1.0in} {\includegraphics{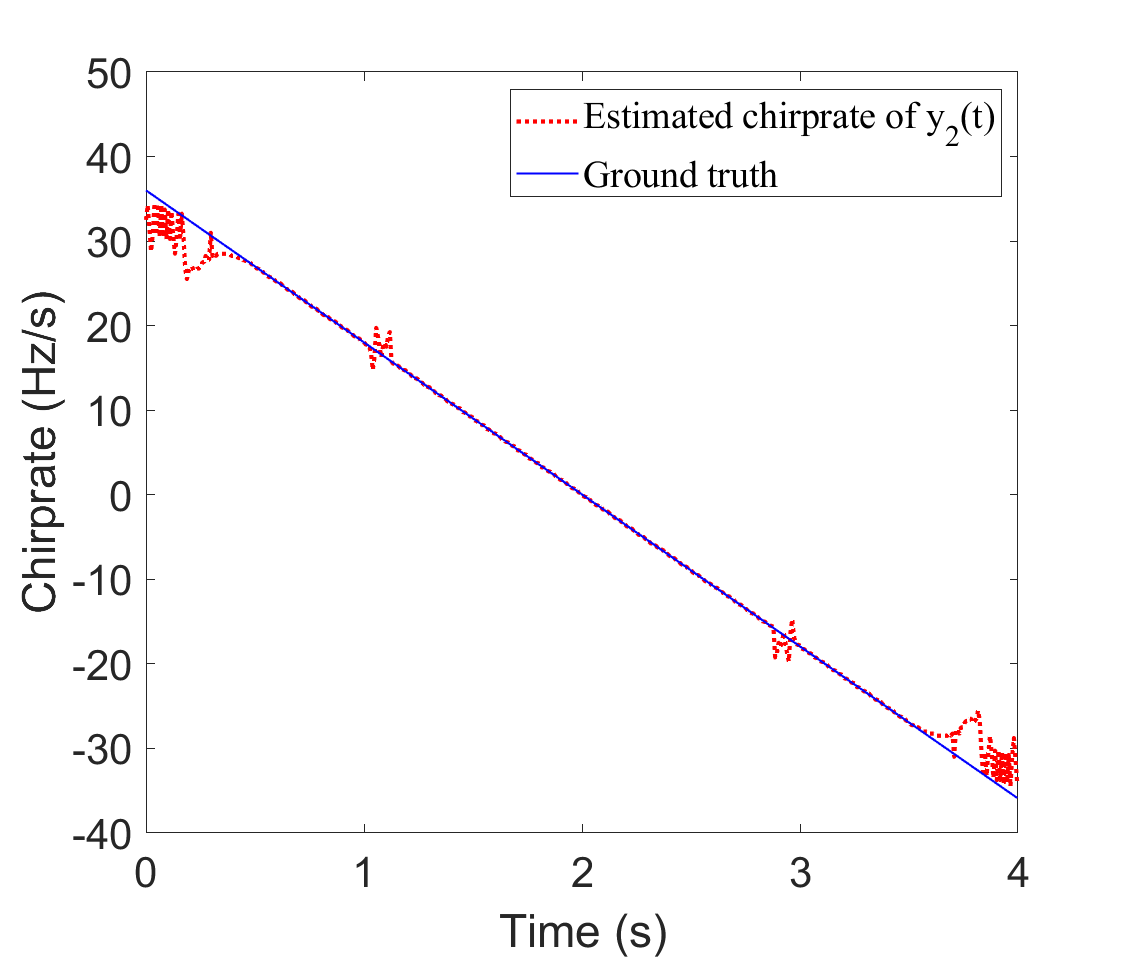}}
&
\resizebox {1.5in}{1.0in} {\includegraphics{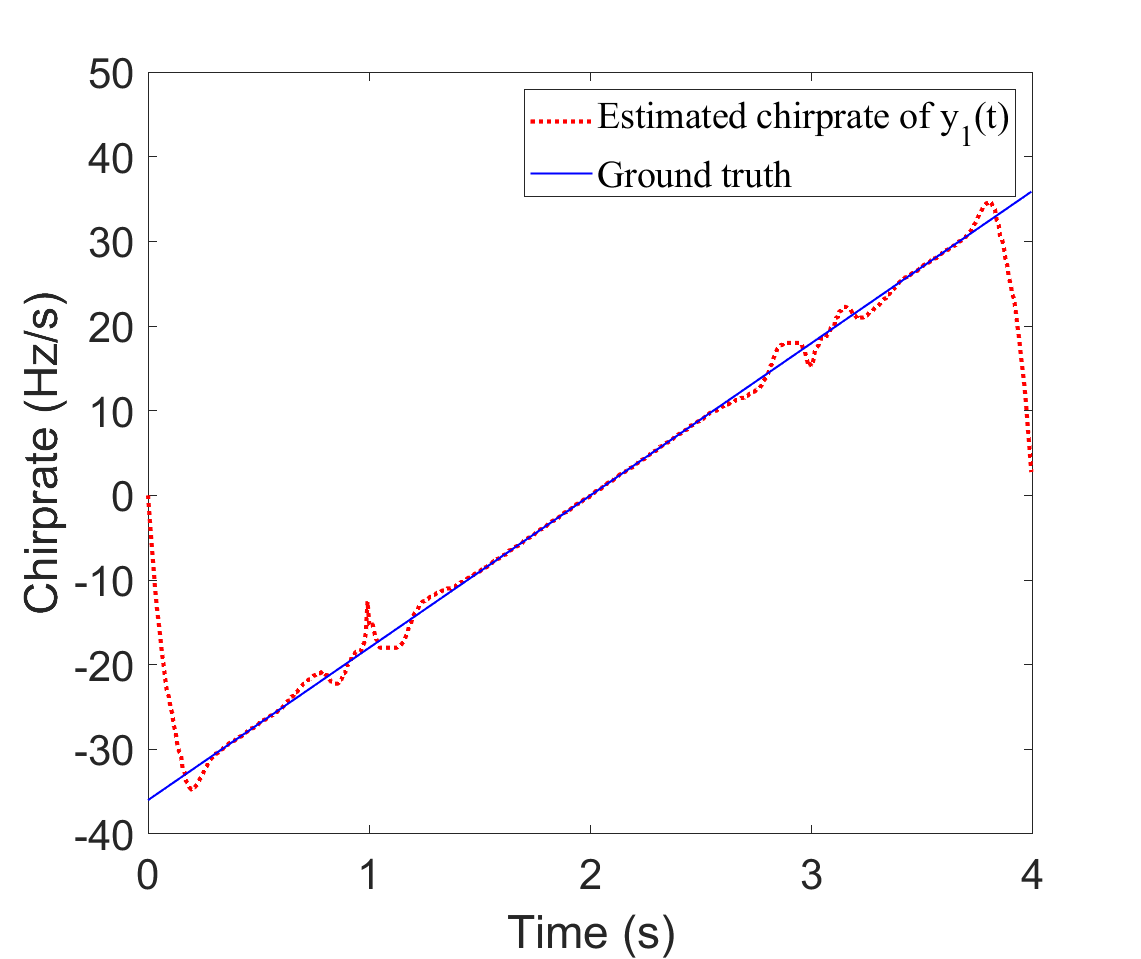}}
&  
\resizebox {1.5in}{1.0in} {\includegraphics{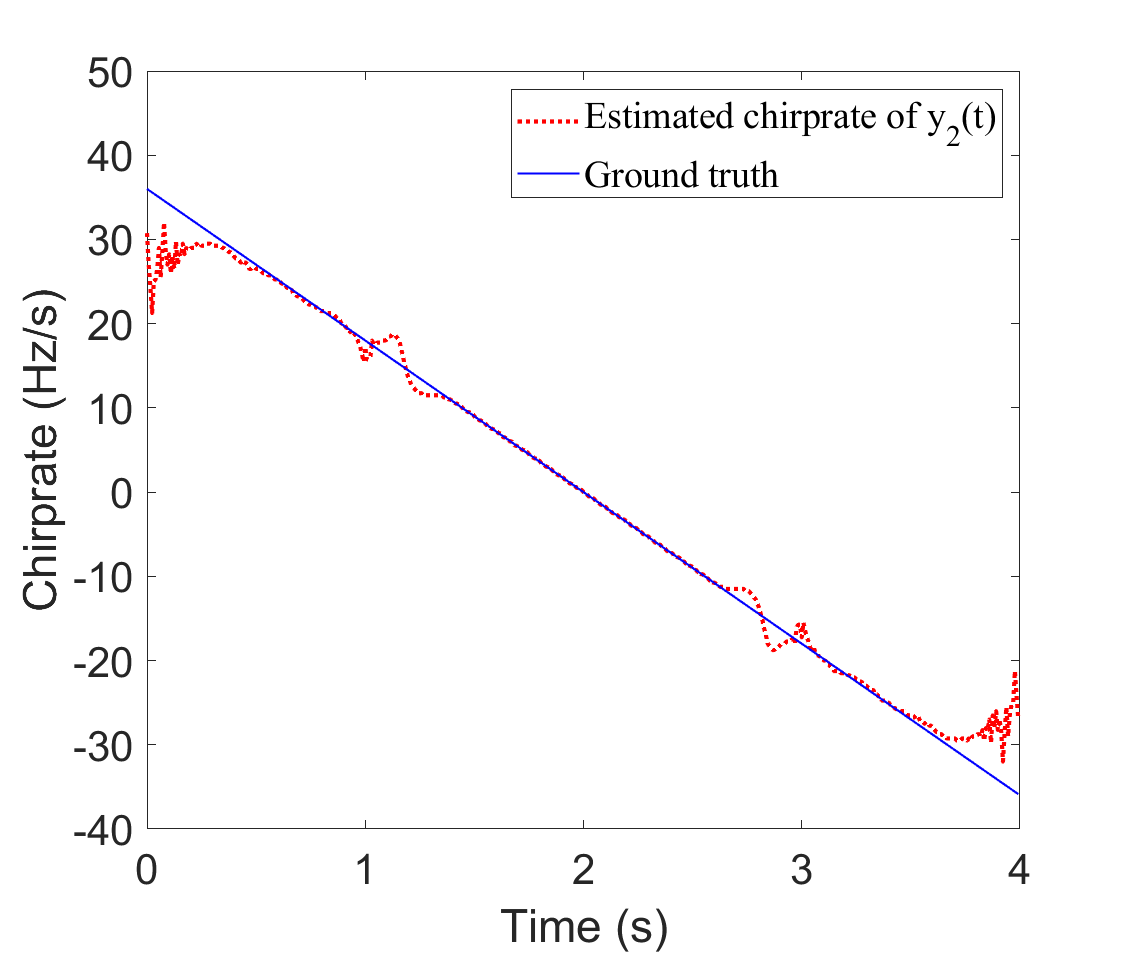}}
\\
\resizebox {1.5in}{1.0in} {\includegraphics{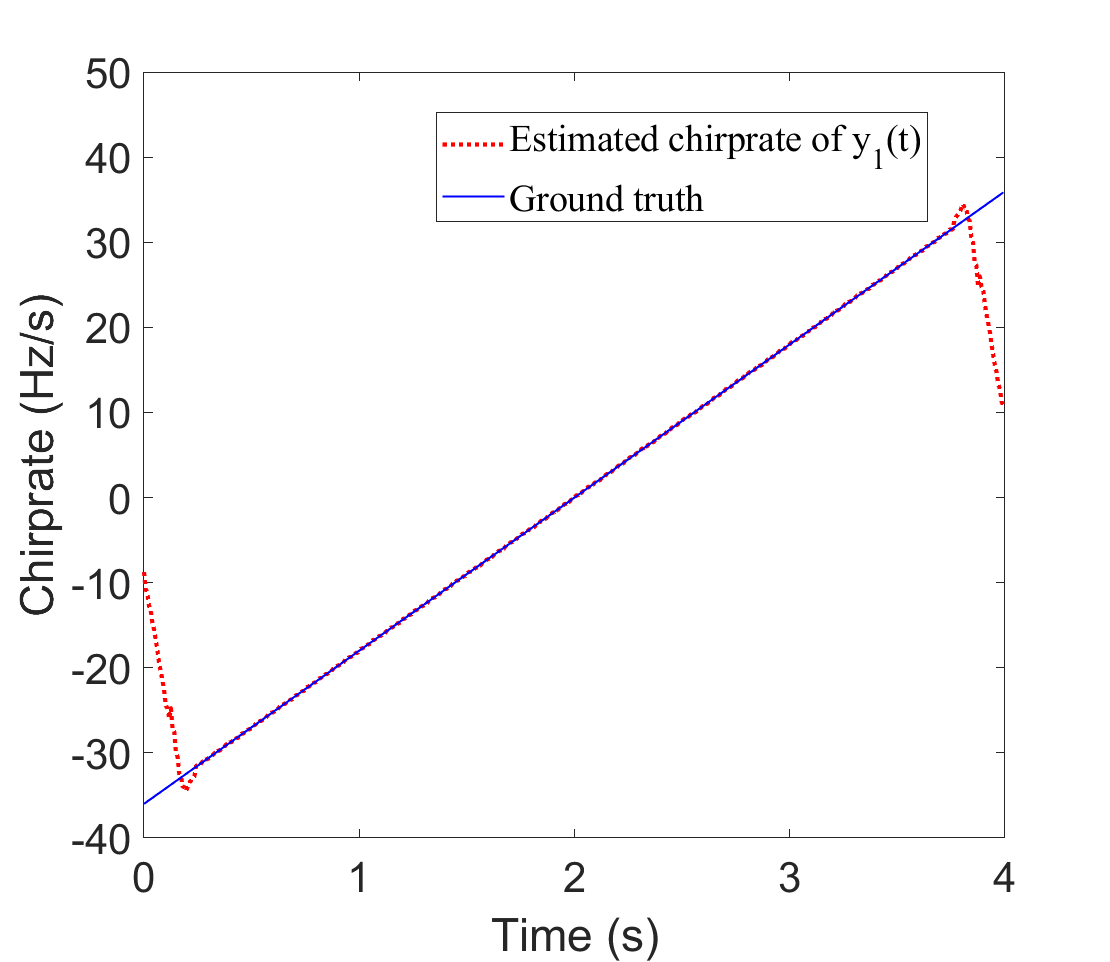}}	
& \resizebox {1.5in}{1.0in} {\includegraphics{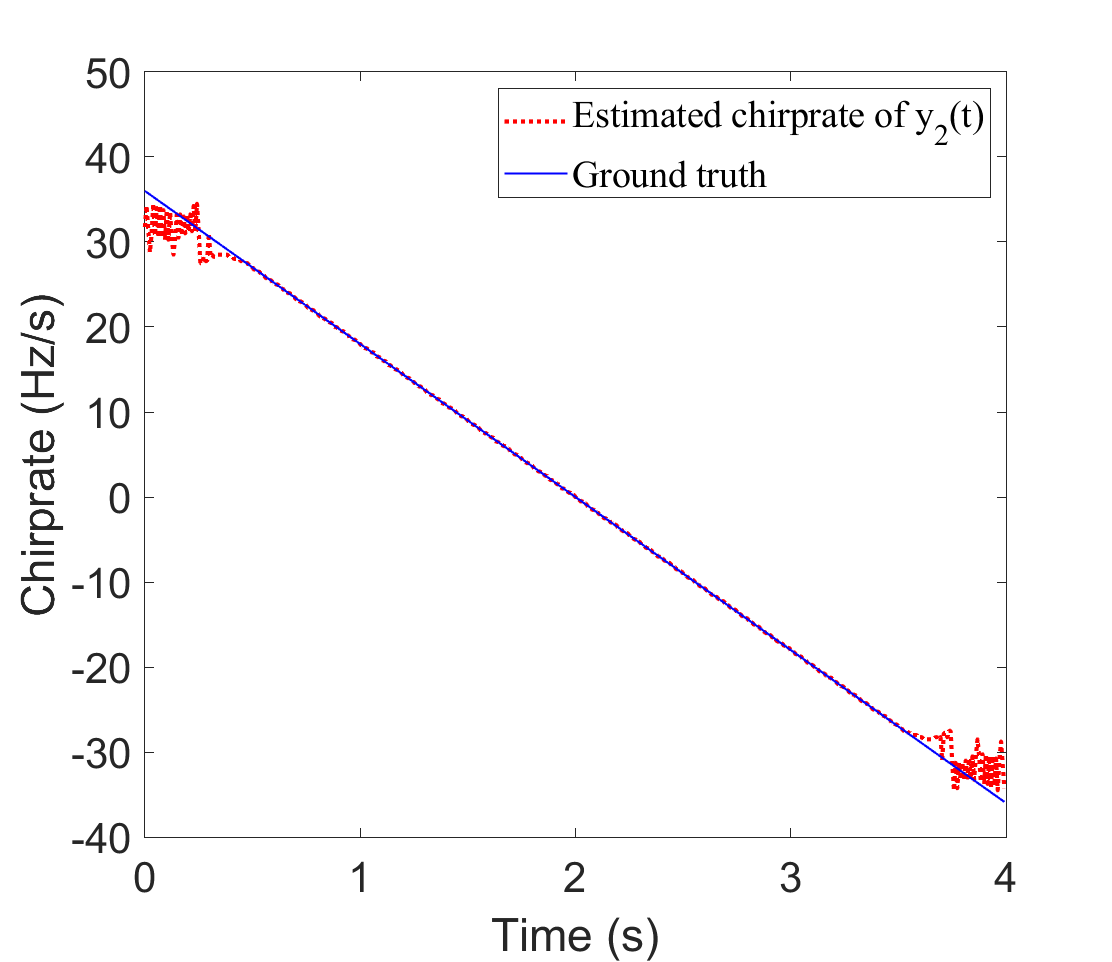}}
&\resizebox {1.5in}{1.0in} {\includegraphics{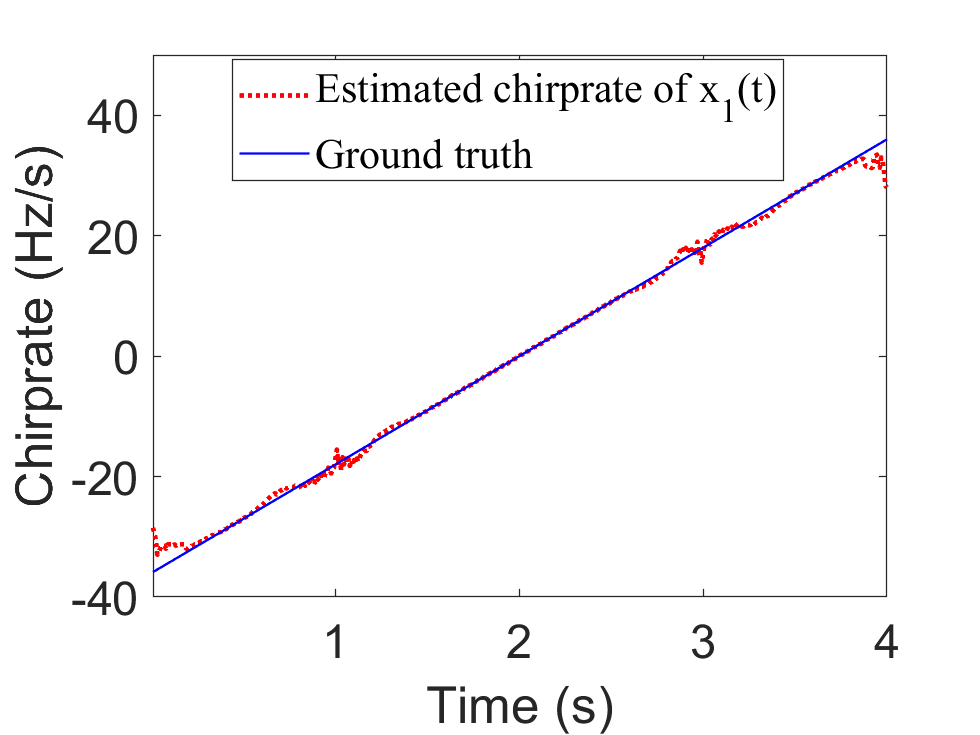}}	&\resizebox {1.5in}{1.0in} {\includegraphics{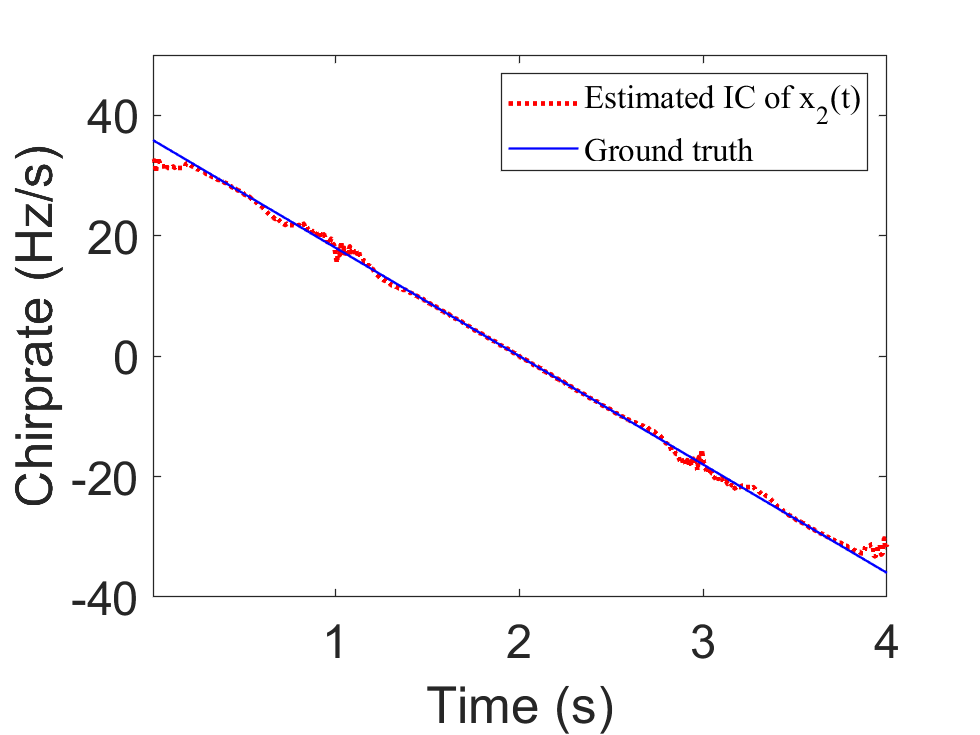}}
	\end{tabular}
	%\vskip -0.3cm 
\caption{\small Chirprate estimation of $y_k(t)$. Top row (from left to right): by SWCT (1st and 2nd panels); by MSWCT (3rd and 4th panels).
Bottom row (from left to right): by SXWCT  (1st and 2nd panels); by MSCT (3rd and 4th panels).}
	\label{figure:example2_chirprates_est}
\end{figure}

Fig.\ref{figure:example2_chirprates_est} shows chirprate estimates by SWCT  (1st and 2nd panels in the top row), MSWCT  (3rd and 4th panels in the top row), SXWCT (1st and 2nd panels in the bottom row) and MSCT  (3rd and 4th panels in the bottom row). 
For \(y_1(t)\), the SWCT exhibits significant chirprate estimation errors near \(t = 1, 3\), whereas the MSWCT introduces even larger errors for \(y_2(t)\) in the same time vicinity. By contrast, the MSCT shows noticeable chirprate estimation errors for both \(y_1(t)\) and \(y_2(t)\) near \(t = 1, 3\), while the SXWCT demonstrates essentially no chirprate estimation errors—except in the vicinity of endpoint times.

%%%%%%%%%%%%%%%%%%%the beginning of figure 13 %%%%%%%%%%%%%%%
%%%%%%%%%%%%%%%%the end of figure 13  %%%%%%%%%%%%%%%%%%%%%

\begin{figure}[H]
	\centering
	\begin{tabular}{cccc}
\resizebox {1.5in}{1.0in} {\includegraphics{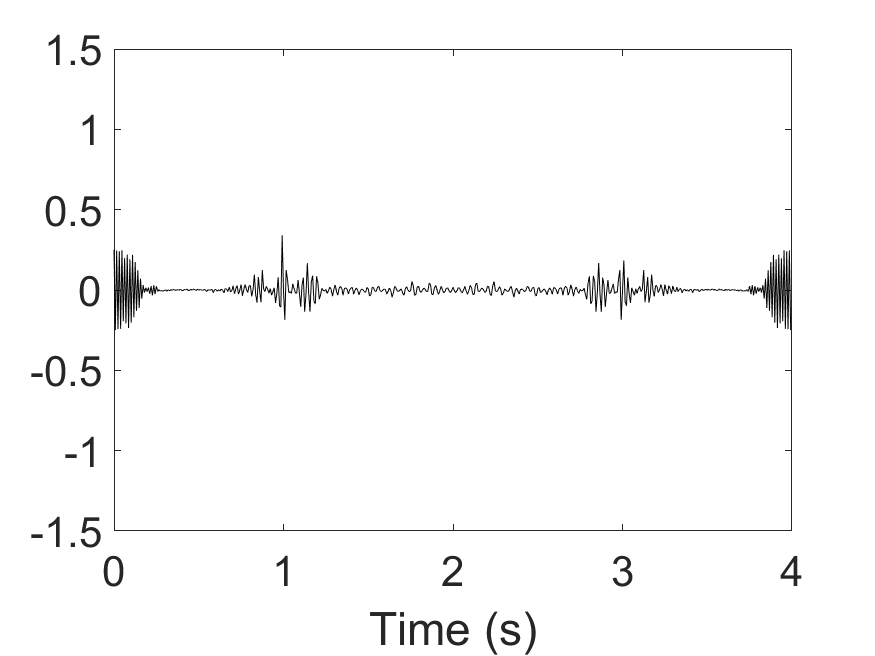}}
& 
\resizebox {1.5in}{1.0in} {\includegraphics{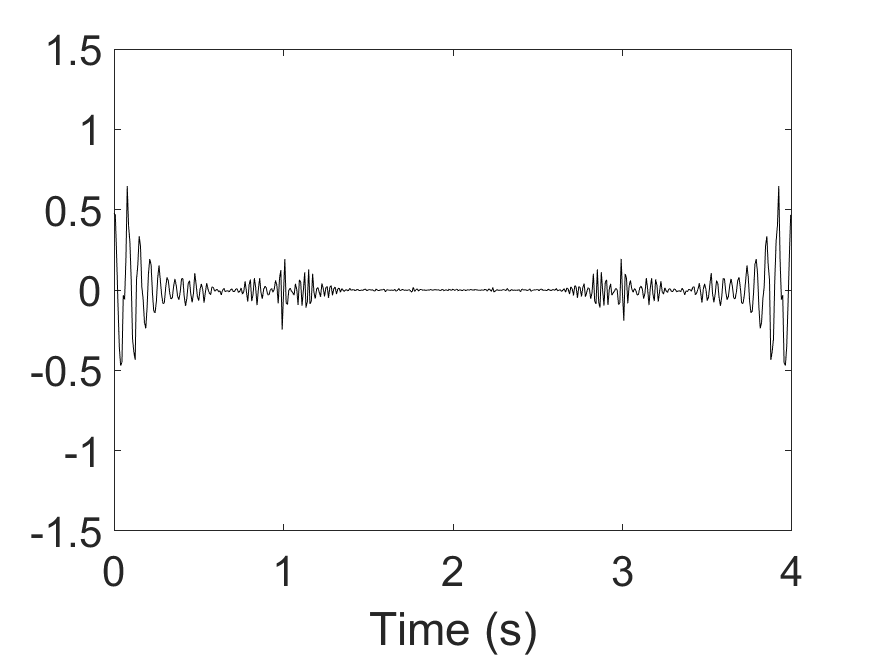}}
&
\resizebox {1.5in}{1.0in} {\includegraphics{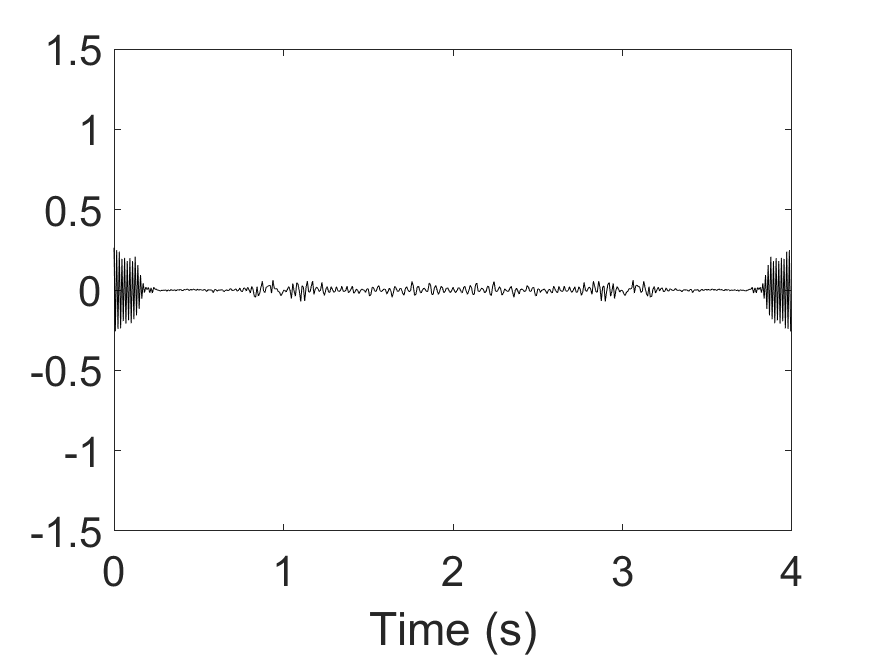}}
& \resizebox {1.5in}{1.0in} {\includegraphics{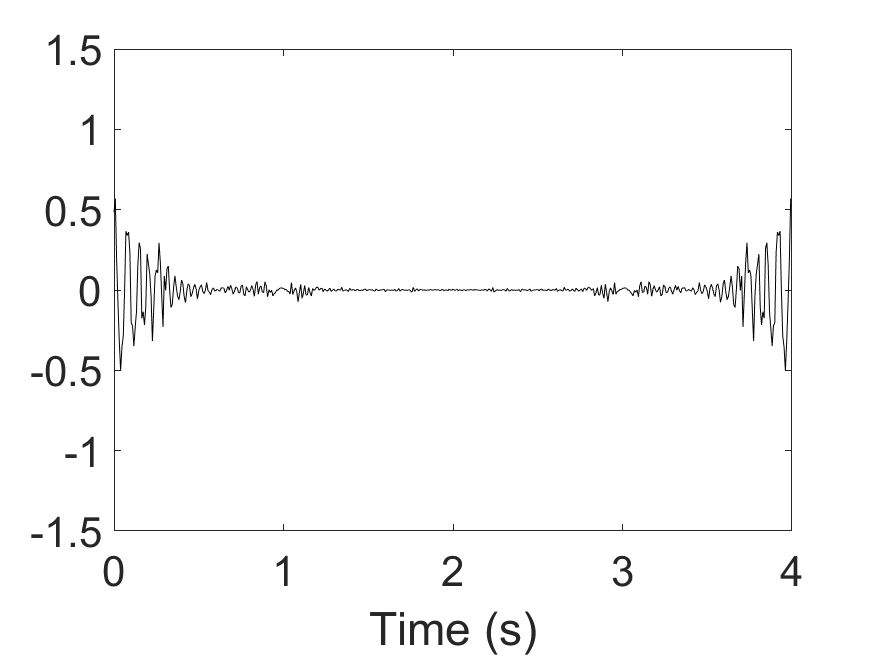}}\\
\resizebox {1.5in}{1.0in} {\includegraphics{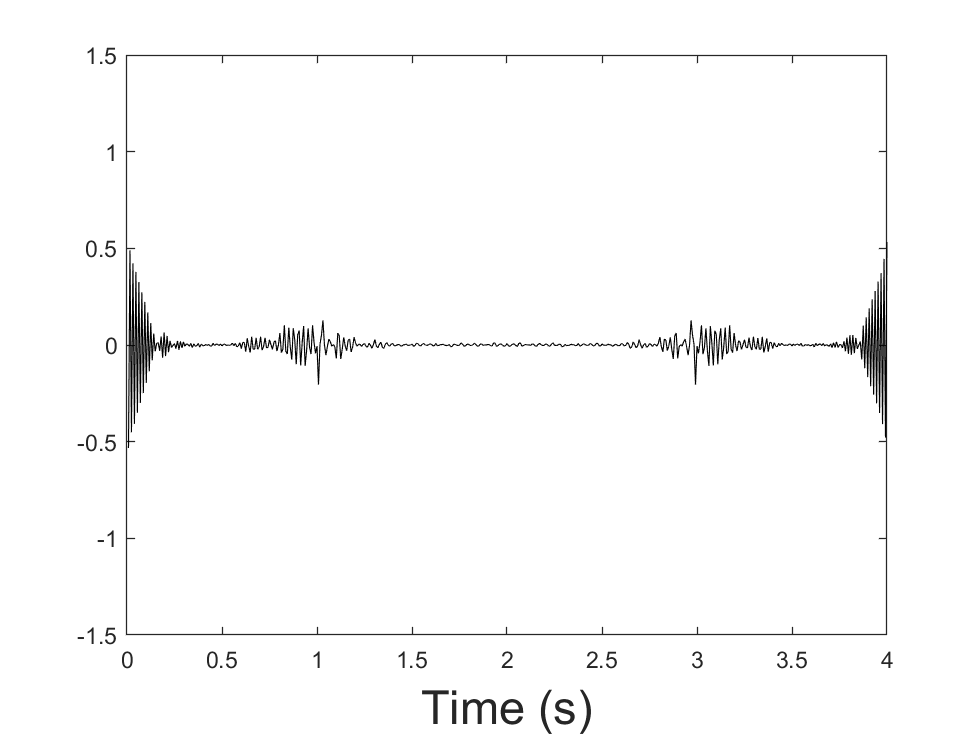}}	
&\resizebox {1.5in}{1.0in} {\includegraphics{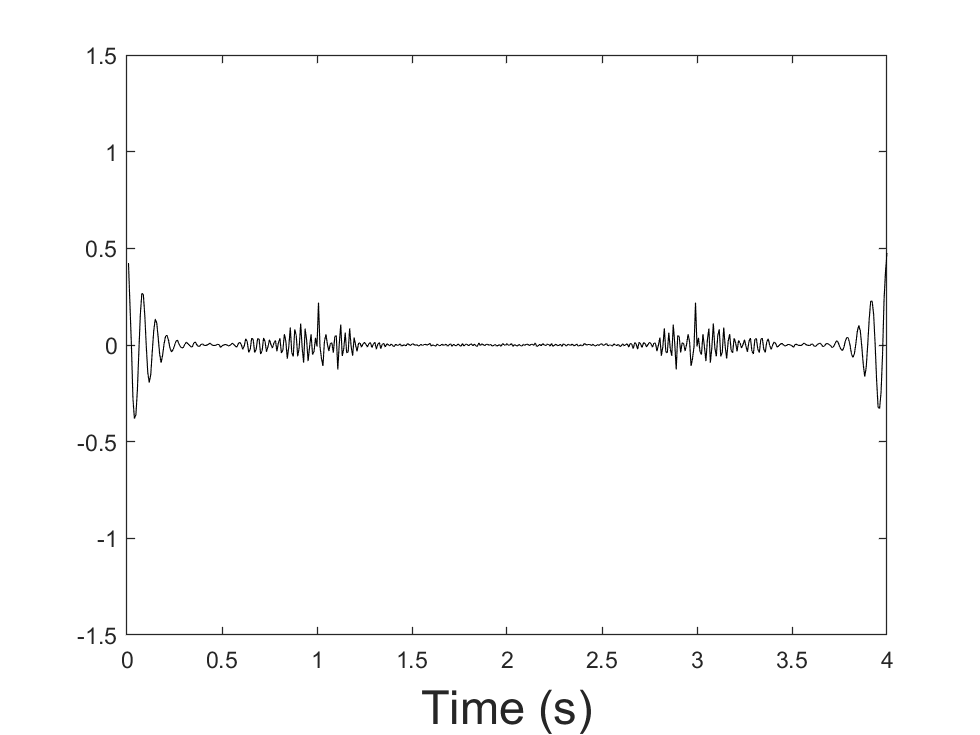}}
&&
\end{tabular} 
\caption{\small Top row (from left to right): recover residual by MSWCT 
real$(y_1(t)-\wt y_1(t))$ (1st planel) and  real$(y_2(t)-\wt y_2(t))$ (2nd panel); 
recover residual by SXWCT  real$(y_1(t)-\wt y_1(t))$ (3rd panel) and  real$(y_2(t)-\wt y_2(t))$ (4th panel).
Bottom row (from left to right): recover residual by MSCT 
real$(y_1(t)-\wt y_1(t))$ (1st planel) and  real$(y_2(t)-\wt y_2(t))$ (2nd panel).
}
\label{figure:Example2_recover}
\end{figure}

The top row of Fig.\ref{figure:Example2_recover} shows the real part of the mode retrieval residual 
by the 3rd-order MSWCT: real$(y_1(t)-\wt y_1(t))$ (1st panel) and  real$(y_2(t)-\wt y_2(t))$ (2nd panel) and that by the 3rd-order SXWCT:  real$(y_1(t)-\wt y_1(t))$ (3rd panel) and  real$(y_2(t)-\wt y_2(t))$ (4th panel). 
The bottom row of Fig.\ref{figure:Example2_recover} provides the real part of the mode retrieval residual 
by MSCT: real$(y_1(t)-\wt y_1(t))$ (1st panel) and  real$(y_2(t)-\wt y_2(t))$ (2nd panel).  
In addition, for signal $y(t)$, RMSEs of IFs and CRs estimation and mode retrieval with  SWCT,  MSWCT   SXWCT and MSCT are provided in Table \ref{table:RMSE_results}.  
While MSWCT and MSCT demonstrate similar performance in mode retrieval, SXWCT outperforms both in terms of chirprate estimation and mode retrieval.

%%%%%%%%%%%%%%%%%%%the beginning of figure 14 %%%%%%%%%%%%%%%

%%%%%%%%%%%%%%%%the end of figure 14  %%%%%%%%%%%%%%%%%%%%%

\begin{table}[H]
    \centering
    \caption{\small RMSE of IFs and CRs estimation and mode retrieval  of \( y(t) \) with different synchrosqueezed transforms.}
    \label{table:RMSE_results}
    \begin{tabular}{lcccccc}
        \toprule
        \textbf{Method} & \textbf{IF1 } & \textbf{IF2 }&\textbf{CR1 } & \textbf{CR2 } & \textbf{error1}& \textbf{error2 }\\
         SWCT &0.2555  &0.1403   &2.2492  &0.4831  &0.1191 &0.0759 \\
        MSWCT & 0.1021 & 0.1094  &0.8319  &0.8643  &0.0581 &0.0546\\   
		%SCT   & 0.1558 &0.1558   &2.2447  &2.2447  &0.1796 &0.1799\\
		MSCT   & 0.1438 &0.1438   &0.5512  &0.5512  &0.0472 &0.0471\\
 SXWCT & 0.0357 &0.0357   &0.0727  &0.0727  &0.0276 &0.0229\\
        \bottomrule
    \end{tabular}
\end{table}
%% SCT entroy l=2.5 sigma=0.12

%%%%%%%%%%%%%%%%%%%the beginning of figure 15 %%%%%%%%%%%%%%%
\begin{figure}[H]
	\centering
	\begin{tabular}{cccc}
	\resizebox {1.5in}{1.0in} {\includegraphics{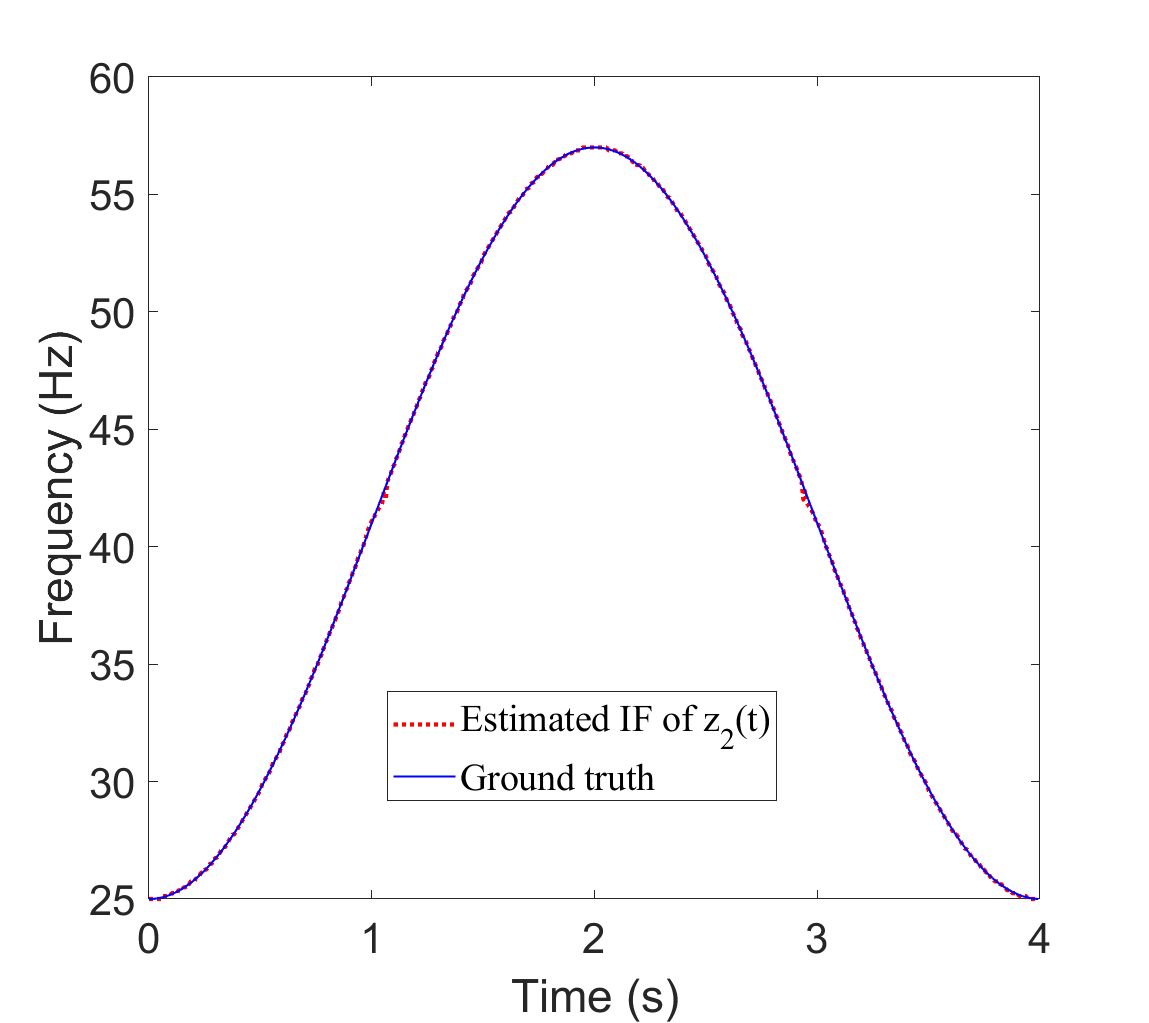}}	
& \resizebox {1.5in}{1.0in} {\includegraphics{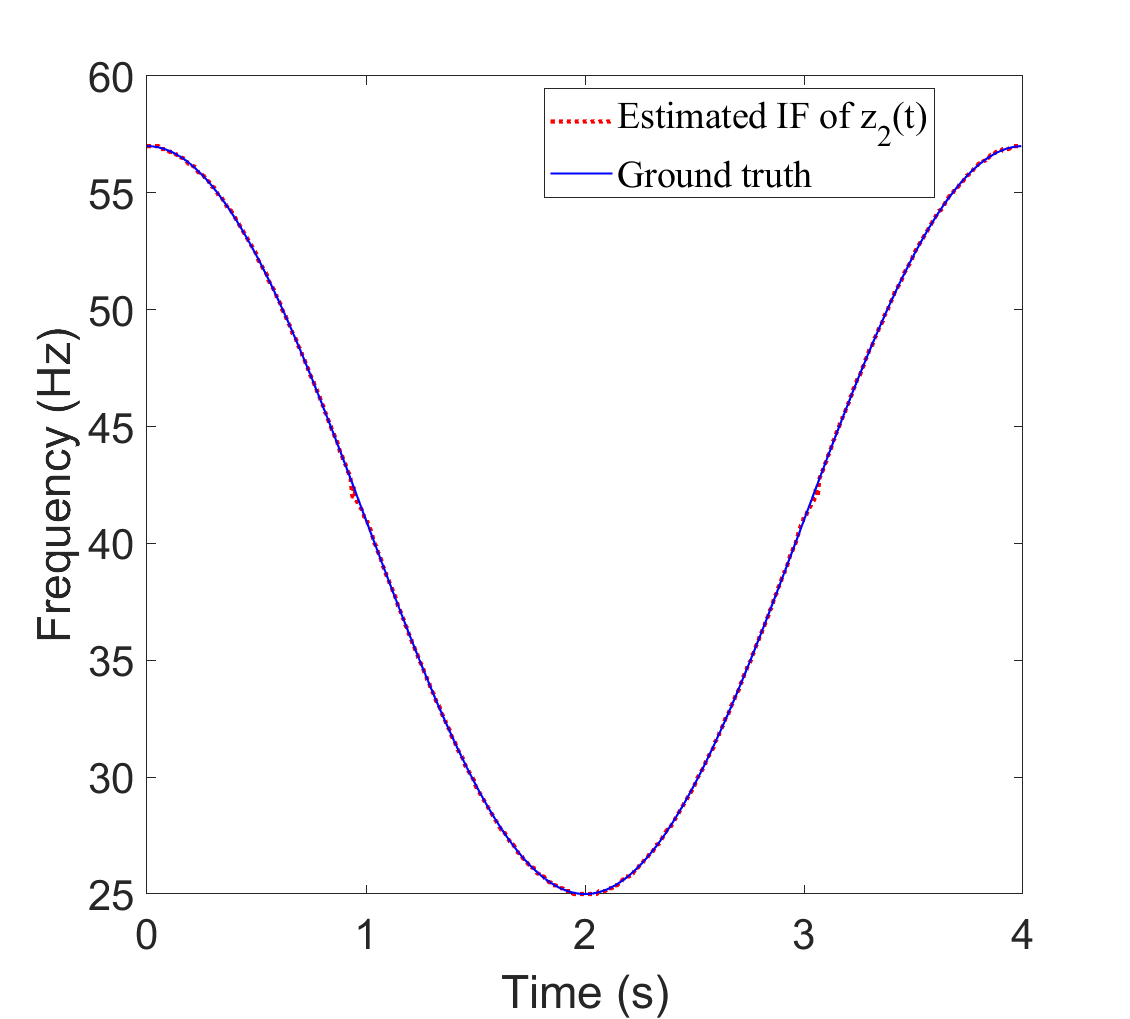}}
&
\resizebox {1.5in}{1.0in} {\includegraphics{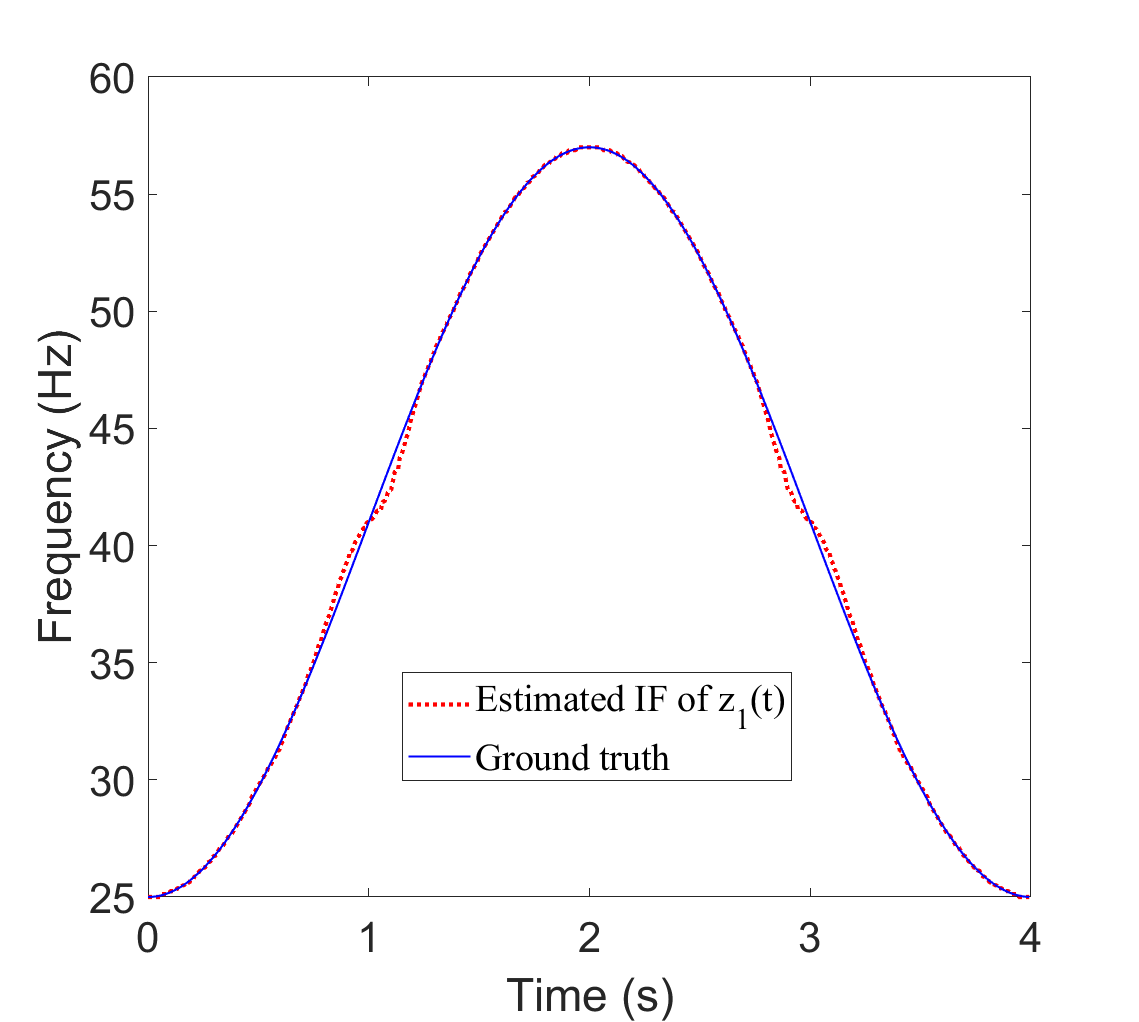}}	
& \resizebox {1.5in}{1.0in} {\includegraphics{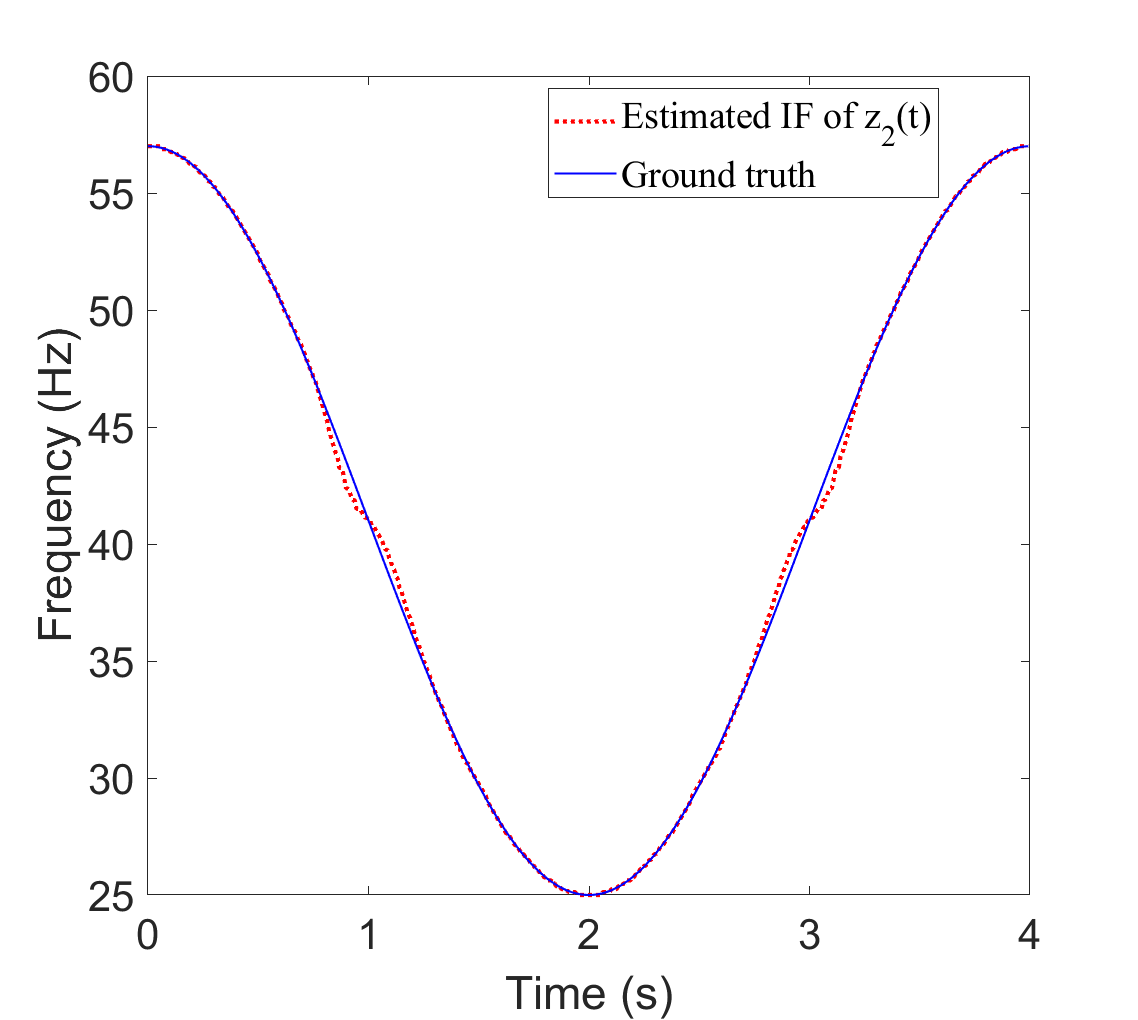}}\\
\resizebox {1.5in}{1.0in} {\includegraphics{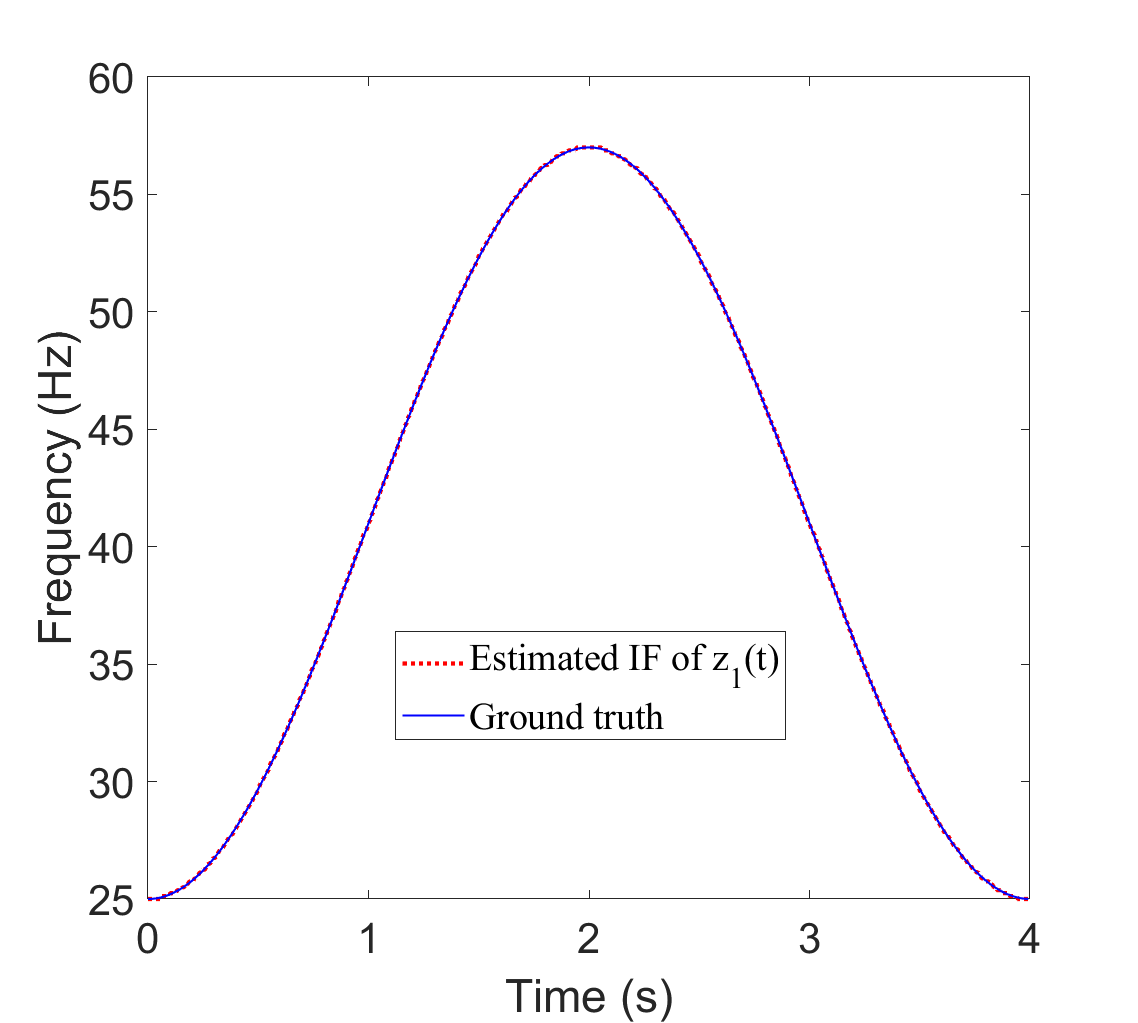}}	
&\resizebox {1.5in}{1.0in} {\includegraphics{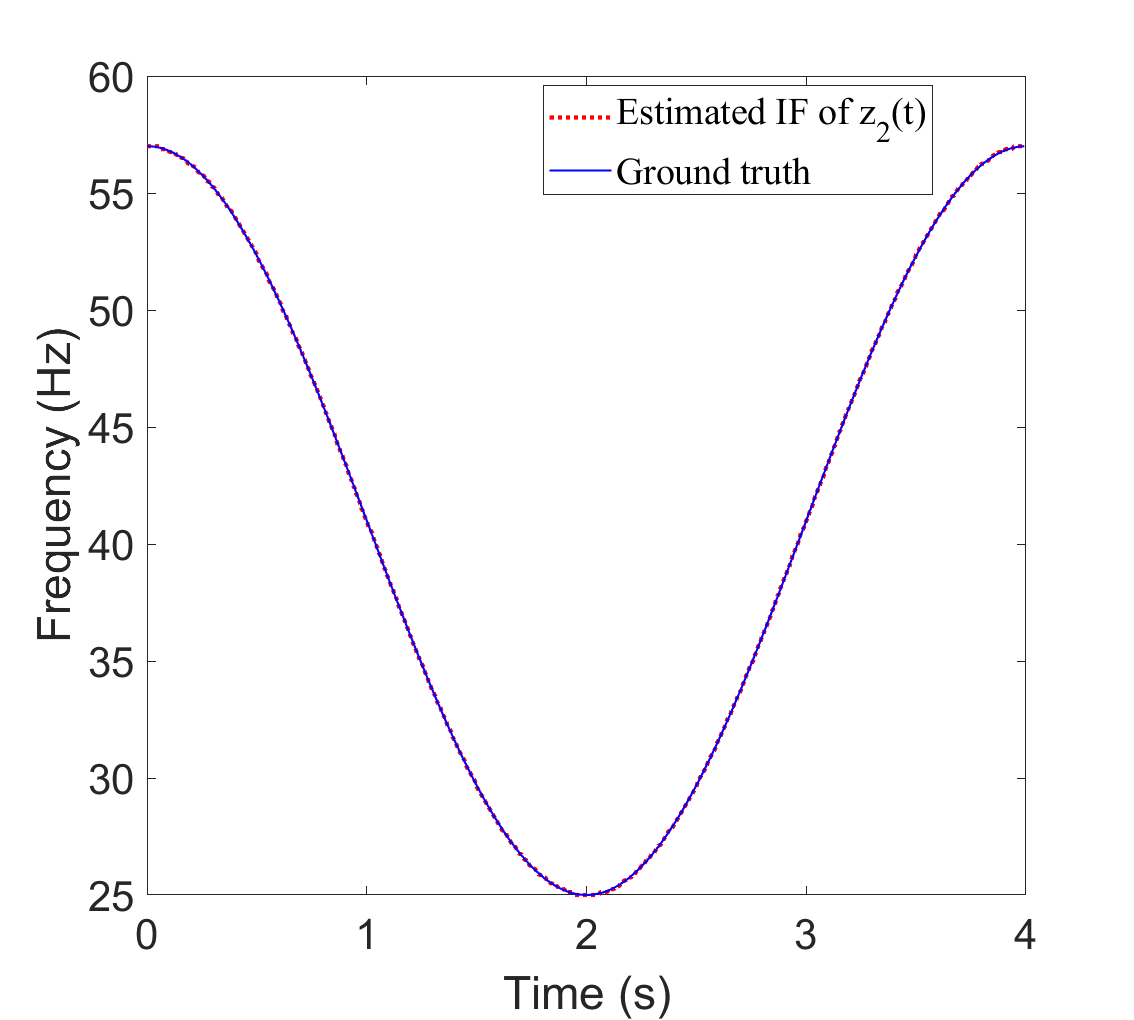}}
&&
	\end{tabular}
\caption{\small IF estimation of $z_k(t)$. Top row (from left to right): by SWCT (1st and 2nd panels); by MSWCT (3rd and 4th panels). Bottom row:  by SXWCT.}
	\label{figure:example3_IFs_est}
\end{figure}
%%%%%%%%%%%%%%%%the end of figure 15  %%%%%%%%%%%%%%%%%%%%%
 
%\bigskip 
{\bf Example 3 (Continued).} Let $z(t)$ be the signal given by \eqref{def_z}. Fig.\ref{figure:example3_IFs_est} presents IF estimation results with SWCT, MSWCT and SXWCT. 
For this signal, SWCT provides accurate IF estimates except for at $t=1, 3$, where the IFs of $z_1(t)$ and $z_2(t)$ is crossover. 
However for $z(t)$, MSWCT yields worse  accurate IF estimates. 
From the bottom row of Fig.\ref{figure:example3_IFs_est}, one can see SXWCT provides very accurate IF  estimates.  
%%%%%%%%%%%%%%%%%%%the beginning of figure 16 %%%%%%%%%%%%%%%
\begin{figure}[H]
	\centering
	\begin{tabular}{cccc}
	\resizebox {1.5in}{1.0in} {\includegraphics{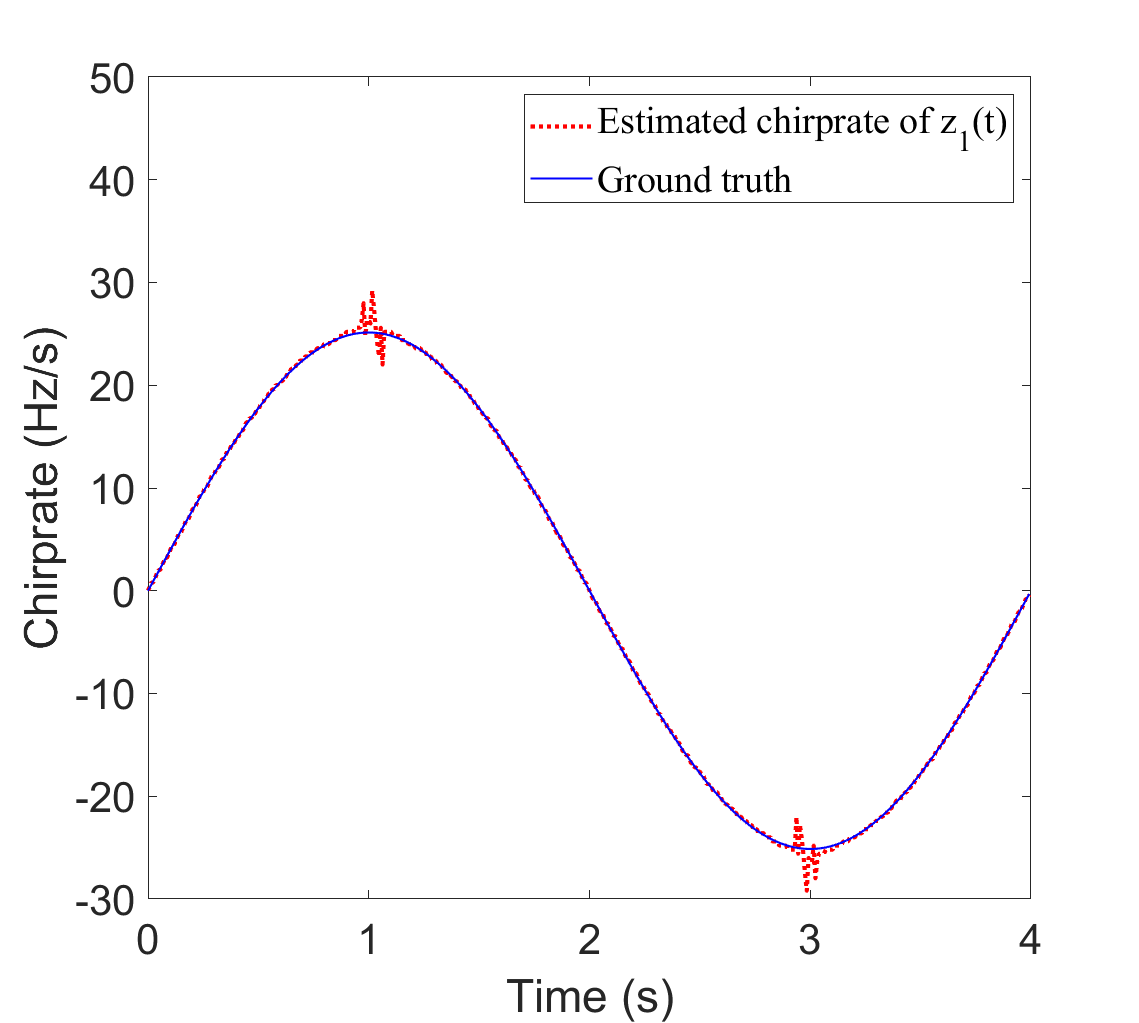}}
&
\resizebox {1.5in}{1.0in} {\includegraphics{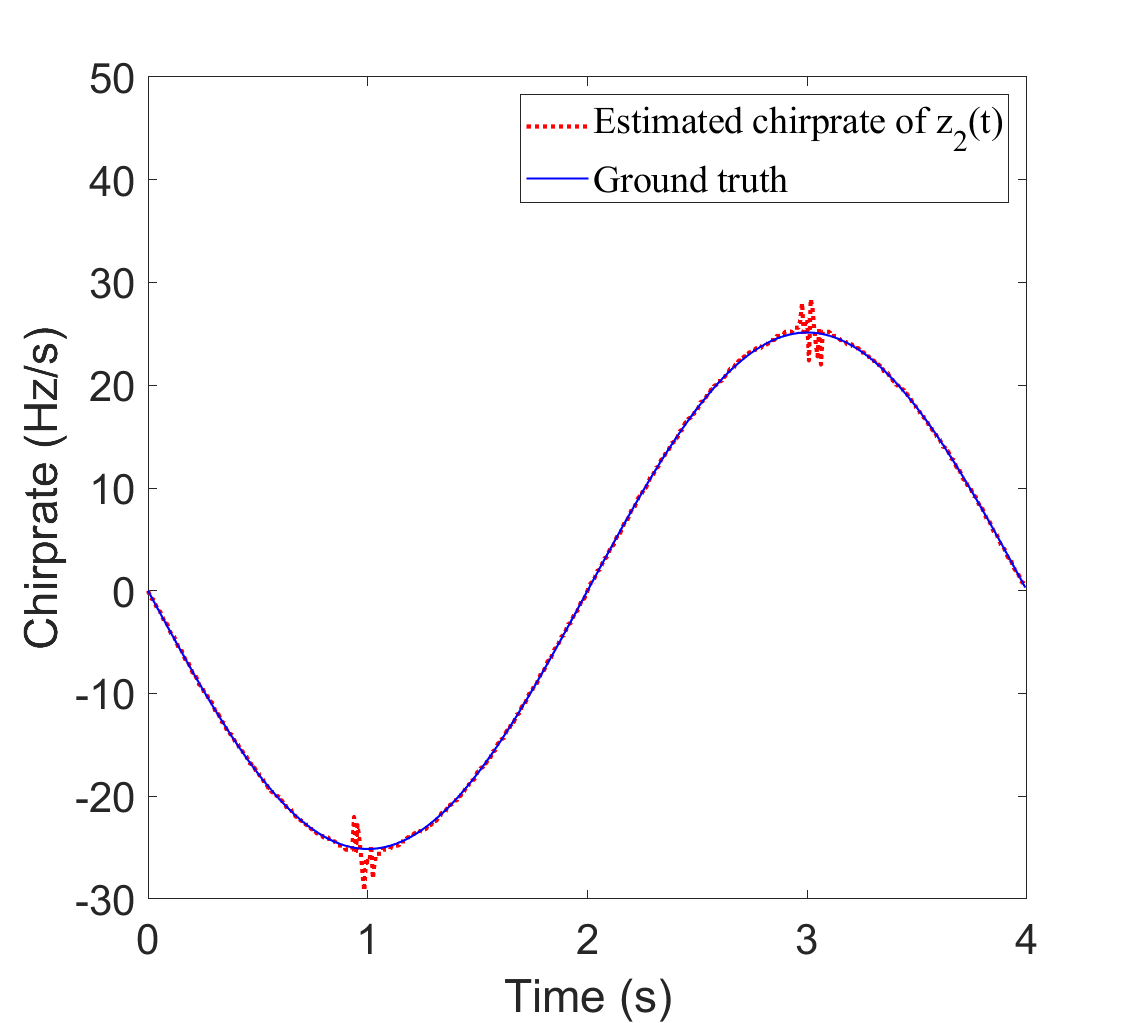}}
&\resizebox {1.5in}{1.0in} {\includegraphics{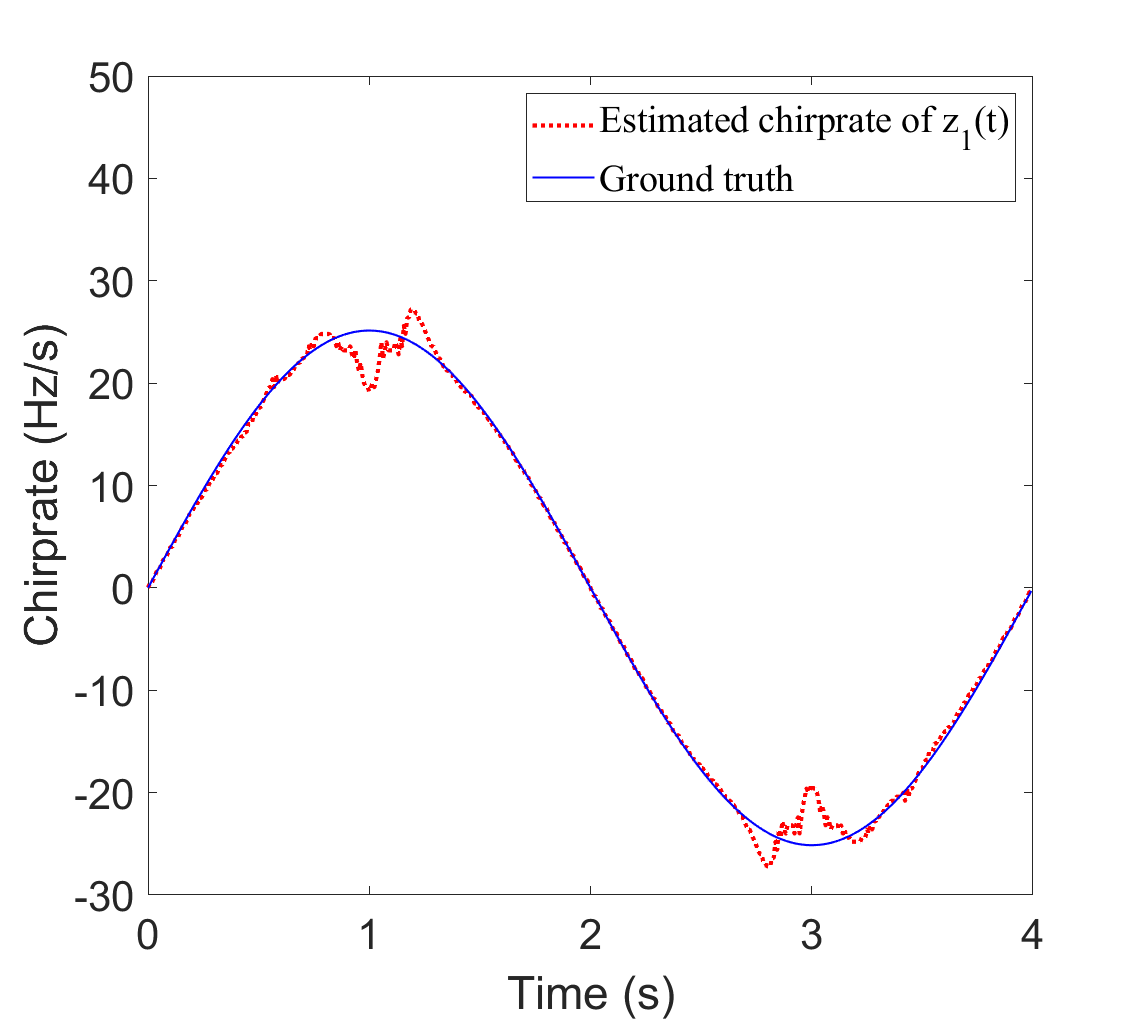}}
&
\resizebox {1.5in}{1.0in} {\includegraphics{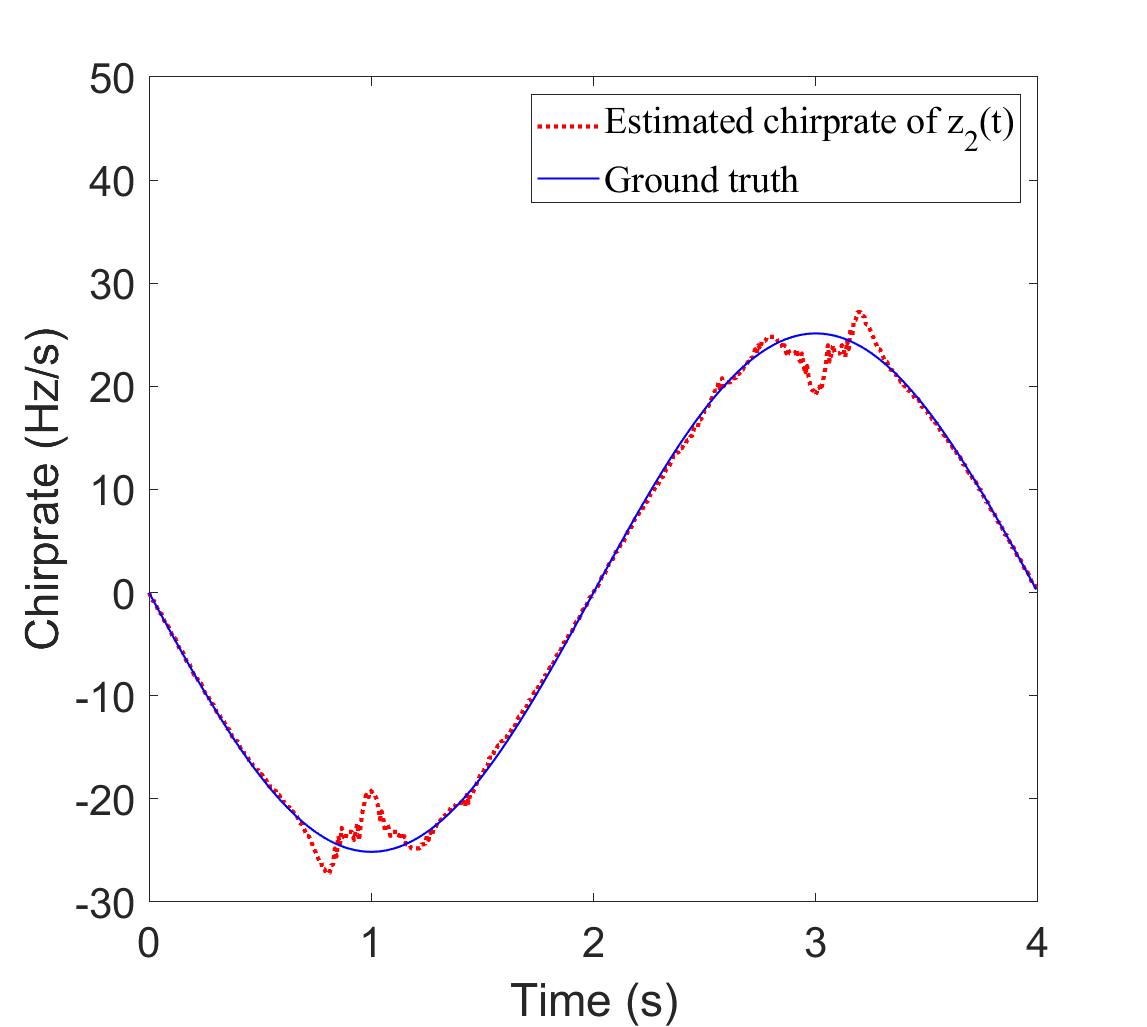}}
\\
\resizebox {1.5in}{1.0in} {\includegraphics{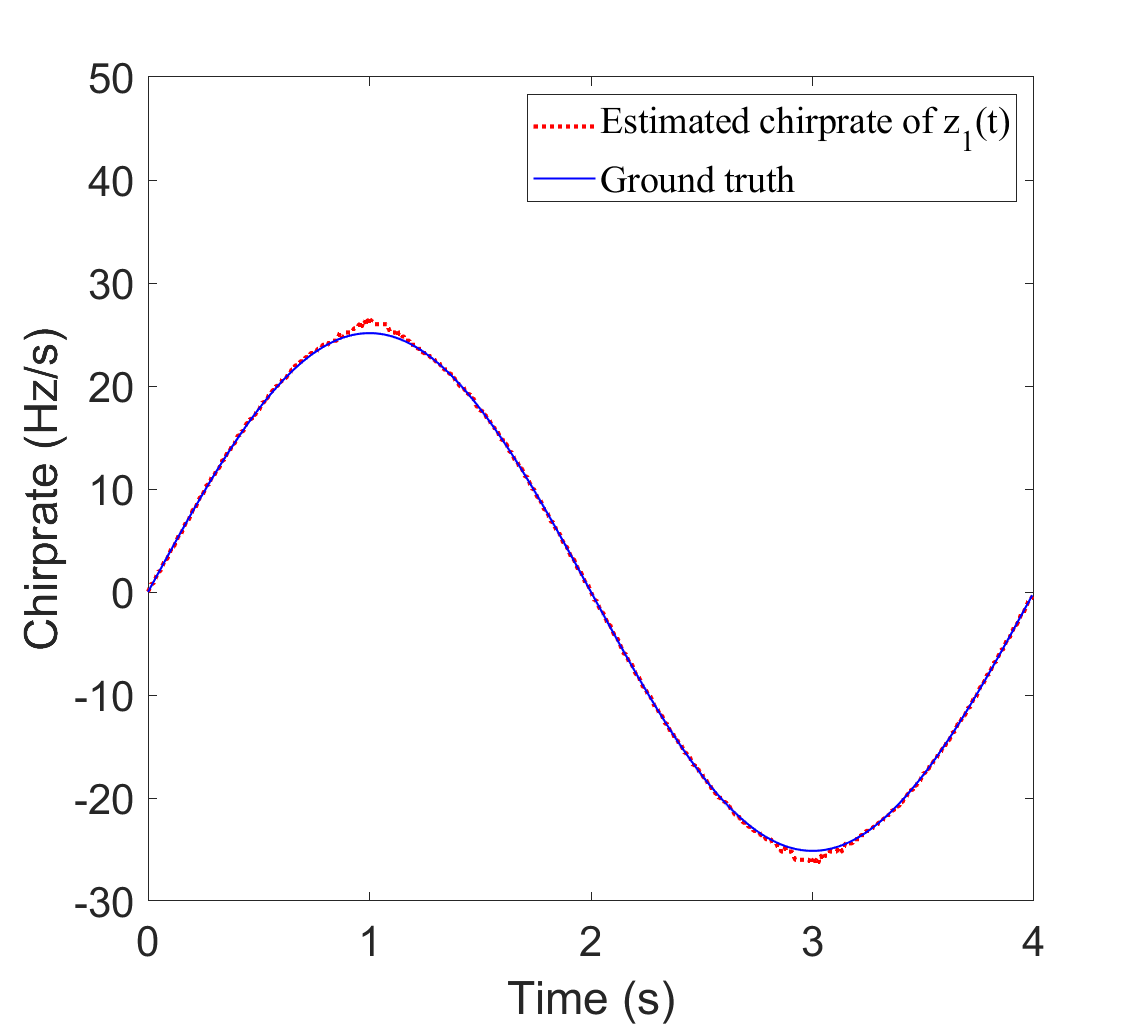}}	
& \resizebox {1.5in}{1.0in} {\includegraphics{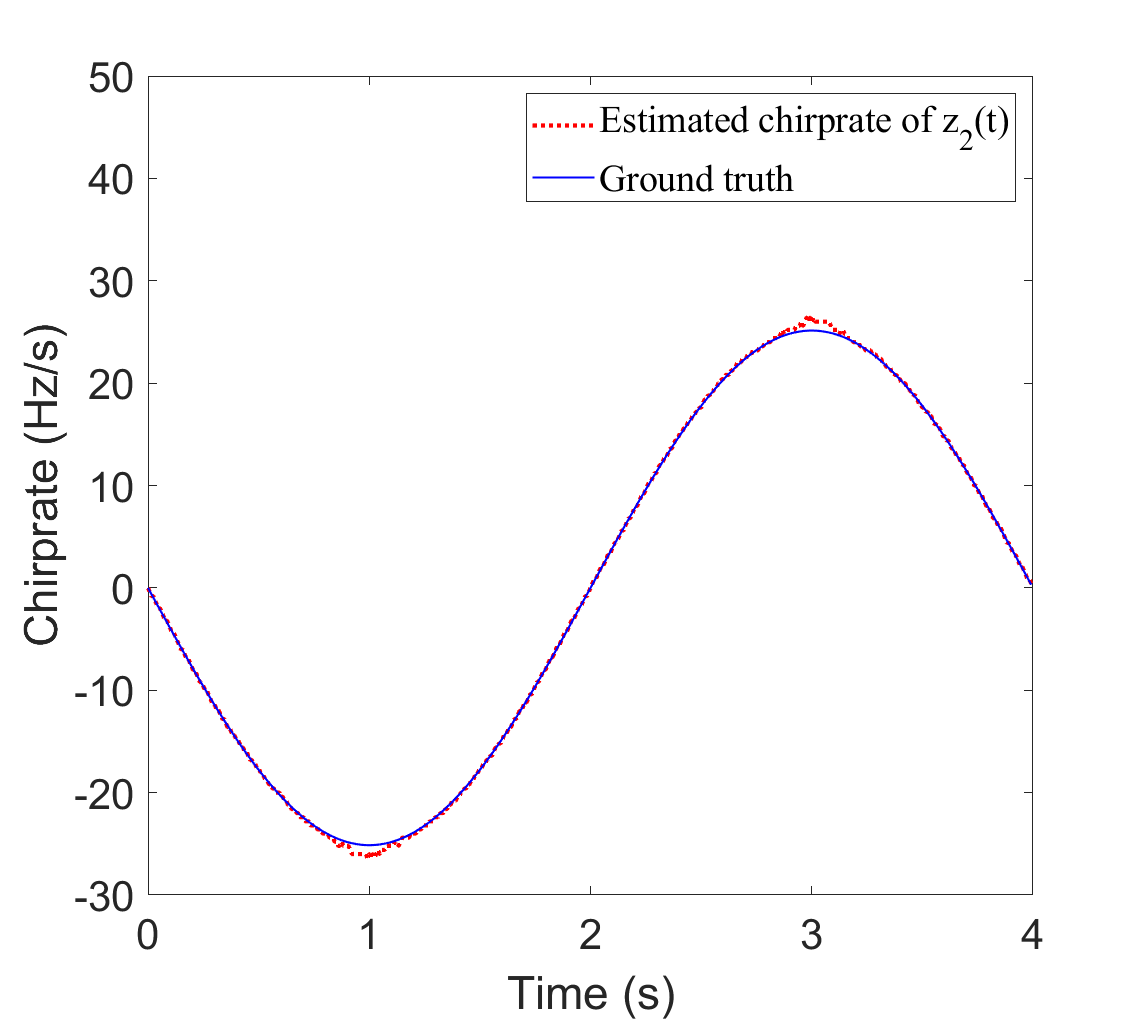}}
&&
	\end{tabular}
\caption{\small Chirprate estimation of $z_k(t)$. Top row (from left to right): by SWCT (1st and 2nd panels); by MSWCT (3rd and 4th panels). Bottom row:  by SXWCT.}
	\label{figure:example3_chirprates_est}
\end{figure}
%%%%%%%%%%%%%%%%the end of figure 16  %%%%%%%%%%%%%%%%%%%%%

\begin{figure}[H]
	\centering
	\begin{tabular}{cccc}
	\resizebox {1.5in}{1.0in} {\includegraphics{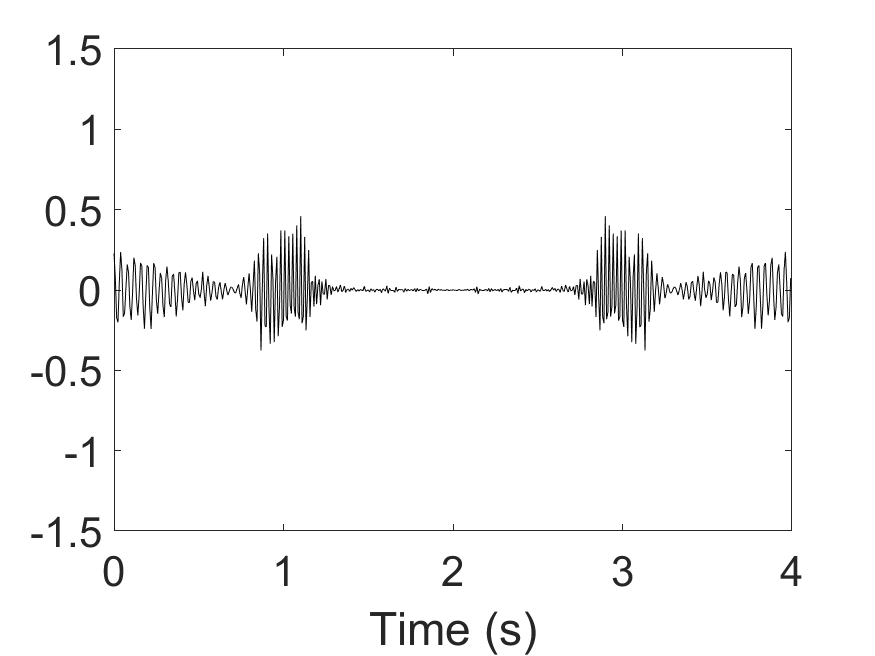}}
&  
\resizebox {1.5in}{1.0in} {\includegraphics{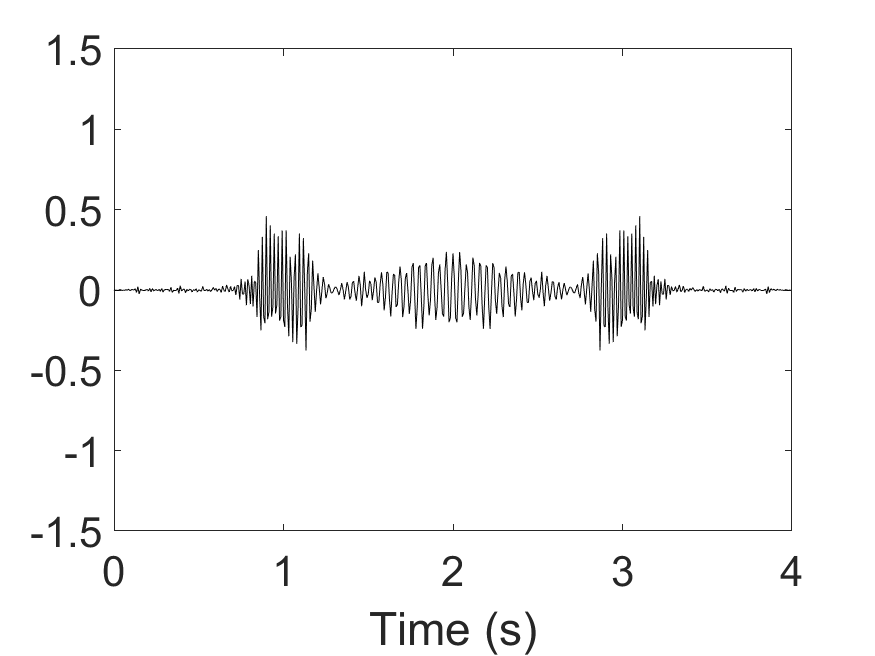}}
&
\resizebox {1.5in}{1.0in} {\includegraphics{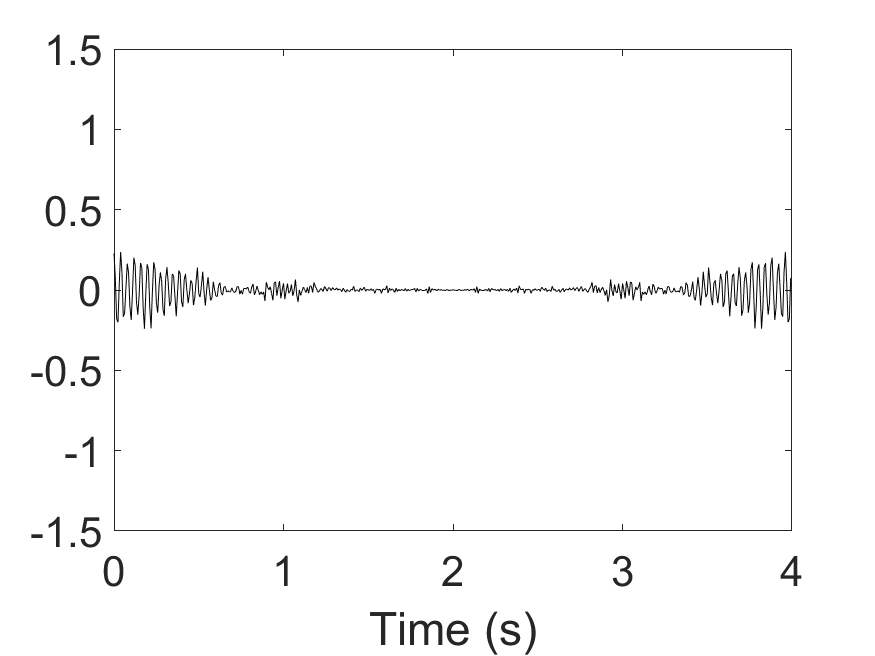}}
& \resizebox {1.5in}{1.0in} {\includegraphics{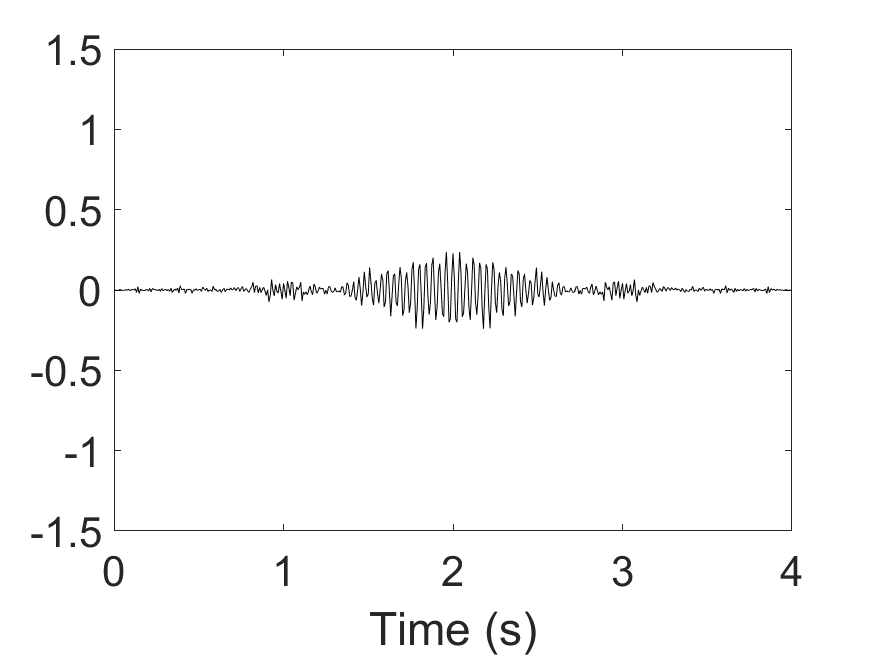}}
	\end{tabular}
\caption{\small From left to right: recover residual by MSWCT 
real$(z_1(t)-\wt z_1(t))$ (1st planel) and  real$(z_2(t)-\wt z_2(t))$ (2nd panel); 
recover residual by SXWCT real$(z_1(t)-\wt z_1(t))$ (3rd panel) and  real$(z_2(t)-\wt z_2(t))$ (4th panel).}
	\label{figure:Example3_recover}
\end{figure}

Fig.\ref{figure:example3_chirprates_est}  shows chirprate estimates by SWCT  (in 1st and 2nd panels in the top row), MSWCT  (3rd and 4th panels in the top row), and SXWCT (in the bottom row). 
There are big chirprate estimate errors near $t=1, 3$ with SWCT. The MSWCT yields larger errors in chirprate estimates than SWCT.  SXWCT provides overall accurate chirprate estimates except for at $t=1, 3$, where there are small estimate errors.  The RMSEs for mode retrieval with MSWCT:
$$
E_{z_1}=0.1699, \; E_{z_2} =0.1981,  
$$
while the errors with SXWCT are 
$$\wt E_{z_1} =0.0360, \wt E_{z_2}=0.1059. 
$$ 
The real part of the mode retrieval residual by 3rd-order MSWCT and 3rd-order SXWCT are shown in Fig.\ref{figure:Example3_recover}. 
To summarize, for $z(t)$, SXWCT also performs better than SWCT and MSWCT in IF estimation, chirprate estimation and mode retrieval.

%%%%%%%%%%%%%%%%%%%the beginning of figure 17 %%%%%%%%%%%%%%%
%%%%%%%%%%%%%%%%the end of figure 17  %%%%%%%%%%%%%%%%%%%%%

\section{Conclusion and  future work}
In this paper, the authors propose a novel transform called X-ray wavelet-chirplet transform (XWCT) by incorprating a special X-ray transform with the wavelet-chirplet transform (WCT). XWCT has a better decay along the chirprate direction than WCT. In addition, the 3rd-order synchrosqueezed  WCT (SWCT) and the 3rd-order synchrosqueezed XWCT (SXWCT) are established to have a sharp time-frequency-chirprate representation of a nonstationary signal. Our proposed 3rd-order instantaneous frequency (IF) reference function and chirprate reference function have simpler expressions. The implementations of WCT, SWCT and mutiple SWCT are present. The experimental results conducted in this paper show that the proposed XWCT indeed provides a transform with a fairly decay along the chirprate direction, and  SXWCT is able to provide  accurate IF and chirprate estimate and mode retrieval without multiple synchrosqueezing opertation. 

SXWCT is derived from XWCT, which computes the WCT along lines with specific directions and averages the results. Consequently, SXWCT requires more computations than WCT, SWCT, and multiple SWCT. 

SXWCT is derived from XWCT, which involves integrating WCT with a window function. As a result, SXWCT has a computational trade-off: it requires more calculations than existing methods such as SCT and SWCT. However, a key advantage of SXWCT is that it eliminates the need for repeated squeezing operations, in contrast to multiple SCT and multiple SWCT approaches, which require iterative squeezing.
In future work, we will explore optimized algorithms to accelerate the computation of XWCT,  and even WCT, chirplet transform in general. 

WCT has been applied to gear condition monitoring \cite{hartono2019gear}. In contrast, the proposed XWCT, 3rd-order SWCT, and SXWCT offer richer three-dimensional signal representations than WCT. We will leverage these features and/or retrieved modes for machinery vibration fault diagnosis and condition monitoring. Specifically, we will develop a triple-head ensemble Transformer algorithm that extends the existing dual-head ensemble Transformer (DHET) \cite{snyder2025integrating} by incorporating a third head—either XWCT or SXWCT.

%\section*{Declaration of competing interest}
%{\bf Declaration of competing interest:} The authors declare that they have no known competing financial interests or personal relationships that could have appeared to influence the work reported in this paper.

\section*{Acknowledgments} 
%The authors gratefully acknowledge the anonymous reviewers for their insightful comments and constructive suggestions, which have significantly contributed to the improvement of this manuscript.  
This work was partially supported by the National Key Research and Development Program of China  under Grants 2022YFA1005703  and the National Natural Science Foundation of China under Grants U21A20455. 

\section*{ Appendix }
Here the expressions of the 3rd-order IF reference function $\gT^g_x(a, b, \gl)$ and chirprate reference function $\Xi^g_x(a, b, \gl)$ by \eqref{3rd_IF} and \eqref{3rd_chirprate} are provided respectively:
\begin{equation}
\label{d_phi}
\gT^g_x(a, b, \gl)
= \frac{\mu}{a}  -\frac 1a {\rm Re}\Big( \frac{1}{i 2\pi } \frac{U^{g'}_x}{U^g_x} +\frac{1}{i2\pi}  \frac{{\rm num}_1}{{\rm denom}}\Big), \quad 
 \Xi^g_x(a, b, \gl) 
=\lambda+ \frac1{a^2}{\rm Re}\Big( \frac{1}{i 2\pi }  \frac{\rm num_2}{\rm denom}\Big),
\end{equation}
where 
\begin{eqnarray*}
&&   \text{denom}:=\frac{i \pi}{2}  
    (U^{b g}_x U^{b^2 g}_x - U^{b^3 g}_x U^{g}_x)(U^{b^2 g}_x U^{g'}_x - U^{b^2 g'}_x U^{g}_x) 
%\\&&  
 -\frac{i \pi}{2}   (U^{b^2 g}_x U^{b^2 g}_x - U^{g}_x U^{b^4 g}_x)(U^{g'}_x U^{bg}_x - U^{bg'}_x U^{g}_x) \\
    &&  + \gl \pi^2 a^2 (U^{b^3 g}_x U^{g}_x - U^{b g}_x U^{b^2 g}_x)^2 + \gl \pi^2 a^2  (U^{b^2 g}_x U^{b^2 g}_x - U^{g}_x U^{b^4 g}_x)(U^{b^2 g}_x U^{g}_x - U^{bg}_x U^{bg}_x), \\
    && \text{num1}:=\frac{i \pi}{2}(U^{b^2 g'}_x U^{ g}_x - U^{ g'}_x U^{b^2 g}_x+2 U^{b g}_x U^{ g}_x )(U^{b^2 g}_x U^{b g'}_x - U^{b^2 g'}_x U^{b g}_x) \\
&&   +\frac{i \pi}{2} (U^{ g''}_x U^{ g}_x - U^{g'}_x U^{ g'}_x)(U^{ b^4 g}_x U^{b g}_x - U^{b^3 g}_x U^{b^2 g}_x) + \gl \pi^2 a^2(U^{b^2 g'}_x U^{g}_x - U^{ g'}_x U^{b^2 g}_x+2U^{ bg}_x U^{ g}_x)(U^{ b^2 g}_x U^{b^2 g}_x - U^{b^3 g}_x U^{b g}_x) \\
    && + \gl \pi^2 a^2 (U^{b g}_x U^{g'}_x - U^{bg'}_x U^{ g}_x-U^{g}_x U^{ g}_x)(U^{b^3 g}_x U^{b^2 g}_x - U^{bg}_x U^{b^4 g}_x), 
\\ %  \end{eqnarray*}
%\begin{align*} 
&&\text{num2}:=\frac{i \pi}{2} 
    (U^{b^2 g'}_x U^{ g}_x - U^{ g'}_x U^{b^2 g}_x+2 U^{b g}_x U^{ g}_x )(U^{b^2 g}_x U^{ g'}_x - U^{b^2 g'}_x U^{ g}_x)\\
&& + 
\frac{i \pi}{2} (U^{ g''}_x U^{ g}_x - U^{g'}_x U^{ g'}_x)(U^{ b^4 g}_x U^{ g}_x - U^{b^2 g}_x U^{b^2 g}_x)   
  + \gl \pi^2 a^2 (U^{b^2 g'}_x U^{g}_x - U^{ g'}_x U^{b^2 g}_x+2U^{ bg}_x U^{ g}_x)(U^{ b g}_x U^{b^2 g}_x - U^{b^3 g}_x U^{ g}_x) \\
&& + \gl \pi^2 a^2 (U^{b g}_x U^{g'}_x - U^{bg'}_x U^{ g}_x-U^{g}_x U^{ g}_x)(U^{b^2 g}_x U^{b^2 g}_x - U^{g}_x U^{b^4 g}_x).  
\end{eqnarray*}

When $g(t)=g_\gs(t)$, by direct caluculations, one can obtain that \eqref{d_phi}  
is reduced to 
\begin{equation}
\label{def_gT_simple}
\gT^{g_\gs}_x(a,b,\gl) = \frac \mu a  + \frac{1}{a} \, \mathrm{Re} \Big( \frac{1}{i2\pi}  \frac{{\rm num}_3}{{\rm denom_1}}\Big), \quad 
%{def_Xi_simple}
\Xi^{g_\gs}_x(a,b,\gl) = \gl - \frac{1}{a^2} \, \mathrm{Re} \Big( \frac{1}{i2\pi}  \frac{{\rm num}_4}{{\rm denom_1}}\Big), 
\end{equation}
\begin{eqnarray*}
&&{\rm num}_3:=2U^{g_{\gs}}_x U^{bg_{\gs}}_x (U^{b^3g_{\gs}}_x U^{bg_{\gs}}_x - U^{b^2g_{\gs}}_x U^{b^2g_{\gs}}_x) + U^{g_{\gs}}_x U^{g_{\gs}}_x (U^{b^2g_{\gs}}_x U^{b^3g_{\gs}}_x - U^{bg_{\gs}}_x U^{b^4g_{\gs}}_x),\\
&&{\rm num}_4:=2U^{g_{\gs}}_x U^{bg_{\gs}}_x (U^{b^3g_{\gs}}_x U^{g_{\gs}}_x - U^{bg_{\gs}}_x U^{b^2g_{\gs}}_x) + U^{g_{\gs}}_x U^{g_{\gs}}_x (U^{b^2g_{\gs}}_x U^{b^2g_{\gs}}_x - U^{g_{\gs}}_x U^{b^4g_{\gs}}_x),\\
&&{\rm denom}_1:=(U_x^{b^3g_{\gs}} U_x^{g_{\gs}} - U_x^{b^2g_{\gs}} U_x^{bg_{\gs}})^2 + (U^{b^2g_{\gs}}_x U^{b^2g_{\gs}}_x - U^{b^4g_{\gs}}_x U_x^{g_{\gs}}) (U^{b^2g_{\gs}}_x U^{g_{\gs}}_x - U^{bg_{\gs}}_x U_x^{bg_{\gs}}). 
\end{eqnarray*}





% %% For authoryear reference style

% \bibitem[Lamport(1994)]{lamport94}
%   \textit{\LaTeX: a document preparation system},
%   Addison Wesley, Massachusetts,

% \end{thebibliography}

% \bibitem{Chen_S17} S. Chen, X. Dong, Z. Peng, W. Zhang, and G. Meng, ``Nonlinear chirp mode decomposition: A variational method," {\it IEEE Trans. Signal Proc.}, vol. 65, no. 22, pp. 6024--6037, Nov. 2017.

\bibliographystyle{elsarticle-num-names} 
\bibliography{references_Mar29_2025}

\begin{thebibliography}{72}
\expandafter\ifx\csname natexlab\endcsname\relax\def\natexlab#1{#1}\fi
\providecommand{\url}[1]{\texttt{#1}}
\providecommand{\href}[2]{#2}
\providecommand{\path}[1]{#1}
\providecommand{\DOIprefix}{doi:}
\providecommand{\ArXivprefix}{arXiv:}
\providecommand{\URLprefix}{URL: }
\providecommand{\Pubmedprefix}{pmid:}
\providecommand{\doi}[1]{\href{http://dx.doi.org/#1}{\path{#1}}}
\providecommand{\Pubmed}[1]{\href{pmid:#1}{\path{#1}}}
\providecommand{\bibinfo}[2]{#2}
\ifx\xfnm\relax \def\xfnm[#1]{\unskip,\space#1}\fi
%Type = Article
\bibitem[{Huang et~al.(1998)Huang, Shen, and et~al.}]{huang1998empirical}
\bibinfo{author}{N.~E. Huang}, \bibinfo{author}{Z.~Shen}, \bibinfo{author}{et~al.},
\newblock \bibinfo{title}{The empirical mode decomposition and the {H}ilbert spectrum for nonlinear and non-stationary time series analysis},
\newblock \bibinfo{journal}{Proc. R. Soc. Lond. A.} \bibinfo{volume}{454} (\bibinfo{year}{1998}) \bibinfo{pages}{903--995}.
%Type = Article
\bibitem[{Daubechies et~al.(2011)Daubechies, Lu, and Wu}]{daubechies2011synchrosqueezed}
\bibinfo{author}{I.~Daubechies}, \bibinfo{author}{J.~Lu}, \bibinfo{author}{H.-T. Wu},
\newblock \bibinfo{title}{Synchrosqueezed wavelet transforms: An empirical mode decomposition-like tool},
\newblock \bibinfo{journal}{Appl. Comput. Harmon. Anal.} \bibinfo{volume}{30} (\bibinfo{year}{2011}) \bibinfo{pages}{243--261}.
%Type = Article
\bibitem[{Thakur and Wu(2011)}]{thakur2011synchrosqueezing}
\bibinfo{author}{G.~Thakur}, \bibinfo{author}{H.-T. Wu},
\newblock \bibinfo{title}{Synchrosqueezing-based recovery of instantaneous frequency from nonuniform samples},
\newblock \bibinfo{journal}{SIAM J. Math. Anal.} \bibinfo{volume}{43} (\bibinfo{year}{2011}) \bibinfo{pages}{2078--2095}.
%Type = Inproceedings
\bibitem[{Oberlin et~al.(2014)Oberlin, Meignen, and Perrier}]{oberlin2014fourier}
\bibinfo{author}{T.~Oberlin}, \bibinfo{author}{S.~Meignen}, \bibinfo{author}{V.~Perrier},
\newblock \bibinfo{title}{The {F}ourier-based synchrosqueezing transform},
\newblock in: \bibinfo{booktitle}{2014 IEEE Int. Conf. on Acoustics, Speech and Signal Proc. (ICASSP)}, \bibinfo{organization}{IEEE}, \bibinfo{year}{2014}, pp. \bibinfo{pages}{315--319}.
%Type = Article
\bibitem[{Oberlin et~al.(2015)Oberlin, Meignen, and Perrier}]{oberlin2015second}
\bibinfo{author}{T.~Oberlin}, \bibinfo{author}{S.~Meignen}, \bibinfo{author}{V.~Perrier},
\newblock \bibinfo{title}{Second-order synchrosqueezing transform or invertible reassignment? towards ideal time-frequency representations},
\newblock \bibinfo{journal}{IEEE Trans. Signal Proc.} \bibinfo{volume}{63} (\bibinfo{year}{2015}) \bibinfo{pages}{1335--1344}.
%Type = Inproceedings
\bibitem[{Oberlin and Meignen(2017)}]{oberlin2017second}
\bibinfo{author}{T.~Oberlin}, \bibinfo{author}{S.~Meignen},
\newblock \bibinfo{title}{The second-order wavelet synchrosqueezing transform},
\newblock in: \bibinfo{booktitle}{2017 IEEE Int. Conf. on Acoustics, Speech and Signal Proc. (ICASSP)}, \bibinfo{organization}{IEEE}, \bibinfo{year}{2017}, pp. \bibinfo{pages}{3994--3998}.
%Type = Article
\bibitem[{Behera et~al.(2018)Behera, Meignen, and Oberlin}]{behera2018theoretical}
\bibinfo{author}{R.~Behera}, \bibinfo{author}{S.~Meignen}, \bibinfo{author}{T.~Oberlin},
\newblock \bibinfo{title}{Theoretical analysis of the second-order synchrosqueezing transform},
\newblock \bibinfo{journal}{Appl. Comput. Harmon. Anal.} \bibinfo{volume}{45} (\bibinfo{year}{2018}) \bibinfo{pages}{379--404}.
%Type = Article
\bibitem[{Lu et~al.(2021)Lu, Alzahrani, and Jiang}]{lu2021second}
\bibinfo{author}{J.~Lu}, \bibinfo{author}{J.~H. Alzahrani}, \bibinfo{author}{Q.~Jiang},
\newblock \bibinfo{title}{A second-order synchrosqueezing transform with a simple form of phase transformation},
\newblock \bibinfo{journal}{Numer. Math.: Theory, Methods Appl.}  (\bibinfo{year}{2021}).
%Type = Article
\bibitem[{Pham and Meignen(2017)}]{pham2017high}
\bibinfo{author}{D.-H. Pham}, \bibinfo{author}{S.~Meignen},
\newblock \bibinfo{title}{High-order synchrosqueezing transform for multicomponent signals analysis--with an application to gravitational-wave signal},
\newblock \bibinfo{journal}{IEEE Trans. Signal Proc.} \bibinfo{volume}{65} (\bibinfo{year}{2017}) \bibinfo{pages}{3168--3178}.
%Type = Article
\bibitem[{Yu et~al.(2017)Yu, Yu, and Xu}]{yu2017synchroextracting}
\bibinfo{author}{G.~Yu}, \bibinfo{author}{M.~Yu}, \bibinfo{author}{C.~Xu},
\newblock \bibinfo{title}{Synchroextracting transform},
\newblock \bibinfo{journal}{IEEE Trans. Ind. Electron.} \bibinfo{volume}{64} (\bibinfo{year}{2017}) \bibinfo{pages}{8042--8054}.
%Type = Article
\bibitem[{Yu et~al.(2018)Yu, Wang, and Zhao}]{yu2018multisynchrosqueezing}
\bibinfo{author}{G.~Yu}, \bibinfo{author}{Z.~Wang}, \bibinfo{author}{P.~Zhao},
\newblock \bibinfo{title}{Multisynchrosqueezing transform},
\newblock \bibinfo{journal}{IEEE Trans. Ind. Electron.} \bibinfo{volume}{66} (\bibinfo{year}{2018}) \bibinfo{pages}{5441--5455}.
%Type = Article
\bibitem[{Li and Liang(2012)}]{li2012generalized}
\bibinfo{author}{C.~Li}, \bibinfo{author}{M.~Liang},
\newblock \bibinfo{title}{A generalized synchrosqueezing transform for enhancing signal time--frequency representation},
\newblock \bibinfo{journal}{Signal Proc.} \bibinfo{volume}{92} (\bibinfo{year}{2012}) \bibinfo{pages}{2264--2274}.
%Type = Article
\bibitem[{Wang et~al.(2013)Wang, Chen, Cai, Chen, Li, and He}]{wang2013matching}
\bibinfo{author}{S.~Wang}, \bibinfo{author}{X.~Chen}, \bibinfo{author}{G.~Cai}, \bibinfo{author}{B.~Chen}, \bibinfo{author}{X.~Li}, \bibinfo{author}{Z.~He},
\newblock \bibinfo{title}{Matching demodulation transform and synchrosqueezing in time-frequency analysis},
\newblock \bibinfo{journal}{IEEE Trans. Signal Proc.} \bibinfo{volume}{62} (\bibinfo{year}{2013}) \bibinfo{pages}{69--84}.
%Type = Article
\bibitem[{Meignen et~al.(2017)Meignen, Pham, and McLaughlin}]{meignen2017demodulation}
\bibinfo{author}{S.~Meignen}, \bibinfo{author}{D.-H. Pham}, \bibinfo{author}{S.~McLaughlin},
\newblock \bibinfo{title}{On demodulation, ridge detection, and synchrosqueezing for multicomponent signals},
\newblock \bibinfo{journal}{IEEE Trans. Signal Proc.} \bibinfo{volume}{65} (\bibinfo{year}{2017}) \bibinfo{pages}{2093--2103}.
%Type = Article
\bibitem[{Jiang and Suter(2017)}]{jiang2017instantaneous}
\bibinfo{author}{Q.~Jiang}, \bibinfo{author}{B.~W. Suter},
\newblock \bibinfo{title}{Instantaneous frequency estimation based on synchrosqueezing wavelet transform},
\newblock \bibinfo{journal}{Signal Proc.} \bibinfo{volume}{138} (\bibinfo{year}{2017}) \bibinfo{pages}{167--181}.
%Type = Article
\bibitem[{Wang et~al.(2018)Wang, Chen, Selesnick, Guo, Tong, and Zhang}]{wang2018matching}
\bibinfo{author}{S.~Wang}, \bibinfo{author}{X.~Chen}, \bibinfo{author}{I.~W. Selesnick}, \bibinfo{author}{Y.~Guo}, \bibinfo{author}{C.~Tong}, \bibinfo{author}{X.~Zhang},
\newblock \bibinfo{title}{Matching synchrosqueezing transform: A useful tool for characterizing signals with fast varying instantaneous frequency and application to machine fault diagnosis},
\newblock \bibinfo{journal}{Mech. Syst. Signal Proc.} \bibinfo{volume}{100} (\bibinfo{year}{2018}) \bibinfo{pages}{242--288}.
%Type = Article
\bibitem[{Jiang et~al.(2022)Jiang, Prater-Bennette, Suter, and Zeyani}]{jiang2022instantaneous}
\bibinfo{author}{Q.~Jiang}, \bibinfo{author}{A.~Prater-Bennette}, \bibinfo{author}{B.~W. Suter}, \bibinfo{author}{A.~Zeyani},
\newblock \bibinfo{title}{Instantaneous frequency-embedded synchrosqueezing transform for signal separation},
\newblock \bibinfo{journal}{Front. Appl. Math. Stat.-Math. of Comput. Data Sci.} \bibinfo{volume}{138} (\bibinfo{year}{2022}) \bibinfo{pages}{167--181}.
%Type = Article
\bibitem[{Sheu et~al.(2017)Sheu, Hsu, Chou, and Wu}]{sheu2017entropy}
\bibinfo{author}{Y.-L. Sheu}, \bibinfo{author}{L.-Y. Hsu}, \bibinfo{author}{P.-T. Chou}, \bibinfo{author}{H.-T. Wu},
\newblock \bibinfo{title}{Entropy-based time-varying window width selection for nonlinear-type time--frequency analysis},
\newblock \bibinfo{journal}{Int. J. Data Sci. Anal.} \bibinfo{volume}{3} (\bibinfo{year}{2017}) \bibinfo{pages}{231--245}.
%Type = Inproceedings
\bibitem[{Berrian and Saito(2017)}]{berrian2017adaptive}
\bibinfo{author}{A.~Berrian}, \bibinfo{author}{N.~Saito},
\newblock \bibinfo{title}{Adaptive synchrosqueezing based on a quilted short-time {F}ourier transform},
\newblock in: \bibinfo{booktitle}{Wavelets and Sparsity XVII}, volume \bibinfo{volume}{10394}, \bibinfo{organization}{SPIE}, \bibinfo{year}{2017}, pp. \bibinfo{pages}{413--432}.
%Type = Article
\bibitem[{Li et~al.(2020{\natexlab{a}})Li, Cai, and Jiang}]{li2020adaptive}
\bibinfo{author}{L.~Li}, \bibinfo{author}{H.~Cai}, \bibinfo{author}{Q.~Jiang},
\newblock \bibinfo{title}{Adaptive synchrosqueezing transform with a time-varying parameter for non-stationary signal separation},
\newblock \bibinfo{journal}{Appl. Comput. Harmon. Anal.} \bibinfo{volume}{49} (\bibinfo{year}{2020}{\natexlab{a}}) \bibinfo{pages}{1075--1106}.
%Type = Article
\bibitem[{Li et~al.(2020{\natexlab{b}})Li, Cai, Han, Jiang, and Ji}]{li2020adaptivestft}
\bibinfo{author}{L.~Li}, \bibinfo{author}{H.~Cai}, \bibinfo{author}{H.~Han}, \bibinfo{author}{Q.~Jiang}, \bibinfo{author}{H.~Ji},
\newblock \bibinfo{title}{Adaptive short-time {F}ourier transform and synchrosqueezing transform for non-stationary signal separation},
\newblock \bibinfo{journal}{Signal Proc.} \bibinfo{volume}{166} (\bibinfo{year}{2020}{\natexlab{b}}) \bibinfo{pages}{107231}.
%Type = Inproceedings
\bibitem[{Li et~al.(2018)Li, Wang, Cai, Jiang, and Ji}]{li2018time}
\bibinfo{author}{L.~Li}, \bibinfo{author}{Z.~Wang}, \bibinfo{author}{H.~Cai}, \bibinfo{author}{Q.~Jiang}, \bibinfo{author}{H.~Ji},
\newblock \bibinfo{title}{Time-varying parameter-based synchrosqueezing wavelet transform with the approximation of cubic phase functions},
\newblock in: \bibinfo{booktitle}{2018 14th IEEE Int. Conf. on Signal Proc. (ICSP)}, \bibinfo{organization}{IEEE}, \bibinfo{year}{2018}, pp. \bibinfo{pages}{844--848}.
%Type = Article
\bibitem[{Lu et~al.(2020)Lu, Jiang, and Li}]{lu2020analysis}
\bibinfo{author}{J.~Lu}, \bibinfo{author}{Q.~Jiang}, \bibinfo{author}{L.~Li},
\newblock \bibinfo{title}{Analysis of adaptive synchrosqueezing transform with a time-varying parameter},
\newblock \bibinfo{journal}{Adv. Comput. Math.} \bibinfo{volume}{46} (\bibinfo{year}{2020}) \bibinfo{pages}{72}.
%Type = Article
\bibitem[{Cai et~al.(2021)Cai, Jiang, Li, and Suter}]{cai2021analysis}
\bibinfo{author}{H.~Cai}, \bibinfo{author}{Q.~Jiang}, \bibinfo{author}{L.~Li}, \bibinfo{author}{B.~W. Suter},
\newblock \bibinfo{title}{Analysis of adaptive short-time {F}ourier transform-based synchrosqueezing transform},
\newblock \bibinfo{journal}{Anal. Appl.} \bibinfo{volume}{19} (\bibinfo{year}{2021}) \bibinfo{pages}{71--105}.
%Type = Article
\bibitem[{Long et~al.(2017)Long, Wang, Li, and Fan}]{long2017applications}
\bibinfo{author}{J.~Long}, \bibinfo{author}{H.~Wang}, \bibinfo{author}{P.~Li}, \bibinfo{author}{H.~Fan},
\newblock \bibinfo{title}{Applications of fractional lower order time-frequency representation to machine bearing fault diagnosis.},
\newblock \bibinfo{journal}{IEEE CAA J. Autom. Sinica} \bibinfo{volume}{4} (\bibinfo{year}{2017}) \bibinfo{pages}{734--750}.
%Type = Article
\bibitem[{Li et~al.(2022)Li, Yu, Jiang, Zang, and Jiang}]{li2022synchrosqueezing}
\bibinfo{author}{L.~Li}, \bibinfo{author}{X.~Yu}, \bibinfo{author}{Q.~Jiang}, \bibinfo{author}{B.~Zang}, \bibinfo{author}{L.~Jiang},
\newblock \bibinfo{title}{Synchrosqueezing transform meets $\alpha$-stable distribution: An adaptive fractional lower-order sst for instantaneous frequency estimation and non-stationary signal recovery},
\newblock \bibinfo{journal}{Signal Proc.} \bibinfo{volume}{201} (\bibinfo{year}{2022}) \bibinfo{pages}{108683}.
%Type = Article
\bibitem[{Meignen et~al.(2012)Meignen, Oberlin, and McLaughlin}]{meignen2012new}
\bibinfo{author}{S.~Meignen}, \bibinfo{author}{T.~Oberlin}, \bibinfo{author}{S.~McLaughlin},
\newblock \bibinfo{title}{A new algorithm for multicomponent signals analysis based on synchrosqueezing: With an application to signal sampling and denoising},
\newblock \bibinfo{journal}{IEEE Trans. Signal Proc.} \bibinfo{volume}{60} (\bibinfo{year}{2012}) \bibinfo{pages}{5787--5798}.
%Type = Article
\bibitem[{Auger et~al.(2013)Auger, Flandrin, Lin, McLaughlin, Meignen, Oberlin, and Wu}]{auger2013time}
\bibinfo{author}{F.~Auger}, \bibinfo{author}{P.~Flandrin}, \bibinfo{author}{Y.-T. Lin}, \bibinfo{author}{S.~McLaughlin}, \bibinfo{author}{S.~Meignen}, \bibinfo{author}{T.~Oberlin}, \bibinfo{author}{H.-T. Wu},
\newblock \bibinfo{title}{Time--frequency reassignment and synchrosqueezing: An overview},
\newblock \bibinfo{journal}{IEEE Signal Proc. Mag.} \bibinfo{volume}{30} (\bibinfo{year}{2013}) \bibinfo{pages}{32--41}.
%Type = Article
\bibitem[{Snyder et~al.(2025)Snyder, Jiang, and Tripp}]{snyder2025integrating}
\bibinfo{author}{Q.~Snyder}, \bibinfo{author}{Q.~Jiang}, \bibinfo{author}{E.~Tripp},
\newblock \bibinfo{title}{Integrating self-attention mechanisms in deep learning: A novel dual-head ensemble transformer with its application to bearing fault diagnosis},
\newblock \bibinfo{journal}{Signal Proc.} \bibinfo{volume}{227} (\bibinfo{year}{2025}) \bibinfo{pages}{109683}.
%Type = Article
\bibitem[{Wu(2020)}]{wu2020current}
\bibinfo{author}{H.-T. Wu},
\newblock \bibinfo{title}{Current state of nonlinear-type time--frequency analysis and applications to high--frequency biomedical signals},
\newblock \bibinfo{journal}{Curr. Opin. Syst. Biol.} \bibinfo{volume}{23} (\bibinfo{year}{2020}) \bibinfo{pages}{8--21}.
%Type = Article
\bibitem[{Chui and Mhaskar(2016)}]{chui2016signal}
\bibinfo{author}{C.~K. Chui}, \bibinfo{author}{H.~Mhaskar},
\newblock \bibinfo{title}{Signal decomposition and analysis via extraction of frequencies},
\newblock \bibinfo{journal}{Appl. Comput. Harmon. Anal.} \bibinfo{volume}{40} (\bibinfo{year}{2016}) \bibinfo{pages}{97--136}.
%Type = Article
\bibitem[{Li et~al.(2022)Li, Chui, and Jiang}]{li2022direct}
\bibinfo{author}{L.~Li}, \bibinfo{author}{C.~K. Chui}, \bibinfo{author}{Q.~Jiang},
\newblock \bibinfo{title}{Direct signal separation via extraction of local frequencies with adaptive time-varying parameters},
\newblock \bibinfo{journal}{IEEE Trans. Signal Proc.} \bibinfo{volume}{70} (\bibinfo{year}{2022}) \bibinfo{pages}{2321--2333}.
%Type = Article
\bibitem[{Chui et~al.(2021{\natexlab{a}})Chui, Jiang, Li, and Lu}]{chui2021analysis}
\bibinfo{author}{C.~K. Chui}, \bibinfo{author}{Q.~Jiang}, \bibinfo{author}{L.~Li}, \bibinfo{author}{J.~Lu},
\newblock \bibinfo{title}{Analysis of an adaptive short-time {F}ourier transform-based multicomponent signal separation method derived from linear chirp local approximation},
\newblock \bibinfo{journal}{J. Comput. Appl. Math.} \bibinfo{volume}{396} (\bibinfo{year}{2021}{\natexlab{a}}) \bibinfo{pages}{113607}.
%Type = Article
\bibitem[{Chui et~al.(2021{\natexlab{b}})Chui, Jiang, Li, and Lu}]{chui2021signal}
\bibinfo{author}{C.~K. Chui}, \bibinfo{author}{Q.~Jiang}, \bibinfo{author}{L.~Li}, \bibinfo{author}{J.~Lu},
\newblock \bibinfo{title}{Signal separation based on adaptive continuous wavelet-like transform and analysis},
\newblock \bibinfo{journal}{Appl. Comput. Harmon. Anal.} \bibinfo{volume}{53} (\bibinfo{year}{2021}{\natexlab{b}}) \bibinfo{pages}{151--179}.
%Type = Article
\bibitem[{He et~al.(2019)He, Cao, Wang, and Chen}]{he2019time}
\bibinfo{author}{D.~He}, \bibinfo{author}{H.~Cao}, \bibinfo{author}{S.~Wang}, \bibinfo{author}{X.~Chen},
\newblock \bibinfo{title}{Time-reassigned synchrosqueezing transform: The algorithm and its applications in mechanical signal processing},
\newblock \bibinfo{journal}{Mech. Syst. Signal Proc.} \bibinfo{volume}{117} (\bibinfo{year}{2019}) \bibinfo{pages}{255--279}.
%Type = Inproceedings
\bibitem[{Fourer and Auger(2019)}]{fourer2019second}
\bibinfo{author}{D.~Fourer}, \bibinfo{author}{F.~Auger},
\newblock \bibinfo{title}{Second-order time-reassigned synchrosqueezing transform: Application to {D}raupner wave analysis},
\newblock in: \bibinfo{booktitle}{2019 27th European Signal Proc. Conf. (EUSIPCO)}, \bibinfo{organization}{IEEE}, \bibinfo{year}{2019}, pp. \bibinfo{pages}{1--5}.
%Type = Article
\bibitem[{He et~al.(2020)He, Tu, Bao, Hu, and Li}]{he2020gaussian}
\bibinfo{author}{Z.~He}, \bibinfo{author}{X.~Tu}, \bibinfo{author}{W.~Bao}, \bibinfo{author}{Y.~Hu}, \bibinfo{author}{F.~Li},
\newblock \bibinfo{title}{Gaussian-modulated linear group delay model: Application to second-order time-reassigned synchrosqueezing transform},
\newblock \bibinfo{journal}{Signal Proc.} \bibinfo{volume}{167} (\bibinfo{year}{2020}) \bibinfo{pages}{107275}.
%Type = Article
\bibitem[{Dong et~al.(2023)Dong, Yu, and Jiang}]{dong2023time}
\bibinfo{author}{H.~Dong}, \bibinfo{author}{G.~Yu}, \bibinfo{author}{Q.~Jiang},
\newblock \bibinfo{title}{Time--frequency--multisqueezing transform},
\newblock \bibinfo{journal}{IEEE Trans. Ind. Electron.} \bibinfo{volume}{71} (\bibinfo{year}{2023}) \bibinfo{pages}{4151--4161}.
%Type = Article
\bibitem[{Ma et~al.(2024{\natexlab{a}})Ma, Yu, Lin, and Sun}]{ma2024high}
\bibinfo{author}{Y.~Ma}, \bibinfo{author}{G.~Yu}, \bibinfo{author}{T.~Lin}, \bibinfo{author}{M.~Sun},
\newblock \bibinfo{title}{A high--resolution time--frequency analysis technique based on bi--directional squeezing and its application in fault diagnosis of rotating machinery},
\newblock \bibinfo{journal}{ISA Trans.} \bibinfo{volume}{147} (\bibinfo{year}{2024}{\natexlab{a}}) \bibinfo{pages}{382--402}.
%Type = Article
\bibitem[{Ma et~al.(2024{\natexlab{b}})Ma, Yu, Lin, and Jiang}]{ma2024synchro}
\bibinfo{author}{Y.~Ma}, \bibinfo{author}{G.~Yu}, \bibinfo{author}{T.~Lin}, \bibinfo{author}{Q.~Jiang},
\newblock \bibinfo{title}{Synchro--transient--extracting transform for the analysis of signals with both harmonic and impulsive components},
\newblock \bibinfo{journal}{IEEE Trans. Ind. Electron.} \bibinfo{volume}{71} (\bibinfo{year}{2024}{\natexlab{b}}) \bibinfo{pages}{13020--13030}.
%Type = Article
\bibitem[{Zhu et~al.(2025)Zhu, Yang, Zhang, and Li}]{zhu2025if}
\bibinfo{author}{X.~Zhu}, \bibinfo{author}{K.~Yang}, \bibinfo{author}{Z.~Zhang}, \bibinfo{author}{W.~Li},
\newblock \bibinfo{title}{If equation: A feature extractor for high-concentration time--frequency representation and application to mixed signals analysis},
\newblock \bibinfo{journal}{Measurement} \bibinfo{volume}{244} (\bibinfo{year}{2025}) \bibinfo{pages}{116423}.
%Type = Article
\bibitem[{Chen et~al.(2006)Chen, Li, Ho, and Wechsler}]{chen2006micro}
\bibinfo{author}{V.~C. Chen}, \bibinfo{author}{F.~Li}, \bibinfo{author}{S.-S. Ho}, \bibinfo{author}{H.~Wechsler},
\newblock \bibinfo{title}{Micro-{D}oppler effect in radar: phenomenon, model, and simulation study},
\newblock \bibinfo{journal}{IEEE Trans. Aerosp. Electron. Syst.} \bibinfo{volume}{42} (\bibinfo{year}{2006}) \bibinfo{pages}{2--21}.
%Type = Article
\bibitem[{Stankovi{\'c} et~al.(2013)Stankovi{\'c}, Orovi{\'c}, Stankovi{\'c}, and Amin}]{stankovic2013compressive}
\bibinfo{author}{L.~Stankovi{\'c}}, \bibinfo{author}{I.~Orovi{\'c}}, \bibinfo{author}{S.~Stankovi{\'c}}, \bibinfo{author}{M.~Amin},
\newblock \bibinfo{title}{Compressive sensing based separation of nonstationary and stationary signals overlapping in time-frequency},
\newblock \bibinfo{journal}{IEEE Trans. Signal Proc.} \bibinfo{volume}{61} (\bibinfo{year}{2013}) \bibinfo{pages}{4562--4572}.
%Type = Article
\bibitem[{Li et~al.(2022)Li, Han, Jiang, and Chui}]{li2022chirplet}
\bibinfo{author}{L.~Li}, \bibinfo{author}{N.~Han}, \bibinfo{author}{Q.~Jiang}, \bibinfo{author}{C.~K. Chui},
\newblock \bibinfo{title}{A chirplet transform-based mode retrieval method for multicomponent signals with crossover instantaneous frequencies},
\newblock \bibinfo{journal}{Digital Signal Proc.} \bibinfo{volume}{120,103262} (\bibinfo{year}{2022}).
%Type = Article
\bibitem[{Chui et~al.(2023)Chui, Jiang, Li, and Lu}]{chui2023analysis}
\bibinfo{author}{C.~K. Chui}, \bibinfo{author}{Q.~Jiang}, \bibinfo{author}{L.~Li}, \bibinfo{author}{J.~Lu},
\newblock \bibinfo{title}{Analysis of a direct separation method based on adaptive chirplet transform for signals with crossover instantaneous frequencies},
\newblock \bibinfo{journal}{Appl. Comput. Harmon. Anal.} \bibinfo{volume}{62} (\bibinfo{year}{2023}) \bibinfo{pages}{24--40}.
%Type = Article
\bibitem[{Chui et~al.(2021)Chui, Jiang, Li, and Lu}]{chui2021time}
\bibinfo{author}{C.~K. Chui}, \bibinfo{author}{Q.~Jiang}, \bibinfo{author}{L.~Li}, \bibinfo{author}{J.~Lu},
\newblock \bibinfo{title}{Time--scale--chirp\_rate operator for recovery of non-stationary signal components with crossover instantaneous frequency curves},
\newblock \bibinfo{journal}{Appl. Comput. Harmon. Anal.} \bibinfo{volume}{54} (\bibinfo{year}{2021}) \bibinfo{pages}{323--344}.
%Type = Article
\bibitem[{Yu and Zhou(2016)}]{yu2016general}
\bibinfo{author}{G.~Yu}, \bibinfo{author}{Y.~Zhou},
\newblock \bibinfo{title}{General linear chirplet transform},
\newblock \bibinfo{journal}{Mech. Syst. Signal Proc.} \bibinfo{volume}{70} (\bibinfo{year}{2016}) \bibinfo{pages}{958--973}.
%Type = Article
\bibitem[{Abratkiewicz(2020)}]{abratkiewicz2020double}
\bibinfo{author}{K.~Abratkiewicz},
\newblock \bibinfo{title}{Double--adaptive chirplet transform for radar signature extraction},
\newblock \bibinfo{journal}{IET Radar, Sonar \& Navigation} \bibinfo{volume}{14} (\bibinfo{year}{2020}) \bibinfo{pages}{1463--1474}.
%Type = Article
\bibitem[{Zhu et~al.(2019)Zhu, Zhang, Li, Gao, Huang, and Wen}]{zhu2019multiple}
\bibinfo{author}{X.~Zhu}, \bibinfo{author}{Z.~Zhang}, \bibinfo{author}{Z.~Li}, \bibinfo{author}{J.~Gao}, \bibinfo{author}{X.~Huang}, \bibinfo{author}{G.~Wen},
\newblock \bibinfo{title}{Multiple squeezes from adaptive chirplet transform},
\newblock \bibinfo{journal}{Signal Proc.} \bibinfo{volume}{163} (\bibinfo{year}{2019}) \bibinfo{pages}{26--40}.
%Type = Article
\bibitem[{Zhu et~al.(2021)Zhu, Zhang, and Gao}]{zhu2021three}
\bibinfo{author}{X.~Zhu}, \bibinfo{author}{Z.~Zhang}, \bibinfo{author}{J.~Gao},
\newblock \bibinfo{title}{Three--dimension extracting transform},
\newblock \bibinfo{journal}{Signal Proc.} \bibinfo{volume}{179} (\bibinfo{year}{2021}) \bibinfo{pages}{107830}.
%Type = Article
\bibitem[{Bruni et~al.(2020)Bruni, Tartaglione, and Vitulano}]{bruni2020radon}
\bibinfo{author}{V.~Bruni}, \bibinfo{author}{M.~Tartaglione}, \bibinfo{author}{D.~Vitulano},
\newblock \bibinfo{title}{Radon spectrogram--based approach for automatic ifs separation},
\newblock \bibinfo{journal}{EURASIP J. Adv. in Signal Proc.} \bibinfo{volume}{2020} (\bibinfo{year}{2020}) \bibinfo{pages}{1--21}.
%Type = Article
\bibitem[{Bruni et~al.(2021)Bruni, Tartaglione, and Vitulano}]{bruni2021pde}
\bibinfo{author}{V.~Bruni}, \bibinfo{author}{M.~Tartaglione}, \bibinfo{author}{D.~Vitulano},
\newblock \bibinfo{title}{A pde--based analysis of the spectrogram image for instantaneous frequency estimation},
\newblock \bibinfo{journal}{Mathematics} \bibinfo{volume}{9} (\bibinfo{year}{2021}) \bibinfo{pages}{247}.
%Type = Article
\bibitem[{Chen et~al.(2017)Chen, Dong, Peng, Zhang, and Meng}]{chen2017nonlinear}
\bibinfo{author}{S.~Chen}, \bibinfo{author}{X.~Dong}, \bibinfo{author}{Z.~Peng}, \bibinfo{author}{W.~Zhang}, \bibinfo{author}{G.~Meng},
\newblock \bibinfo{title}{Nonlinear chirp mode decomposition: A variational method},
\newblock \bibinfo{journal}{IEEE Trans. Signal Proc.} \bibinfo{volume}{65} (\bibinfo{year}{2017}) \bibinfo{pages}{6024--6037}.
%Type = Article
\bibitem[{Zhu et~al.(2020)Zhu, Yang, Zhang, Gao, and Liu}]{zhu2020frequency}
\bibinfo{author}{X.~Zhu}, \bibinfo{author}{H.~Yang}, \bibinfo{author}{Z.~Zhang}, \bibinfo{author}{J.~Gao}, \bibinfo{author}{N.~Liu},
\newblock \bibinfo{title}{Frequency--chirprate reassignment},
\newblock \bibinfo{journal}{Digital Signal Proc.} \bibinfo{volume}{104} (\bibinfo{year}{2020}) \bibinfo{pages}{102783}.
%Type = Article
\bibitem[{Chen and Wu(2023)}]{chen2023disentangling}
\bibinfo{author}{Z.~Chen}, \bibinfo{author}{H.-T. Wu},
\newblock \bibinfo{title}{Disentangling modes with crossover instantaneous frequencies by synchrosqueezed chirplet transforms, from theory to application},
\newblock \bibinfo{journal}{Appl. Comput. Harmon. Anal.} \bibinfo{volume}{62} (\bibinfo{year}{2023}) \bibinfo{pages}{84--122}.
%Type = Article
\bibitem[{Chen et~al.(2024{\natexlab{a}})Chen, Xie, Cui, and Su}]{chen2024multiple}
\bibinfo{author}{T.~Chen}, \bibinfo{author}{L.~Xie}, \bibinfo{author}{M.~Cui}, \bibinfo{author}{H.~Su},
\newblock \bibinfo{title}{Multiple enhanced synchrosqueezing in the time--frequency--chirprate space},
\newblock \bibinfo{journal}{Signal Proc.} \bibinfo{volume}{222} (\bibinfo{year}{2024}{\natexlab{a}}) \bibinfo{pages}{109541}.
%Type = Article
\bibitem[{Chen et~al.(2024{\natexlab{b}})Chen, Zhang, and Yang}]{chen2024composite}
\bibinfo{author}{X.~Chen}, \bibinfo{author}{Z.~Zhang}, \bibinfo{author}{W.~Yang},
\newblock \bibinfo{title}{Composite signal detection using multisynchrosqueezing wavelet transform},
\newblock \bibinfo{journal}{Digital Signal Proc.} \bibinfo{volume}{149} (\bibinfo{year}{2024}{\natexlab{b}}) \bibinfo{pages}{104482}.
%Type = Inproceedings
\bibitem[{Mann and Haykin(1991)}]{mann1991chirplet}
\bibinfo{author}{S.~Mann}, \bibinfo{author}{S.~Haykin},
\newblock \bibinfo{title}{The chirplet transform: A generalization of {G}abor’s logon transform},
\newblock in: \bibinfo{booktitle}{Vision interface}, volume~\bibinfo{volume}{91}, \bibinfo{organization}{Citeseer Princeton}, \bibinfo{year}{1991}, pp. \bibinfo{pages}{205--212}.
%Type = Article
\bibitem[{Mann and Haykin(1992)}]{mann1992adaptive}
\bibinfo{author}{S.~Mann}, \bibinfo{author}{S.~Haykin},
\newblock \bibinfo{title}{Adaptive ``chirplet'' transform: An adaptive generalization of the wavelet transform},
\newblock \bibinfo{journal}{Optical Engineering} \bibinfo{volume}{31} (\bibinfo{year}{1992}) \bibinfo{pages}{1243--1256}.
%Type = Article
\bibitem[{Mann and Haykin(1995)}]{mann1995chirplet}
\bibinfo{author}{S.~Mann}, \bibinfo{author}{S.~Haykin},
\newblock \bibinfo{title}{The chirplet transform: Physical considerations},
\newblock \bibinfo{journal}{IEEE Trans. Signal Proc.} \bibinfo{volume}{43} (\bibinfo{year}{1995}) \bibinfo{pages}{2745--2761}.
%Type = Book
\bibitem[{Stankovi{\'c} et~al.(2013)Stankovi{\'c}, Dakovi{\'c}, and Thayaparan}]{stankovic2013time}
\bibinfo{author}{L.~Stankovi{\'c}}, \bibinfo{author}{M.~Dakovi{\'c}}, \bibinfo{author}{T.~Thayaparan}, \bibinfo{title}{Time-Frequency Signal Analysis with Applications}, \bibinfo{publisher}{Artech house}, \bibinfo{year}{2013}.
%Type = Article
\bibitem[{Averbuch and Shkolnisky(2004)}]{averbuch20043d}
\bibinfo{author}{A.~Averbuch}, \bibinfo{author}{Y.~Shkolnisky},
\newblock \bibinfo{title}{3{D} discrete {X}-ray transform},
\newblock \bibinfo{journal}{Appl. Comput. Harmon. Anal.} \bibinfo{volume}{17} (\bibinfo{year}{2004}) \bibinfo{pages}{259--276}.
%Type = Article
\bibitem[{Stankovi{\'c}(2001)}]{stankovic2001measure}
\bibinfo{author}{L.~Stankovi{\'c}},
\newblock \bibinfo{title}{A measure of some time--frequency distributions concentration},
\newblock \bibinfo{journal}{Signal Proc.} \bibinfo{volume}{81} (\bibinfo{year}{2001}) \bibinfo{pages}{621--631}.
%Type = Article
\bibitem[{Thakur et~al.(2013)Thakur, Brevdo, Fu{\v{c}}kar, and Wu}]{thakur2013synchrosqueezing}
\bibinfo{author}{G.~Thakur}, \bibinfo{author}{E.~Brevdo}, \bibinfo{author}{N.~S. Fu{\v{c}}kar}, \bibinfo{author}{H.-T. Wu},
\newblock \bibinfo{title}{The synchrosqueezing algorithm for time-varying spectral analysis: Robustness properties and new paleoclimate applications},
\newblock \bibinfo{journal}{Signal Proc.} \bibinfo{volume}{93} (\bibinfo{year}{2013}) \bibinfo{pages}{1079--1094}.
%Type = Article
\bibitem[{Carmona et~al.(1997)Carmona, Hwang, and Torr{\'e}sani}]{carmona1997characterization}
\bibinfo{author}{R.~A. Carmona}, \bibinfo{author}{W.-L. Hwang}, \bibinfo{author}{B.~Torr{\'e}sani},
\newblock \bibinfo{title}{Characterization of signals by the ridges of their wavelet transforms},
\newblock \bibinfo{journal}{IEEE Trans. Signal Proc.} \bibinfo{volume}{45} (\bibinfo{year}{1997}) \bibinfo{pages}{2586--2590}.
%Type = Article
\bibitem[{Chen et~al.(2017)Chen, Dong, Xing, Peng, Zhang, and Meng}]{chen2017separation}
\bibinfo{author}{S.~Chen}, \bibinfo{author}{X.~Dong}, \bibinfo{author}{G.~Xing}, \bibinfo{author}{Z.~Peng}, \bibinfo{author}{W.~Zhang}, \bibinfo{author}{G.~Meng},
\newblock \bibinfo{title}{Separation of overlapped non-stationary signals by ridge path regrouping and intrinsic chirp component decomposition},
\newblock \bibinfo{journal}{IEEE Sensors Journal} \bibinfo{volume}{17} (\bibinfo{year}{2017}) \bibinfo{pages}{5994--6005}.
%Type = Article
\bibitem[{He et~al.(2021)He, Jiang, Hu, and Li}]{he2021local}
\bibinfo{author}{Y.~He}, \bibinfo{author}{Z.~Jiang}, \bibinfo{author}{M.~Hu}, \bibinfo{author}{Y.~Li},
\newblock \bibinfo{title}{Local maximum synchrosqueezing chirplet transform: An effective tool for strongly nonstationary signals of gas turbine},
\newblock \bibinfo{journal}{IEEE Trans. Instrum. Meas.} \bibinfo{volume}{70} (\bibinfo{year}{2021}) \bibinfo{pages}{1--14}.
%Type = Article
\bibitem[{Laurent and Meignen(2021)}]{laurent2021novel}
\bibinfo{author}{N.~Laurent}, \bibinfo{author}{S.~Meignen},
\newblock \bibinfo{title}{A novel ridge detector for nonstationary multicomponent signals: Development and application to robust mode retrieval},
\newblock \bibinfo{journal}{IEEE Trans. Signal Proc.} \bibinfo{volume}{69} (\bibinfo{year}{2021}) \bibinfo{pages}{3325--3336}.
%Type = Article
\bibitem[{Meignen and Colominas(2023)}]{meignen2023new}
\bibinfo{author}{S.~Meignen}, \bibinfo{author}{M.~A. Colominas},
\newblock \bibinfo{title}{A new ridge detector localizing strong interference in multicomponent signals in the time-frequency plane},
\newblock \bibinfo{journal}{IEEE Trans. Signal Proc.} \bibinfo{volume}{71} (\bibinfo{year}{2023}) \bibinfo{pages}{3413--3425}.
%Type = Article
\bibitem[{Su et~al.(2024)Su, Liu, Sheu, and Wu}]{su2024ridge}
\bibinfo{author}{Y.-W. Su}, \bibinfo{author}{G.-R. Liu}, \bibinfo{author}{Y.-C. Sheu}, \bibinfo{author}{H.-T. Wu},
\newblock \bibinfo{title}{Ridge detection for nonstationary multicomponent signals with time-varying wave-shape functions and its applications},
\newblock \bibinfo{journal}{IEEE Trans. Signal Proc.}  (\bibinfo{year}{2024}).
%Type = Article
\bibitem[{Zhang et~al.(2022)Zhang, Liu, Tan, Yang, and Zhang}]{zhang2022two}
\bibinfo{author}{R.~Zhang}, \bibinfo{author}{X.~Liu}, \bibinfo{author}{Y.~Tan}, \bibinfo{author}{X.~Yang}, \bibinfo{author}{L.~Zhang},
\newblock \bibinfo{title}{Two dimensional local maximum synchroextracting chirplet transfrom and application of characterizing micro--{D}oppler signals},
\newblock \bibinfo{journal}{Signal Proc.} \bibinfo{volume}{198} (\bibinfo{year}{2022}) \bibinfo{pages}{108598}.
%Type = Article
\bibitem[{Hartono et~al.(2019)Hartono, Halim, and Roberts}]{hartono2019gear}
\bibinfo{author}{D.~Hartono}, \bibinfo{author}{D.~Halim}, \bibinfo{author}{G.~W. Roberts},
\newblock \bibinfo{title}{Gear fault diagnosis using the general linear chirplet transform with vibration and acoustic measurements},
\newblock \bibinfo{journal}{Journal of Low Frequency Noise, Vibration and Active Control} \bibinfo{volume}{38} (\bibinfo{year}{2019}) \bibinfo{pages}{36--52}.

\end{thebibliography}

\end{document}